\documentclass[%
 aip,
 amsmath,amssymb,
reprint, onecolumn 
]{revtex4-1}

\usepackage[T1]{fontenc}
\usepackage{mathptmx}
\usepackage{etoolbox}

\usepackage{amssymb} 
\usepackage{bm}

\usepackage{graphicx}
\usepackage{dcolumn}
\usepackage{multirow}
\usepackage{array} 
\usepackage{tabularx}
\usepackage{caption}
\usepackage{subcaption}
\usepackage{threeparttable}
\usepackage{booktabs}
\usepackage{adjustbox}

\usepackage{tikz}
\usepackage{pgfplots}
\usetikzlibrary{external, arrows.meta}
\usepgfplotslibrary{fillbetween}
\pgfplotsset{compat=1.9}
\pgfplotsset{
    every axis/.append style={
        line width=0.6pt,
    },
    mygridstyle/.style={
        grid style=solid,
        line width=0.35pt,
        color=gray!50,
    },
}

\usepackage{CJKutf8}

\usepackage{hyperref}
\usepackage{xcolor}
\usepackage{cleveref}
\usepackage{mathrsfs}

\usepackage[normalem]{ulem}
\newcolumntype{L}{>{\raggedright\arraybackslash}X} 
\newcolumntype{C}{>{\centering\arraybackslash}X}   
\definecolor{hcorange}{RGB}{255,127,80}
\definecolor{hcblue}{RGB}{30,144,255}
\definecolor{hcgreen}{RGB}{44, 230, 44}
\definecolor{hcpurple}{RGB}{229,229,255}
\definecolor{hcpurple1}{RGB}{155,110,188}
\definecolor{hcred}{RGB}{130,118,213}
\definecolor{hc03}{RGB}{255,99,72}
\definecolor{hc15}{RGB}{47,53,66}

\crefname{equation}{Eq.}{Eqs.}
\Crefname{equation}{Eq.}{Eqs.}
\crefname{figure}{Fig.}{Figs.}
\Crefname{figure}{Fig.}{Figs.}
\crefname{algorithm}{Algorithm}{Algorithms}
\Crefname{algorithm}{Algorithm}{Algorithms}
\crefname{section}{Sec.}{Sections}
\Crefname{section}{Sec.}{Sections}
\crefname{table}{Tab.}{Tables.}
\Crefname{table}{Tab.}{Tables.}

\makeatletter
\def\@email#1#2{%
 \endgroup
 \patchcmd{\titleblock@produce}
  {\frontmatter@RRAPformat}
  {\frontmatter@RRAPformat{\produce@RRAP{*#1\href{mailto:#2}{#2}}}\frontmatter@RRAPformat}
  {}{}
}%
\makeatother

\begin{document}

\preprint{AIP/123-QED}

\title[Optimization Synthetic Jet Actuator Parameters.]{Effect of Synthetic Jets Actuator Parameters on Deep Reinforcement Learning-Based Flow Control Performance in a Square Cylinder}

\author{Wang Jia (\begin{CJK*}{UTF8}{gbsn}贾旺\end{CJK*})}
\author{Hang Xu (\begin{CJK*}{UTF8}{gbsn}徐航\end{CJK*})}
\email{hangxu@sjtu.edu.cn}
\affiliation{School of Ocean and Civil Engineering, Shanghai Jiao Tong University, Shanghai, 200240, China}

\date{\today}
        
\begin{abstract}

We conduct an active flow control (AFC) study on the mass flow rate of synthetic jets on the upper and lower surfaces of a square cylinder using a deep reinforcement learning (DRL) algorithm, with a focus on investigating the influence of the position and width of the synthetic jets on the flow control performance.
At Reynolds numbers ($Re$) of 100 and 500, it is found that our proposed method significantly reduced the lift and drag coefficients of the square cylinder, and completely suppressed vortex shedding in the wake.
In particular, at $Re=100$, placing the synthetic jets near the tail corner was beneficial for reducing drag, with a maximum drag reduction rate of 14.4\%. 
When $Re=500$, positioning the synthetic jets near the leading edge corner resulted in a maximum optimal drag reduction effect of 65.5\%. 
This indicates that locating the synthetic jet at the main flow separation point can achieve optimal control.
Furthermore, we notice that when the synthetic jets are positioned near the tail corner, vortex shedding can be completely suppressed.
Additionally, a narrower width of the synthetic jets can enhance flow instability and increase the cost of flow control.

\end{abstract}

\maketitle

\section{INTRODUCTION}\label{sec:introduction}

In the backdrop of advancements in sophisticated algorithms, computational hardware, open-source software and the integration of vast amounts of data in the era of big data, scientific inquiry is transitioning from first principles to data-driven methods.\cite{DataDriven,Datadrivenscience}
The field of fluid dynamics is also benefiting from this trend, with machine learning offering sophisticated algorithms capable of handling high-dimensional, large-scale fluid dynamics data,\cite{annurevfluid} establishing mathematical modeling frameworks, and providing integrated, modular packages and libraries as open-source resources.\cite{Brunton2015,Stevendynamical}
Machine learning excels in tasks like feature extraction, modal analysis, and data mining, providing techniques for extracting insights from datasets.\cite{mahesh2020machine,Datadrivenscience}
Experimental fluid dynamics measurements and numerical simulation generate extensive datasets, making them ideal candidates for techniques capable of handling high-dimensional, large-scale data.\cite{SCOTTCOLLIS2004237,DataDriven}
Furthermore, machine learning algorithms can embed physical information from the field of fluid dynamics and adaptively perform tasks related to execution, control, and optimization.\cite{Multifidelity,Bewley2001}
The field of flow control is rapidly advancing, propelled by the development of state-of-the-art machine learning algorithms and unprecedented volumes of data from high-precision numerical simulations across multiple spatiotemporal scales.\cite{annurevfluid,Multifidelity,Bewley2001}

Active flow control (AFC) is an advanced fluid management strategy aimed at enhancing fluid properties and performance through proactive interventions in the flow field. By introducing energy or momentum, AFC alters the fluid's natural state, employing techniques such as fluid injection and suction, synthetic jets, and electromagnetic control.\cite{annurevCattafesta,Aram2018}
Synthetic jet technology produces jets by cyclically inhaling and expelling fluid from a fixed location, requiring no external energy source.\cite{annurevfluidGlezer,SCOTTCOLLIS2004237} 
This method offers several advantages, including independence from external fluid sources, high controllability, low energy consumption, and suitability for operation in narrow or complex spaces.\cite{jahanmiri2010active,Smith1998} 
These advantages render synthetic jets particularly promising tools for present and future flow control technologies.\cite{MAL071,Aerospace2023} AFC is widely applied across various sectors, including aerospace, automotive, energy production, and environmental engineering, where it serves to reduce drag, control vortices, enhance lift, and stabilize flows.\cite{SCOTTCOLLIS2004237,smith2003comparison}
Despite its considerable advantages, the implementation and practical application of AFC face challenges due to the system's strong nonlinearity, high dimensionality, and time delays. Recent advancements in machine learning algorithms and control theory are propelling progress in adaptive, real-time intelligent control, facilitating more effective management of complex flow dynamics and further augmenting the capabilities of AFC technologies.\cite{Brunton2015,Datadrivenscience,Stevendynamical}

Deep Reinforcement Learning (DRL) merges the capabilities of Deep Learning (DL) for processing high-dimensional data with the decision-making and policy optimization strengths of Reinforcement Learning (RL).\cite{mnih2013playing,lecun2015deep} 
DRL is recognized for its ability to learn and adapt autonomously, enabling end-to-end learning that is highly applicable across complex environments.
DRL has demonstrated exceptional performance in a wide range of domains, including game AI, robotic control, autonomous driving, resource management and scheduling, and financial trading. It is capable of autonomously learning optimal decision-making strategies in complex environments, surpassing human-level capabilities.\cite{mnih2015human,lillicrap2019continuous,henderson2019deep}
DRL is a sophisticated blend of DL and RL, two powerful branches of machine learning. 
It combines the feature extraction capabilities of DL for complex high-dimensional data with the decision-making abilities of RL through interactions with the environment.\cite{li2018deep,kaiser2024modelbased}
This synergy allows DRL to excel in environments that require both advanced perception and precise control. 
Despite challenges related to sample efficiency and stability, DRL demonstrates remarkable potential on a variety of challenging decision-making problems.\cite{8103164,kaiser2024modelbased}

Given the significant potential of DRL in control and optimization, a large number of scholars have been applying it in the field of flow control.
\citeauthor{rabault2019artificial}\cite{rabault2019artificial} pioneered the application of DRL to AFC technologies, achieving an 8\% reduction in drag around a cylinder at $Re=100$. 
This seminal work sparked interest in applying DRL techniques within the field of fluid dynamics.
\citeauthor{tang2020robust}\cite{tang2020robust} focused on the robustness across different Reynolds numbers, while 
\citeauthor{parisRobustFlowControl2021}\cite{parisRobustFlowControl2021} explored the impact of probe distribution on control performance. 
\citeauthor{liReinforcementlearning}\cite{liReinforcementlearning} applied DRL-based AFC to cylinders under various blockage ratios, integrating flow physics mechanisms during the control process to achieve significant outcomes. 
\citeauthor{Dixiapnas}\cite{Dixiapnas} developed an active flow control strategy based on DRL through experimental fluid dynamics, demonstrating the effectiveness of applying reinforcement learning in experimental fluid mechanics.
\citeauthor{rabault2019accelerating}\cite{rabault2019accelerating} and \citeauthor{jia2024optimal}\cite{jia2024optimal} addressed issues related to parallel strategies and training when coupling Computational Fluid Dynamics (CFD) codes with DRL framework in parallel computing environments.
\citeauthor{wangDRLinFluids}\cite{wangDRLinFluids} developed an open source library that integrated CFD code \texttt{OpenFoam} with DRL framework \texttt{TensorForce}. 
DRL demonstrates excellent control performance in the field of AFC, sparking increasing exploration. 
There are applications from low to relatively high Reynolds numbers, transitions from 2D to 3D configurations, shifts from numerical simulations to fluid experiments, and geometric changes in the body of interest from circular to square bodies.
While there is a plethora of related work, we will not delve into each individually. 
Instead, we summarize in \cref{tab:tab1} some relevant research on the application of DRL-based AFC methods in the context of flow around bluff bodies.

\begin{table}[ht]
\centering
\caption{Summary of studies on AFC strategies for various bluff bodies.}
\label{tab:tab1}
\vspace{-\baselineskip}
\begin{tabularx}{\textwidth}{
  >{\centering\arraybackslash}p{0.1\linewidth}
  >{\centering\arraybackslash}p{0.1\linewidth}
  >{\centering\arraybackslash}p{0.2\linewidth}
  >{\centering\arraybackslash}p{0.12\linewidth}
  >{\centering\arraybackslash}p{0.12\linewidth}
  >{\centering\arraybackslash}p{0.1\linewidth}
  >{\centering\arraybackslash}p{0.1\linewidth}
  >{\centering\arraybackslash}p{0.08\linewidth}
}
\toprule
\toprule
Re & Bluff Body & Reference & Strategy & Control Algorithm & Solver & Drag Reduction & Vortex Suppressed \\
\midrule
100 & Cylinder & \citeauthor{rabault2019artificial}\cite{rabault2019artificial} & Synthetic Jets & PPO & \texttt{FeniCS} & 8\% & -  \\
100 & Cylinder & \citeauthor{wangDRLinFluids}\cite{wangDRLinFluids} & Synthetic Jets & PPO & \texttt{OpenFOAM} & 8\% & -  \\
100 & Cylinder & \citeauthor{castellanos}\cite{castellanos} & Synthetic Jets & LGPC/PPO & \texttt{FeniCS} & 8\% & -  \\
100 & Cylinder & \citeauthor{jia2024deep}\cite{jia2024deep} & Synthetic Jets & PPO & \texttt{OpenFOAM} & 8\% & YES \\
100 & Cylinder & \citeauthor{ren2021bluff}\cite{ren2021bluff} & WSLB & PPO & \texttt{LBM} & - & - \\
120 & Cylinder & \citeauthor{parisRobustFlowControl2021}\cite{parisRobustFlowControl2021} & Synthetic Jets &S-PPO-CMA & \texttt{FastS} & 18.4\% & - \\
$\leq$200 & Cylinder & \citeauthor{liReinforcementlearning}\cite{liReinforcementlearning} & Synthetic Jets & PPO & \texttt{Nek5000} & - & YES \\
100-300 & Cylinder & \citeauthor{he2023policy}\cite{he2023policy} & Synthetic Jets & PPO & \texttt{OpenFOAM} & 6\%-24\% & - \\
100-400 & Cylinder & \citeauthor{tangRobust}\cite{tangRobust} & Synthetic Jets & PPO & \texttt{FeniCS} & 6\%-39\% & - \\
1,000 & Cylinder & \citeauthor{renApplying}\cite{renApplying} & Synthetic Jets & PPO & \texttt{LBM} & 30\% & - \\
10,160 & Cylinder & \citeauthor{Dixiapnas}\cite{Dixiapnas} & Cylinders & TD3 & \texttt{Lilypad} & 30\% & - \\
100 & Ellipse & \citeauthor{jia2024deep}\cite{jia2024deep} & Synthetic Jets & PPO & \texttt{OpenFOAM} & 16\% & YES \\
100 & Square  & \citeauthor{wangDRLinFluids}\cite{wangDRLinFluids} & Synthetic Jets & SAC & \texttt{OpenFOAM} & 14\% & - \\
100 & Square  & \citeauthor{chen2023deep}\cite{chen2023deep} & Synthetic Jets & SAC & \texttt{OpenFOAM} & 14\% & YES \\
100 & Square  & \citeauthor{Xia2024}\cite{Xia2024} & Synthetic Jets & SAC and TQC  & \texttt{FEniCS} & 17.3\% & - \\
100-400 & Square  & \citeauthor{jia2024robust}\cite{jia2024robust} & Synthetic Jets & SAC & \texttt{OpenFOAM} & 14\%-47\% & YES \\
500-2,000 & Square  & \citeauthor{yan2023stabilizing}\cite{yan2023stabilizing} & Synthetic Jets & SAC & \texttt{OpenFOAM} & 44\%-61\% & - \\
\bottomrule
\bottomrule
\end{tabularx}
\end{table}

The flow around a square cylinder exhibits instabilities similar to those around a circular cylinder, but with different separation mechanisms and variations in lift, drag, and Strouhal number with Reynolds number. For square cylinders, the separation points are fixed at the leading or trailing edge, depending on the Reynolds number.\cite{Sohankar} 
The location of these separation points is crucial for guiding active flow control strategies.
Synthetic jets control flow separation by periodically injecting or extracting fluid into or from the boundary layer, directly altering its momentum distribution.\cite{annurevCattafesta,YOU20081349} This enhances the energy within the boundary layer, delays the separation point, and prevents flow separation. The effectiveness of synthetic jets depends largely on parameters such as amplitude, frequency, and application location.\cite{SCOTTCOLLIS2004237,Aram2018} Extensive parameter studies are necessary to optimize control performance.
The location of the synthetic jet is particularly crucial. Selecting the appropriate location maximizes the jet's impact on the boundary layer, delays the separation point, and improves control performance. By optimizing the amplitude and frequency of the jet and selecting the best application location, optimal flow control can be achieved, significantly enhancing aerodynamic performance and system stability.

\citeauthor{yan2023stabilizing}, \citeauthor{chen2023deep} and \citeauthor{yan2024aero}\cite{yan2023stabilizing,chen2023deep,yan2024aero} optimized the position of synthetic jets to maximize control effectiveness.
Particularly, two studies have reached starkly different conclusions, prompting further reflection and analysis by researchers. 
\citeauthor{chen2023deep}\cite{chen2023deep} utilized DRL-based AFC to mitigate vibrations caused by lift on a square cylinder at $Re = 100$ and investigated the impact of synthetic jet positioning on control performance at this Reynolds number. Their findings indicate that when synthetic jets are positioned near the leading edge, the drag coefficient is reduced by 7.2\% compared to the baseline flow. However, when positioned near the trailing edge, the drag coefficient is reduced by 13.3\%. 
Therefore, at $Re = 100$, placing synthetic jets near the trailing corners of the square cylinder is more advantageous for achieving higher drag reduction and lift coefficient reduction.
\citeauthor{yan2023stabilizing}\cite{yan2023stabilizing} used DRL and zero-flux jet actuators based on CFD to reduce drag and lift on a square cylinder at $Re = 500$, $1,000$, and $2,000$. At $Re = 500$, when synthetic jets are positioned near the leading corners, the drag coefficient is reduced by 44.4\% compared to the baseline flow. When positioned near the trailing corners, the drag coefficient is reduced by only 1.4\%. The results show that placing jets near the leading corners yields better drag reduction than near the trailing corners. Similar conclusions were drawn at $Re = 1,000$ and $ Re=2,000 $. 
We are curious whether these differing conclusions are due to the different Reynolds numbers or differences in the design of the DRL-based AFC framework.

Given the recent advancements in DRL-based AFC for flow around square cylinders, there is a clear lack of parameterization studies on synthetic jets to reconcile the discrepancies in control effectiveness reported in the literature. 
Beyond the aforementioned qualitative differences in drag reduction due to synthetic jets positioning at varying Reynolds numbers, we also note that \citeauthor{chen2023deep}\cite{chen2023deep} achieved both drag reduction and complete suppression of vortex shedding in the wake of a cylinder at $Re = 100$. However, in \citeauthor{yan2023stabilizing}\cite{yan2023stabilizing}, even at Reynolds numbers as high as 500 or more, the control strategies failed to suppress vortex shedding.
To determine whether these divergent conclusions arise from differences in Reynolds numbers or variations in DRL-based AFC framework designs, we employ an AFC framework controlled by a SAC agent, testing at consistent Reynolds numbers. 
This study will conduct a more extensive parametric analysis using two synthetic jet arrangements and four synthetic jet widths to evaluate the sensitivity of these jet parameters to flow control performance.
This will help us gain a comprehensive understanding of how changes in jet parameters affect the effectiveness of flow control strategies.

The structure of this paper is as follows:
In \cref{sec:Methodology}, we detail the numerical simulation methods for flow around a square cylinder, the DRL algorithm, and how the numerical simulation is used as an environment for DRL training to build the DRL-based AFC framework.
In \cref{sec:results}, we describe the research results in detail.
First, we describe the flow patterns of the Baseline and Base flows. 
Then, we discuss the convergence performance of the DRL algorithm using CFD as the environment, and how the flow field stability around the square cylinder is enhanced during training.
Finally, we analyze the effects of synthetic jets positioning and width on flow control performance at $Re = 100$ and $Re = 500$. We delve into the flow control mechanisms of synthetic jets design near the leading and trailing corners of the square cylinder and explain the physical mechanisms underlying the qualitative conclusions of this study.
In \cref{sec:Conclusions}, we summarize the main conclusions of this research and discuss the implications of these findings for the field of flow control.

\section{PROBLEM DESCRIPTION AND METHODOLOGY}\label{sec:Methodology} 

\begin{figure*}[htbp]
    \centering
    \includegraphics{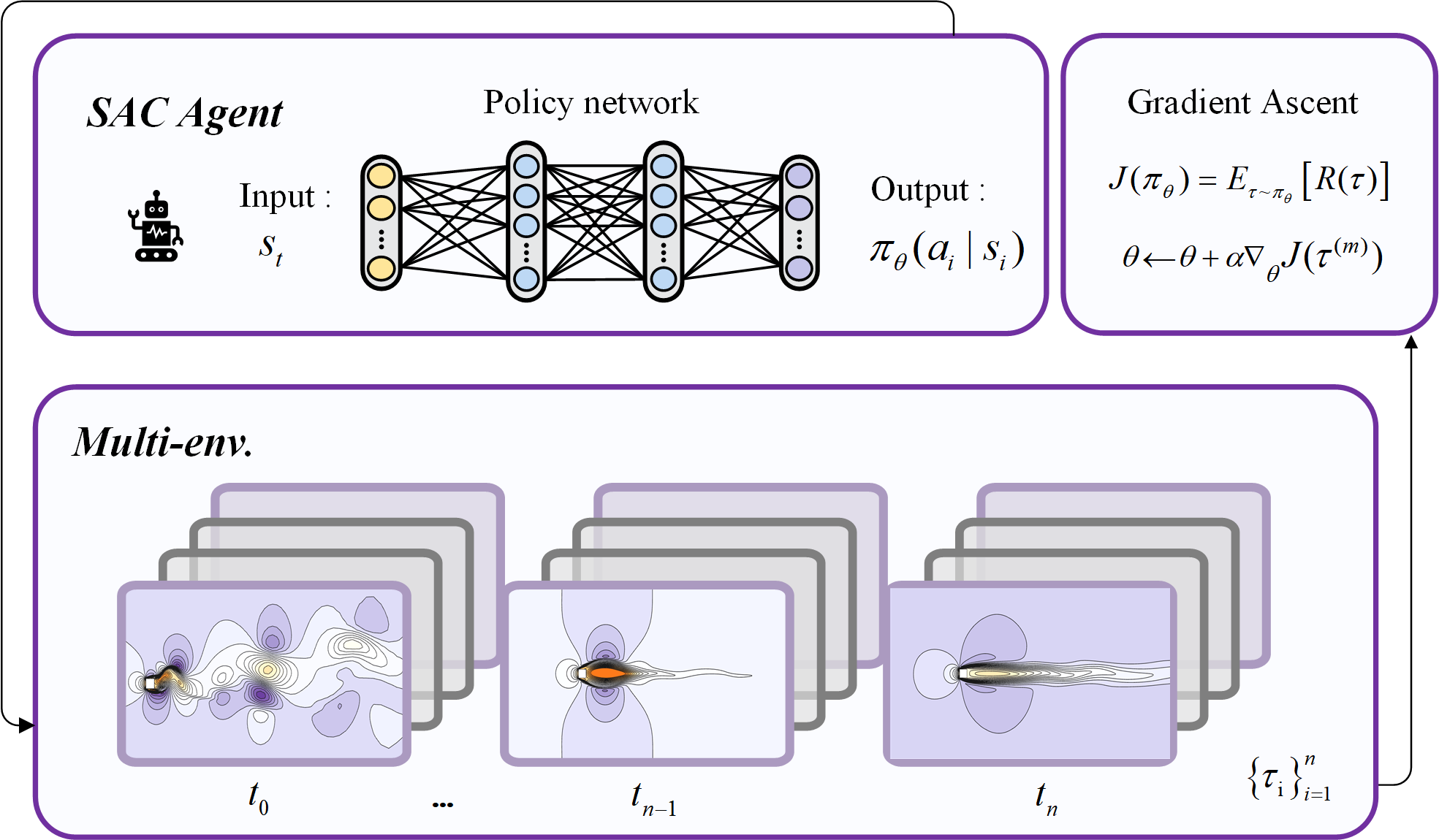}
    \caption{A multi-environment parallel DRL training framework, in which the agent adopts the SAC type and the environment is set to numerical simulation environment. The parallel computation across $n$ environments yields a set of trajectories $\{\tau_i\}_{i=1}^{n}$, with each trajectory $\tau_i$ comprising the sequence of states, actions, and rewards as $\tau_i = \{(s_0, a_0), (s_1, a_1, r_1), \ldots, (s_T, a_T, r_T)\}$, where $(s_t, a_t, r_t)$ represents the state, action , and reward at time-step $t$, respectively, for the $i$-th environment. The parameters $\theta$ of the policy network are updated using gradient ascent, i.e., $\theta \leftarrow \theta + \alpha \nabla_\theta J(\pi_\theta)$, where $\alpha$ denotes the learning rate.
    }
    \label{fig:figure1}
\end{figure*}

We develop a DRL framework that enables interaction between the agent and the numerical simulation environment, validated in flow regimes characterized by two-dimensional vortex shedding. 
In our DRL framework, the agent observes the current state of the CFD environment and interacts with it to adjust the mass flow rates of two synthetic jets on the square cylinder, based on feedback from reward signals. 
Through an iterative optimization process, the SAC agent gradually approaches the optimal flow control strategy by learning and trial-and-error, thereby achieving multiple control objectives such as reducing lift and drag coefficients and suppressing vortex shedding. The specific operational process is depicted in \cref{fig:figure1}.
\Cref{sec:cfd} provides detailed information about the flow environment.
\Cref{sec:ML} elaborates on the ML techniques discussed in this paper, including reinforcement learning, deep learning, and deep reinforcement learning.
\Cref{sec:DRL-Enhanced AFC} introduces the key components for performing active flow control using DRL methods, as well as the software and hardware configurations relevant to training.

\subsection{Flow environment}\label{sec:cfd}

\begin{figure*}[ht]
\centering
\includegraphics{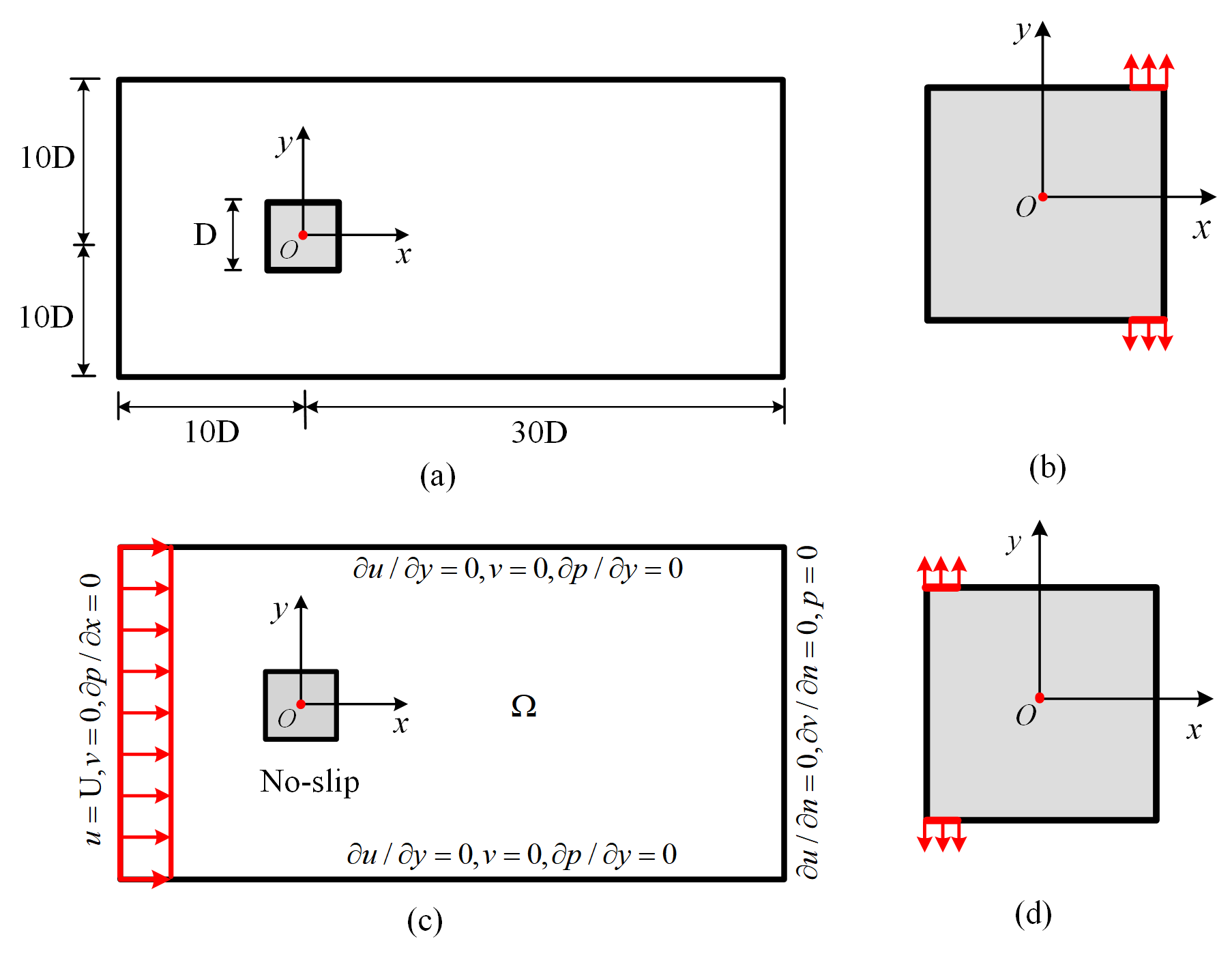}
\caption{A schematic of the computational domain and boundary conditions. (a) Establishment of the coordinate system and detailed dimensions of the computational domain. Note that while the dimensions of the square cylinder and computational domain are accurately labeled, the relative proportions are not to scale and are intended for illustrative purposes only. (b) Placement of the synthetic jets near the trailing corner point. (c) Specification of the boundary conditions for the computational domain. (d) Location of the synthetic jets near the leading corner points.}
\label{fig:figure2}
\end{figure*}

In the present study, we analyze the flow past a two-dimensional square cylinder, as illustrated in \Cref{fig:figure2}. To construct the physical model, we establish a Cartesian coordinate system with the center of the square cylinder as the origin. The flow direction is defined as the positive $x$-axis, while the direction perpendicular to the flow is designated as the positive $y$-axis. The square cylinder has a side length denoted by $D$.
The computational domain extends $30 D$ downstream from the center of the square cylinder to define the outlet boundary, and $10 D$ upstream to establish the inlet. The lateral boundaries extend $10 D$ from the center of the square cylinder on both the left and right sides. This configuration results in a rectangular computational domain with dimensions of $40 D$ in the streamwise direction and 20$D$ in the spanwise direction.

To comprehensively explore the impact of synthetic jets arrangement on control performance, we address two key aspects: the placement and width of the synthetic jets.
Given that the flow separation points around the square cylinder are fixed, occurring at the leading and trailing edge corners, we strategically position the synthetic jets at these corner points in our study. 
Specifically, placement variations include positions near both the leading and trailing edges of the square cylinder. Furthermore, the widths of the synthetic jets are designed to encompass $D/20$, $D/25$, $D/30$, and $D/35$, where $D$ represents the side length of the square cylinder. By systematically varying these parameters, we aim to provide a detailed analysis of their effects on flow control performance.
The synthetic jets on the upper and lower walls of the square cylinder are set up with a uniform velocity distribution, where the magnitude of the jet velocity represents non-dimensional mass flow rates, and the direction of the jet velocity is along the outward normal vector. The total mass flow rate of the two jets is zero, which more accurately reflects real-world applications compared to adding or removing mass from the system. This ensures that any observed drag reduction can be attributed to the effectiveness of the flow control strategy, rather than to any propulsive effect. \Cref{fig:figure2} illustrates the schematic placement of the synthetic jets on the square cylinder, positioned near the leading or trailing edge corners.

The Navier-Stokes equations for an incompressible viscous fluid in a domain $\Omega \subset \mathbb{R}^{nd}$ over a time interval $(0, T)$ are fundamental in fluid dynamics. 
These equations describe how the velocity field $\mathbf{u} = \mathbf{u}(\mathbf{x}, t)$ and pressure field $p = p(\mathbf{x}, t)$ of the fluid evolve over time, where $\mathbf{x}$ represents the spatial coordinates and $t$ represents time.

\begin{subequations}
\begin{equation}
    \frac{\partial \mathbf{u}}{\partial t} + \mathbf{u} \cdot (\nabla \mathbf{u}) = -\nabla p + Re^{-1} \Delta \mathbf{u} \quad \text{on} \quad \Omega \times (0, T),
\end{equation}

\begin{equation}
    \nabla \cdot \mathbf{u} = 0 \quad \text{on} \quad \Omega \times (0, T).
\end{equation}
\end{subequations}

Here, $Re = \frac{\overline{U}D}{\nu}$ is the Reynolds number, where $\overline{U}$ is the mean velocity at the inlet and $\nu$ is the kinematic viscosity. 
Together, these equations articulate the fundamental behaviors of fluid flow, linking the dynamics of velocity changes to the principle of mass conservation in fluid mechanics.

In the simulation setup described, the inlet of the computational domain is characterized by a uniform velocity distribution. Specifically, the velocity at the inlet is set to a magnitude of 2, with the direction aligned along the $x$-$axis$.
The outlet of the computational domain adopts $Neumann$-$type$ boundary conditions to ensure that the stress vector is zero to simulate the natural flow behavior at infinite distance.
The upper and lower boundaries of the domain are designated as far-field boundary conditions, aimed at minimizing their impact on the flow around the square cylinder and approximating an unbounded fluid domain as closely as possible.
The synthetic jets are configured such that their velocity direction is perpendicular to the walls of the square cylinder, directed along the outward normal of the walls. 
Except for the locations of the synthetic jets, all other parts of the square cylinder's walls are treated with no-slip solid wall boundary conditions, ensuring that the fluid does not penetrate these surfaces and accurately reflecting the physical constraints imposed by solid boundaries on fluid flow.
In addition, The boundary conditions are marked in detail in \cref{fig:figure2}.

\begin{figure*}[ht]
    \centering
    \begin{subfigure}{0.465\textwidth}
    \includegraphics[width=\textwidth]{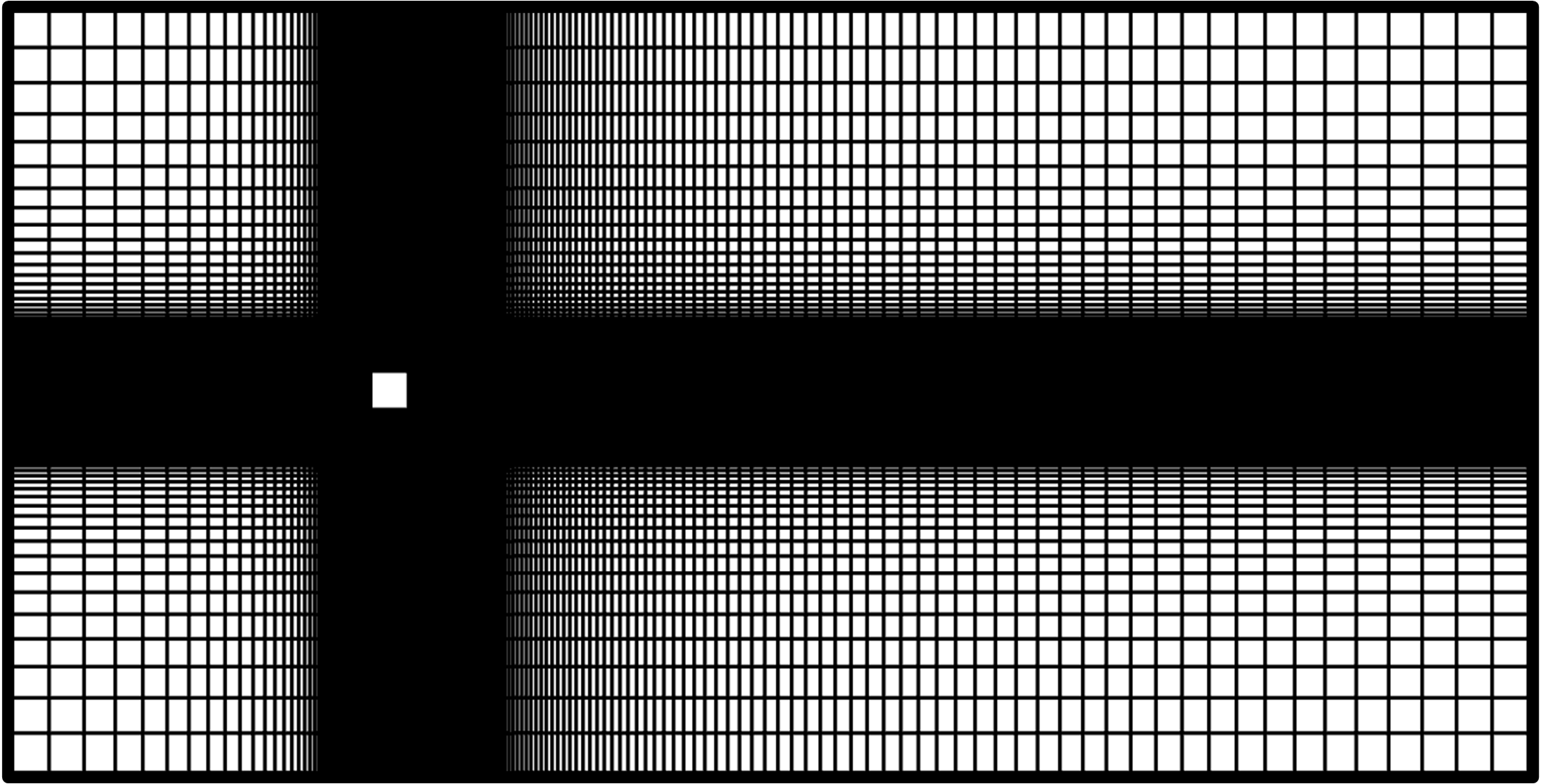}
    \caption{}
    \label{fig:figure301}
    \end{subfigure}
    \begin{subfigure}{0.26\textwidth}
    \includegraphics[width=\textwidth]{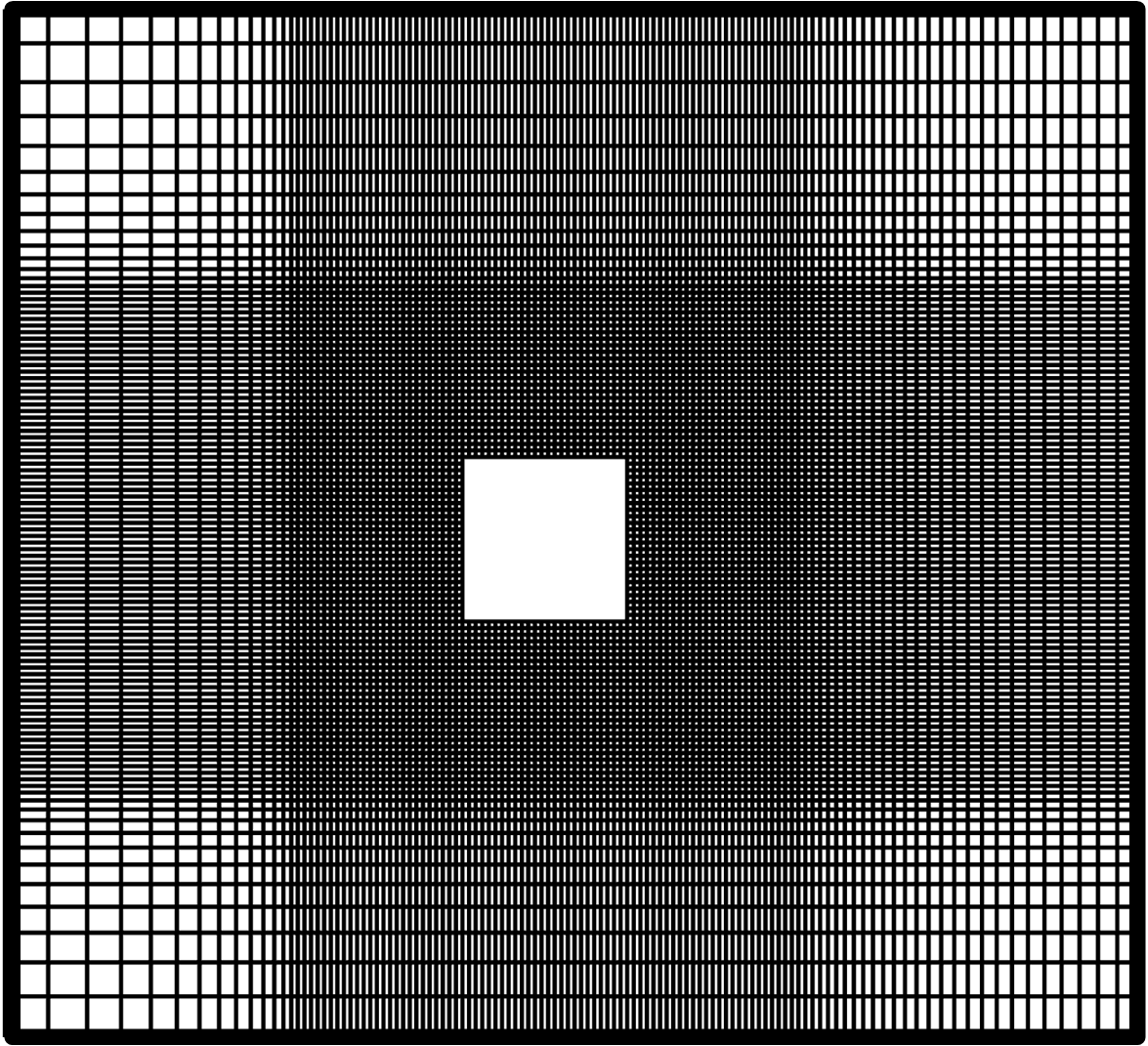}
    \caption{}
    \label{fig:figure302}
    \end{subfigure}
    \begin{subfigure}{0.26\textwidth}
    \includegraphics[width=\textwidth]{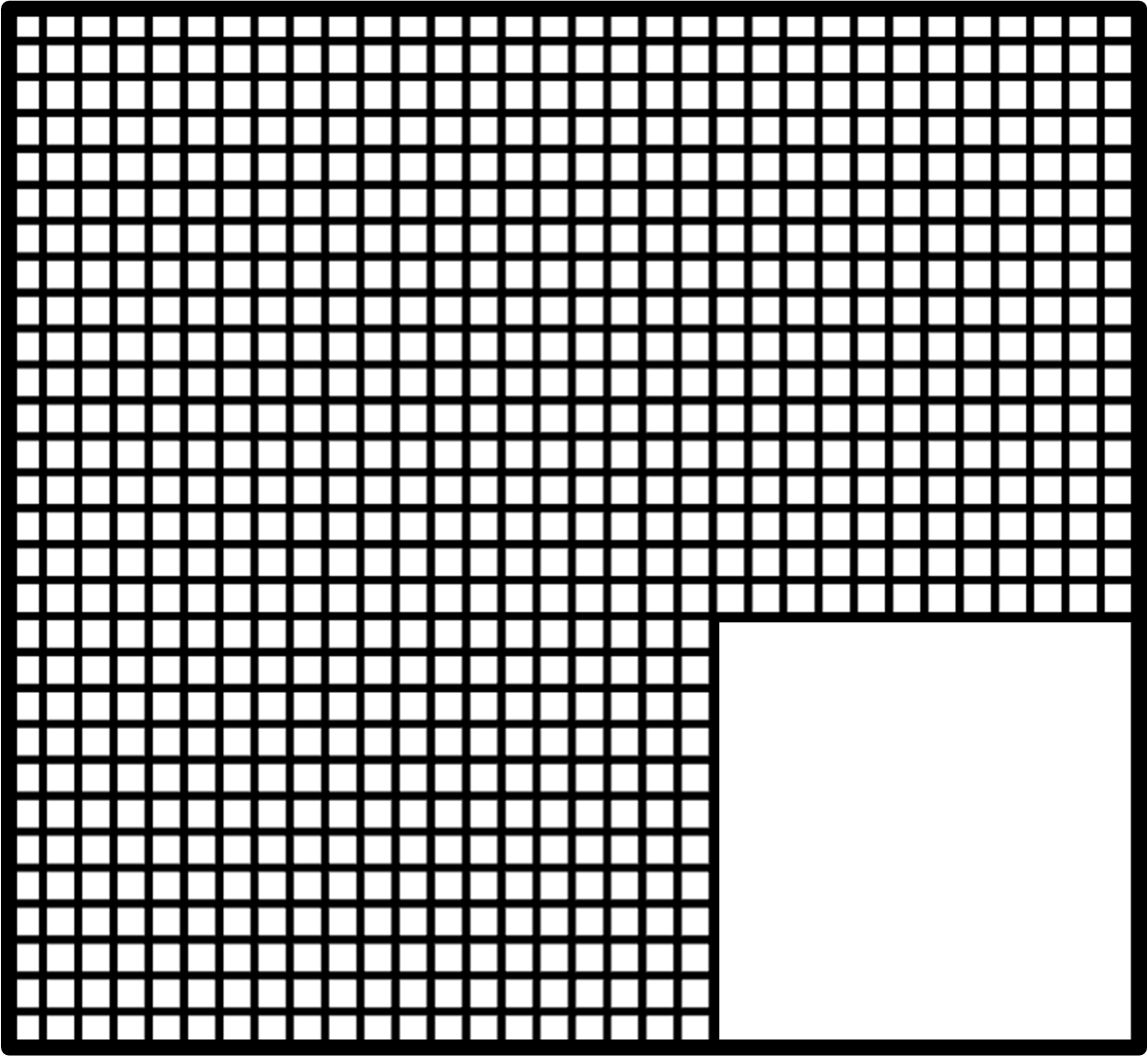}
    \caption{}
    \label{fig:figure303}
    \end{subfigure}  
    \caption{Computational domain discretization. (a) Discretization scheme covering the entire computational domain. (b) Magnified depiction of the grid structure close to the square cylinder. (c) Meshing of a quarter of the square cylinder.}
    \label{fig:figure3}
\end{figure*}

\cref{fig:figure3} illustrates the discretization of the computational domain into 23,264 quadrilateral mesh elements via a structured grid approach. 
\cref{fig:figure3}(a) provides a detailed depiction of the discretization across the entire computational domain, with finer mesh scales near the square cylinder and coarser mesh scales near the boundary walls.
\cref{fig:figure3}(b) exclusively demonstrates the transition of mesh scale from coarse to fine.
\cref{fig:figure3}(c) demonstrates the refinement of the mesh surrounding the square cylinder, where the quadrilateral mesh elements are approximately square, ensuring accurate capture of flow field structures in fluid dynamics calculations.

The bluff body drag coefficient ($C_D$) and lift coefficient ($C_L$) are defined as
\begin{align}
     C_D = \frac{F_D}{\frac{1}{2} \rho_\infty U_\infty^2 D},\quad C_L = \frac{F_L}{\frac{1}{2} \rho_\infty U_\infty^2 D}.
\end{align}
Where $F_D$ and $F_L$ are the drag and lift forces, defined as the surface integrals of the pressure and viscous forces on the bluff body with respect to the $x$-$axis$ and $y$-$axis$, respectively.
$\rho_\infty$ denotes the density of the fluid far from the square cylinder. 
$U_\infty$, the freestream velocity of the fluid, represents the undisturbed velocity far from any obstacles or boundaries, and it is equal to 2.
 
The Strouhal number $(St)$ is used to describe the characteristic frequency of oscillatory flow phenomena and is defined as follows:

\begin{equation}
St = \frac{f_s \cdot D}{U_\infty}.
\end{equation}
Where $f_s$ is the shedding frequency calculated based on the periodic evolution of the $C_L$ and $U_\infty=2$.

In the current study, the incompressible flow dynamics are analyzed using the open-source CFD toolkit, \texttt{OpenFOAM}, as outlined by Jasak \textit{et al.} \cite{jasakLibrary,jasak2009} \texttt{OpenFOAM} stands as a rigorously tested and extensively applied computational framework that offers a comprehensive suite of numerical algorithms for the resolution of the Navier-Stokes equations. The solver implemented within \texttt{OpenFOAM} employs the finite volume method for the discretization of the computational domain, subdividing it into a structured mesh of control volumes.
The numerical simulations use the \texttt{pimpleFoam} solver to solve the unsteady Navier-Stokes equations. The temporal discretization employs the \texttt{Backward} scheme, while the divergence discretization of the velocity field uses the \texttt{Gauss linear} scheme, and the gradient discretization of the velocity utilizes the \texttt{leastSquares} method. These discretization schemes are selected to ensure the requisite mathematical precision and stability for the simulation of fluid flow dynamics. 
To promote numerical stability within the simulations, the time step size is meticulously selected as $\Delta t = 0.0005$.

To verify the accuracy of the grid, this study conducted a grid convergence investigation. We compared results from three different grid scales and contrasted them with computational outcomes from other studies, summarizing the findings in \cref{tab:tab2}. The table lists the values for the mean drag coefficient ($C_{D, Mean}$), the standard deviation of the drag coefficient ($C_{D, Std}$), the standard deviation of the lift coefficient ($C_{L, Std}$), and the $S_t$ at  $Re = 100$ and $Re = 500$. All three grids utilized structured meshes, differing only in the level of refinement, with the number of grid cells detailed in the table. 
The computational results in \cref{tab:tab2} demonstrate that the differences among the three grid resolutions are insignificant and consistent with values obtained in existing literature.
In the subsequent advanced DRL training phase, the main grid scheme is employed to balance computational accuracy and cost-effectiveness, ensuring the required precision is achieved while minimizing computational overhead.

\begin{table}[htbp]
\centering
\caption{Grid independence test for a square cylinder.}
\vspace{-\baselineskip}
\begin{tabular}{
  >{\centering\arraybackslash}p{0.08\textwidth}
  >{\centering\arraybackslash}p{0.28\textwidth}
  >{\centering\arraybackslash}p{0.10\textwidth}
  >{\centering\arraybackslash}p{0.11\textwidth}
  >{\centering\arraybackslash}p{0.10\textwidth}
  >{\centering\arraybackslash}p{0.11\textwidth}
  >{\centering\arraybackslash}p{0.11\textwidth} 
}
\toprule
\toprule
Configuration & Case & Cell number & \(C_{D, Mean}\) & \(C_{D, Std}\) & \(C_{L, Std}\) & $S_t$ \\
\midrule   
         & {Coarse} & 16,400  & 1.471 & 2.023 & 0.196 & 0.148 \\
$Re=100$ & {Main} & 23,264  & 1.549 & 2.020 & 0.179  & 0.145 \\
         & {Fine}   & 28,450  & 1.533 & 1.910 & 0.173 & 0.145 \\
         & \citeauthor{sen2011flow}\cite{sen2011flow} & - & 1.530 & - & - & 0.145 \\     
         & \citeauthor{Sharma}\cite{Sharma} & - & 1.490 & - & - & 0.150 \\    
         & \citeauthor{Singh}\cite{Singh} & - & 1.510 & - & - & 0.147 \\     
\midrule     
         & {Coarse} & 16,400  & 2.070 & 0.481 & 1.164  & 1.130 \\         
$Re=500$ & {Main} & 23,264  & 2.060 & 0.441 & 1.173 & 0.127 \\
         & {Fine}   & 28,450  & 2.067 & 0.419 & 1.181  & 0.128 \\         
         & \citeauthor{Shear}\cite{Shear} & - & 2.050 & - & - & 0.133 \\        
         & \citeauthor{Dependence}\cite{Dependence} & - & 1.940 & - & - & 0.136 \\  
         & \citeauthor{Sohankar}\cite{Sohankar} & - & 1.870 & - & - & 0.126 \\           
\bottomrule
\bottomrule
\end{tabular}
\label{tab:tab2}
\end{table}

\subsection{Machine learning}\label{sec:ML} 

In machine learning, three main paradigms exist: supervised learning, unsupervised learning, and reinforcement learning. 
Supervised learning uses labeled examples to predict labels for unlabeled inputs. 
Unsupervised learning discovers patterns in unlabeled data. 
Reinforcement learning involves interacting with an environment to maximize cumulative rewards.

\begin{figure*}[ht]
    \centering
    \begin{minipage}{0.37\textwidth}
        \begin{subfigure}{\textwidth}
            \includegraphics[width=\textwidth]{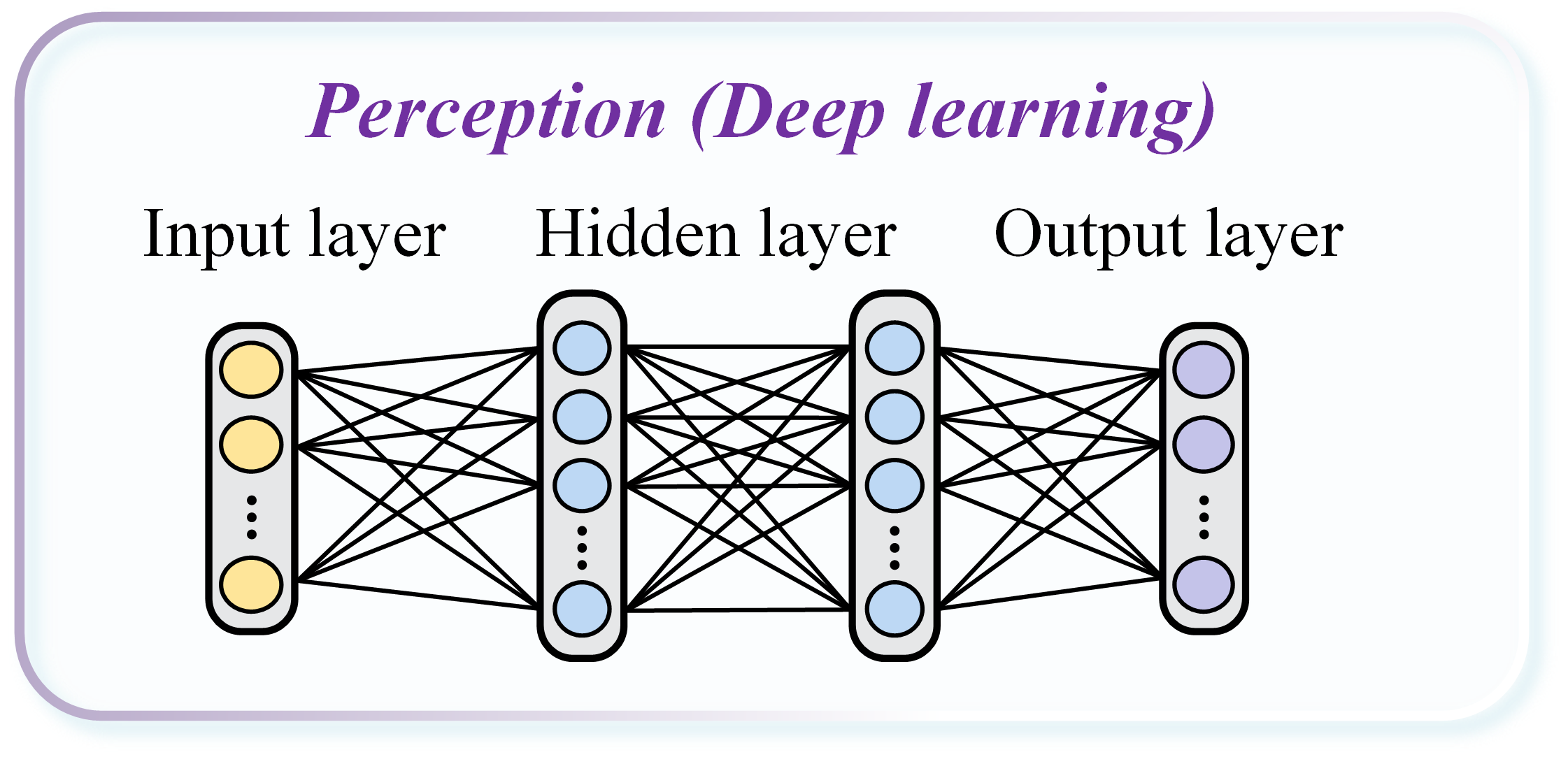}
            \caption{DL}
            \label{fig:figure401}
        \end{subfigure}\\ 
        \begin{subfigure}{\textwidth}
            \includegraphics[width=\textwidth]{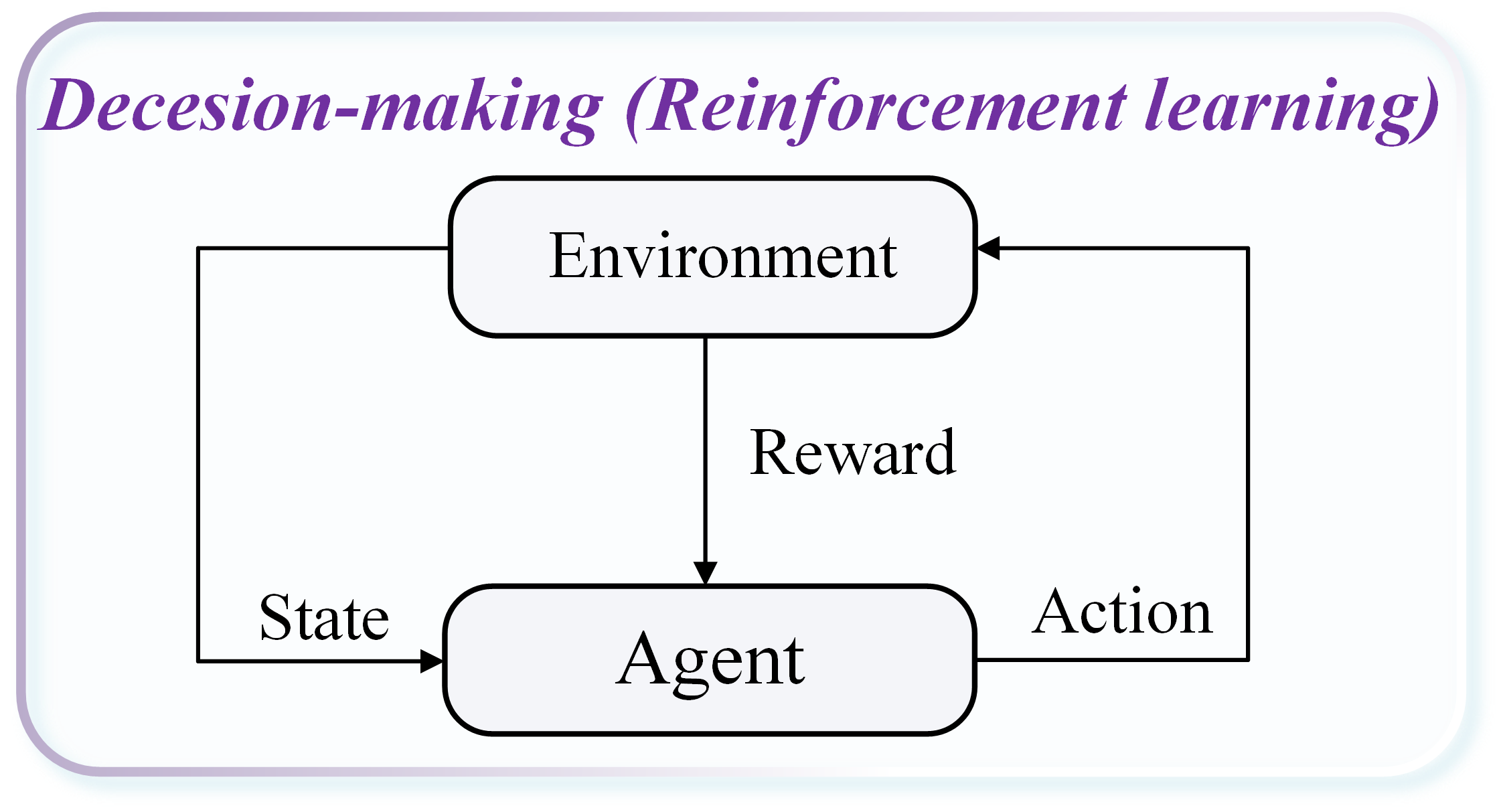}
            \caption{RL}
            \label{fig:figure402}
        \end{subfigure}
    \end{minipage}
    \begin{minipage}{0.45\textwidth}
        \begin{subfigure}{\textwidth}
            \includegraphics[width=\textwidth]{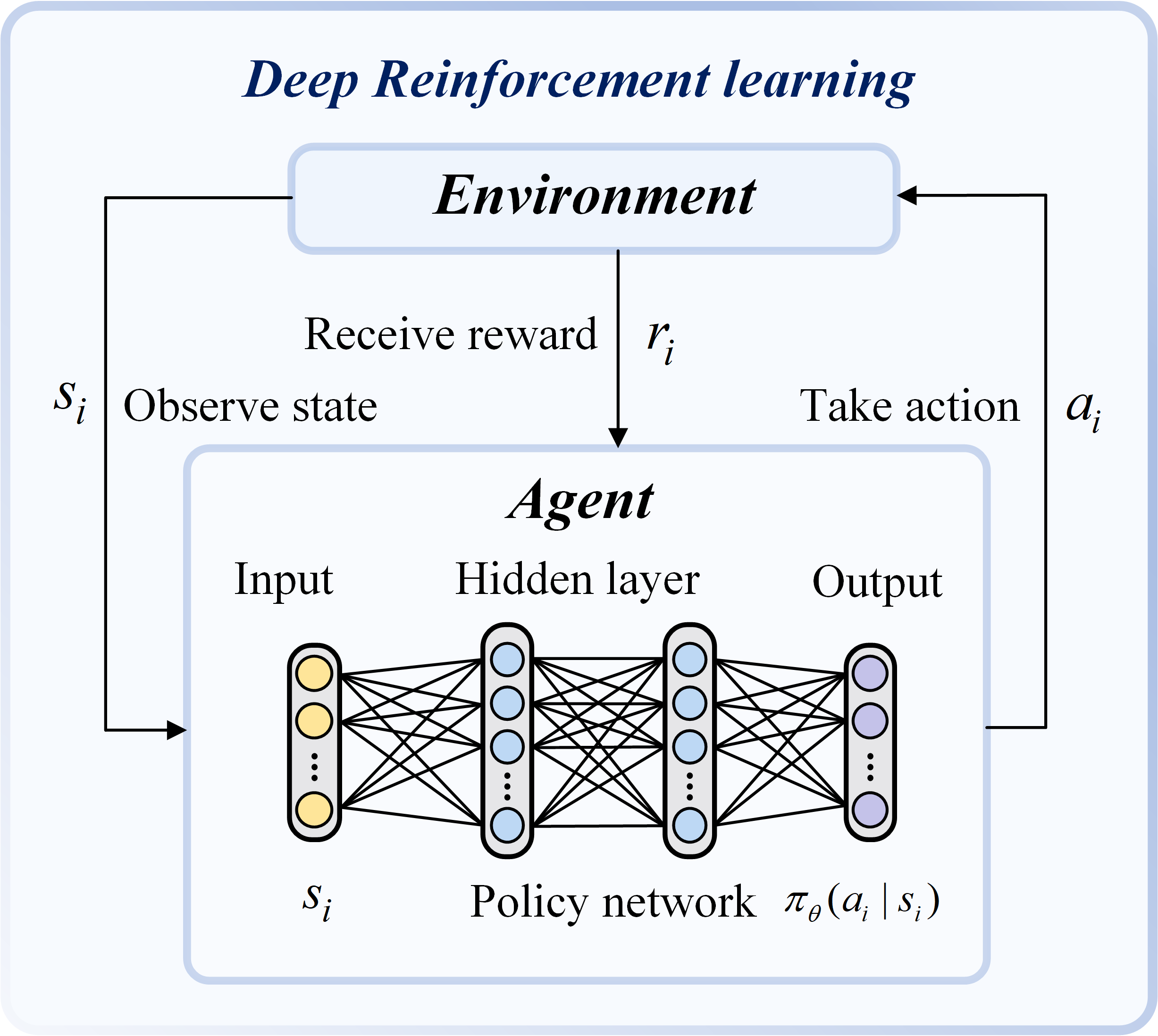}
            \caption{DRL}
            \label{fig:figure403}
        \end{subfigure}
    \end{minipage}
    \caption{Basic architectures of DL, RL, and DRL. (a) DL; (b) RL; (c) DRL.}
    \label{fig:figure4}
\end{figure*}

\paragraph{Deep learning}
DL, a subfield of machine learning, focuses on artificial neural networks inspired by biological neural networks. These networks, depicted in \cref{fig:figure4}(a), comprise interconnected layers of nodes or neurons. Each layer performs specific transformations on its inputs before passing them to the next layer. The term 'depth' in deep learning refers to the multiple layers of the neural network, comprising numerous linear and nonlinear processing units. Multiple layers enable these networks to perform complex transformations at various levels of abstraction, enhancing their ability to model intricate relationships in data. DNNs typically utilize backpropagation for training, adjusting neuron weights based on error rates from previous epochs. This process, combined with advanced optimization algorithms like Adam or stochastic gradient descent, efficiently minimizes the loss function.

\paragraph{Reinforcement learning}
RL is a crucial branch of machine learning that focuses on the interaction between an agent and its environment. 
As illustrated in \cref{fig:figure4}(b), the environment is the entity with which the agent interacts. 
RL aims to maximize cumulative rewards by learning optimal behavior or policies through trial and error during these interactions.
A finite Markov Decision Process (MDP) is defined by the tuple $($S, A, P, R, $\gamma$$)$, where $S$ is the finite state space, $A$ is the finite action space, $P$ is the state transition probability function, and $R$ is the reward function. $\gamma$ is the discount factor, representing the importance given to future rewards relative to immediate rewards.
At each time-step $i$, the agent observes the environment in state $s_i$, selects an action $a_i$, resulting in a transition to state $s_{i+1}$ with a probability $p(s_{i+1} | s_i, a_i)$. The agent receives a reward $R$ associated with this transition. The agent's decision-making process is governed by a policy denoted as $\pi(a_i | s_i)$.

\paragraph{Deep Reinforcement learning}
As shown in \cref{fig:figure4}(c), embedding DNNs into the basic RL framework constitutes the most fundamental DRL structure. DRL integrates DL with RL to manage complex decision-making tasks in high-dimensional environments. Within this integration, DL plays a pivotal role in estimating several key functions critical to the RL process.
DRL utilizes DNNs as function approximators to model complex value functions and policy functions, which are otherwise challenging to address with conventional RL methods. 
The state value function ($V_{\pi}(s)$) estimates the expected cumulative reward that can be obtained from a particular state $s$ under a given policy \( \pi \). The formula for $V_{\pi}(s)$ is expressed as:
\begin{equation}
V_{\pi}(s) = \mathbb{E}_{\pi} \left[ \sum_{t=0}^{\infty} \gamma^t R_{i} \mid s_i = s \right],
\end{equation}
where \( \mathbb{E}_{\pi} \) denotes the expectation under policy $\pi$, $\gamma$ is the discount factor, $R_{i}$ represents the reward at time step $t$, and $s_i$ denotes the state at time step $t$. DNNs approximate $V_{\pi}(s)$, enabling the handling of large and complex state spaces.
The action-value function (\(Q_{\pi}(s, a)\)) estimates the expected cumulative reward from taking action \(a_t\) in state \(s_t\) and subsequently following policy \( \pi \). The formula for \(Q_{\pi}(s, a)\) is expressed as:
\begin{equation}
Q_{\pi}(s, a) = \mathbb{E}_{\pi} \left[ \sum_{t=0}^{\infty} \gamma^t R_{i} \mid s_i = s, a_i = a \right],
\end{equation}
where \( a_i \) denotes the action at time step \( t \). The remaining symbols keep the same meaning as in $V_{\pi}(s)$.
DNNs are employed to approximate \( Q_{\pi}(s, a) \), facilitating the evaluation of action values in extensive action spaces.
Also, DNNs are commonly used as function approximators to model policy functions. The policy $\pi(a_i | s_i)$ is represented by a neural network, which outputs the probabilities of selecting each possible action given a state. Additionally, the transition function $P(s_{i+1} \mid s_i, a_i)$ predicts the next state $s_{i+1}$ given the current state $s_i$ and action $a_i$. DNNs can approximate this function to help the agent simulate and plan future actions more effectively.

\paragraph{Soft Actor-Critic Algorithm}

\begin{figure*}[htbp]
    \centering
    \includegraphics{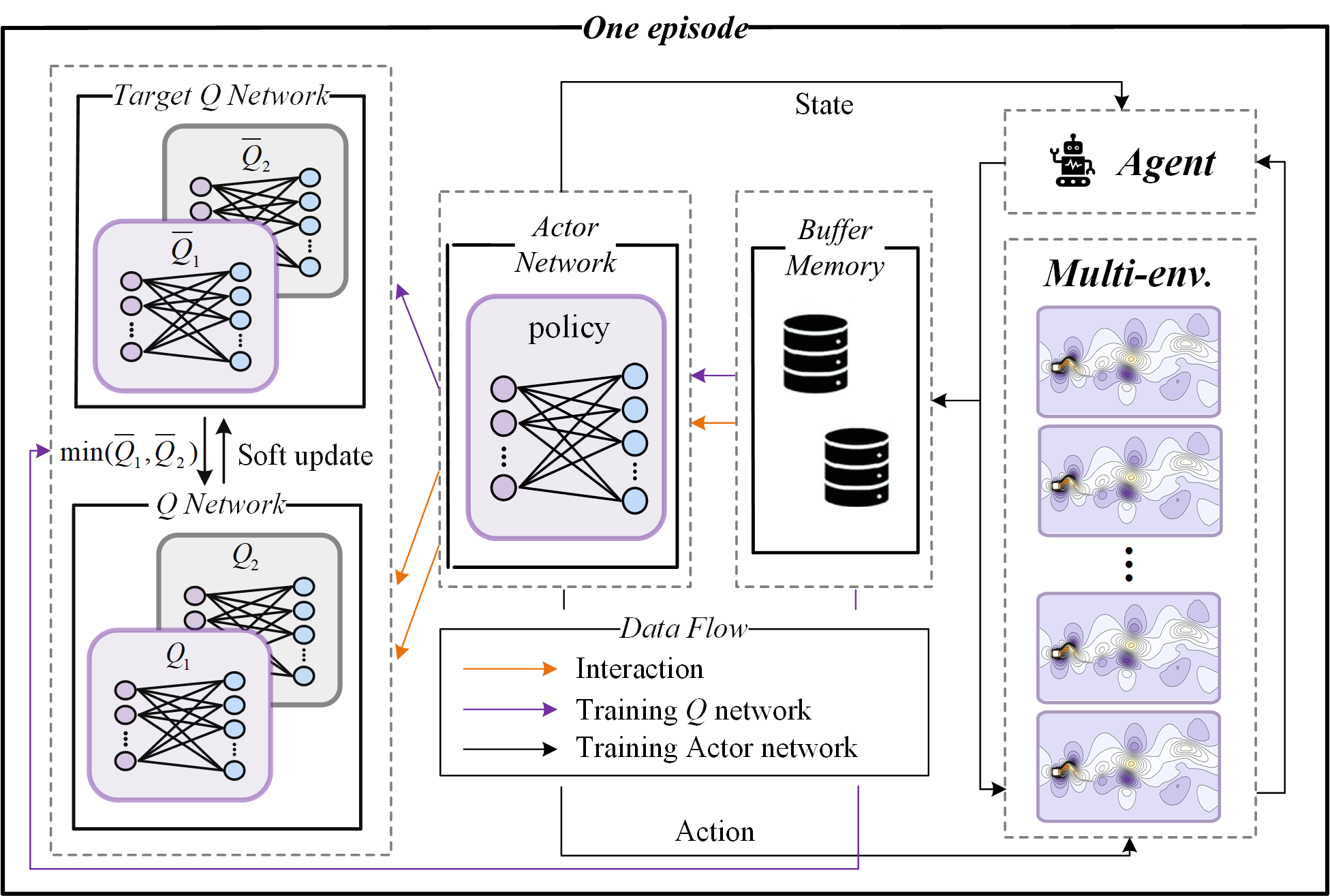}
    \caption{The SAC algorithm collects experience data through interaction with the environment, storing it in a replay buffer.
    It updates the value network, policy network, and target value network, using entropy regularization to encourage exploration, resulting in an efficient and stable reinforcement learning policy.}
    \label{fig:figure5}
\end{figure*}

The SAC algorithm is a state-of-the-art deep reinforcement learning method that combines Actor-Critic techniques with off-policy learning. SAC is highly efficient, stable, and robust, particularly for tasks with continuous action spaces. 
Its performance relies on experience replay, Actor-Critic architecture, and entropy regularization, which together foster accelerated learning and balanced policy exploration.
The data collection and network parameter update process of the SAC algorithm is illustrated in \cref{fig:figure5}.

The agent interacts with the environment, selecting actions according to its current policy $\pi$. The experience tuples $(s_i, a_i, r_i, s_{i+1})$ collected from these interactions are stored in a replay buffer. When updating the network parameters, the algorithm samples batches of experience data from the replay buffer.
Moreover, entropy (\( \mathcal{H} \)) is a metric that quantifies the uncertainty of a random variable. Within the framework of reinforcement learning, the entropy of a policy, \( \mathcal{H}(\pi(\cdot \mid s_i)) \), represents the stochasticity of the policy at a given state \( s_i \).\cite{haarnoja2018soft,pmlrv80haarnoja18b,DBLP}
The expression for this is given by:
\begin{equation}
\mathcal{H}(\pi(\cdot \mid s_i)) = -\sum_{a_i} \pi(a_i \mid s_i) \log \pi(a_i \mid s_i),
\end{equation}
where \( \pi(a_i \mid s_i) \) denotes the probability of taking action \( a_i \) in state \( s_i \) according to policy \( \pi \). The policy is trained with the objective to maximize the expected return and the entropy at the same time:
\begin{equation}
\pi^*=\arg \max _\pi \sum_i \mathbb{E}_{\left(s_i, a_i\right) \sim \rho_\pi}\left[r\left(s_i, a_i\right)+\alpha \mathcal{H}\left(\pi\left(\cdot \mid s_i\right)\right)\right],
\end{equation}
where $\mathcal{H}(\cdot)$ is the entropy measure and the variable \( \alpha \) functions as a temperature parameter that calibrates the relative significance of the entropy term against the reward, thereby modulating the stochasticity of the resulting optimal policy.\cite{pmlrv80haarnoja18b,haarnoja2019soft}

\subsection{DRL-Enhanced Active Flow Control}\label{sec:DRL-Enhanced AFC}

We propose an interaction framework between the agent and the numerical simulation environment, where the agent observes the current state of the CFD environment and adjusts the mass flow rates of the two synthetic jets on the square cylinder based on reward feedback. This iterative optimization process enables the SAC agent to converge to an optimal flow control strategy. 
Moreover, the control objective of the DRL-based AFC framework is to reduce the lift and drag coefficients of the square cylinder while minimizing vortex shedding in the wake flow field. 

The design of the DRL components is centered around this goal. The AFC problem is integrated with the DRL framework by defining three key components: the agent's action \(a_t\) (the mass flow rate of the synthetic jets), the agent's state \(s_t\) (observational data from the environment), and the agent's reward \(r_t\) (the control objectives).
Firstly, the agent's actions are represented by the mass flow rates of the synthetic jets. The jet flow is controlled by dimensionless mass flow rates \(Q_1\) and \(Q_2\), with the condition \(Q_1 + Q_2 = 0\) to maintain zero net mass flow. This condition ensures that drag reduction is achieved through indirect flow control rather than direct momentum injection, avoiding the instability in numerical results caused by mass addition or subtraction. Additionally, the synthetic jet flow rate is limited to not exceeding 2\% of the upstream inflow. 
Furthermore, to ensure the smoothness of continuous actions, a smoothing function \( S \) is utilized between consecutive actions \( a_t \) and \( a_{t+1} \), facilitating a gradual transition and mitigating abrupt changes in jet velocity.
The smoothing function \(S\) is defined as follows:
\begin{equation}
S(V_{\Gamma_1,T_i}, a, V_{\Gamma_1,T_{i-1}}) = V_{\Gamma_1,T_i} + \beta \cdot (a - V_{\Gamma_1,T_{i-1}}),
\end{equation}
where \(V_{\Gamma_1,T_i}'\) is the updated value at time step \(i\), \(V_{\Gamma_1,T_i}\) is the current value at time step \(i\), \(a\) represents the target action magnitude \(a_t\), \(V_{\Gamma_1,T_{i-1}}\) is the value at the previous time step \(i-1\), and \(\beta\) is a coefficient determining the extent of adjustment towards the target \(a\). This function effectively interpolates between the previous value and the target action, with \(\beta\) controlling the smoothness of the transition.

\begin{figure*}[ht]
    \centering
    \begin{subfigure}{0.45\textwidth}
    \includegraphics[width=\textwidth]{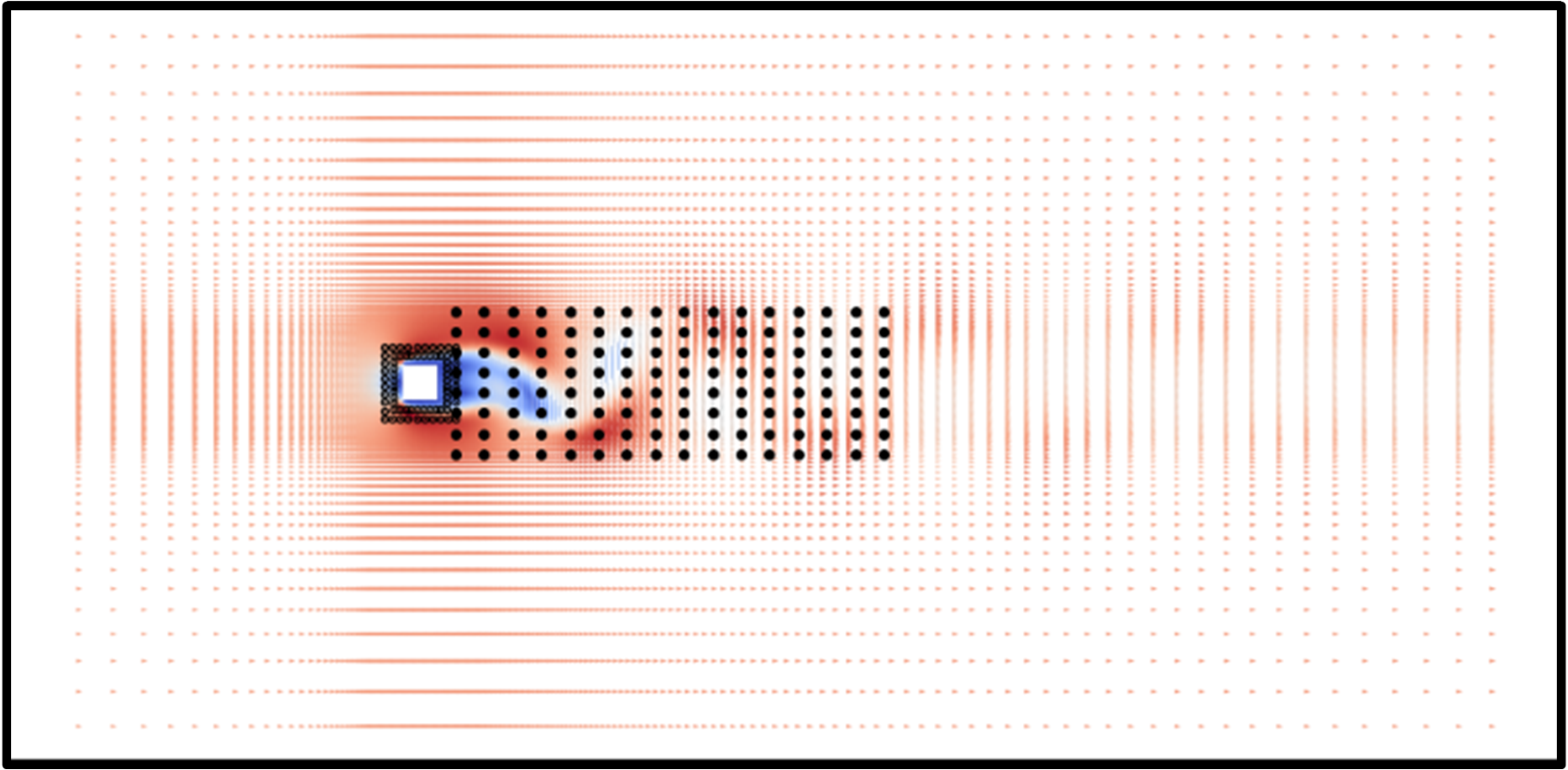}
    \caption{}
    \label{fig:figure6a}
    \end{subfigure}
    \begin{subfigure}{0.45\textwidth}
    \includegraphics[width=\textwidth]{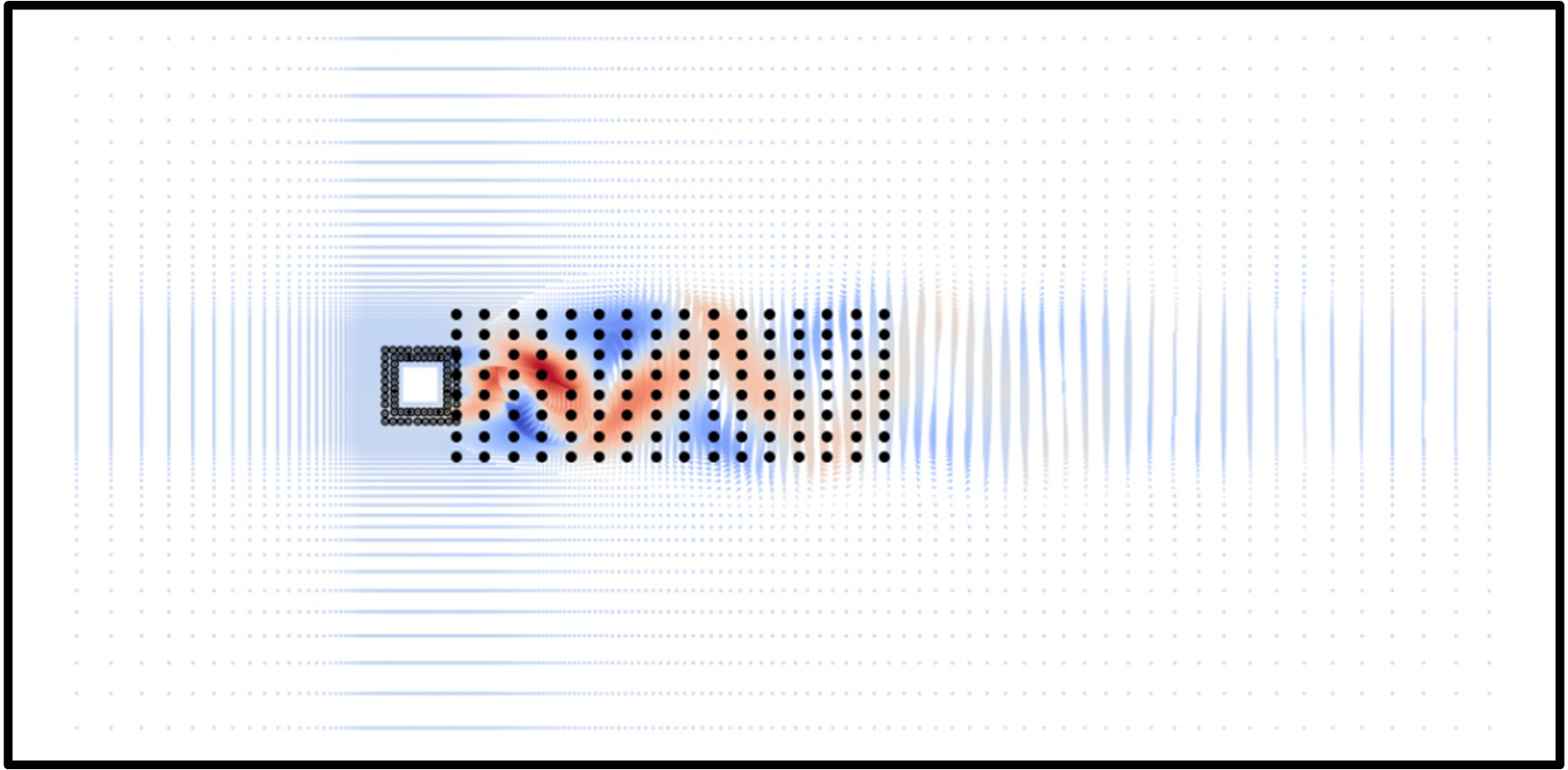}
    \caption{}
    \label{fig:figure6b}
    \end{subfigure}  
    \caption{Arrangement for probes position in flow around a square cylinder. (a) The probes are distributed in the instantaneous velocity field. (b) The probes are distributed in the pulsating velocity field.}
    \label{fig:figure6}
\end{figure*} 

Next, the state \(s_t\) represents the information observed by the agent from the environment. In this study, \(s_t\) is represented by physical information collected from specific locations within the computational domain. Specifically, 201 probes are placed around the square cylinder and in the wake region to capture instantaneous velocity or pressure data.
The probes are strategically distributed to cover the area around the square cylinder, as one of the control objectives is to manage the lift and drag forces acting on the square cylinder. This necessitates the agent observing the physical field information in this region. Additionally, the probes are positioned to cover the recirculation zone in the cylinder's wake and capture areas with the highest fluctuating values, which is crucial for controlling the wake's instability. The specific distribution of the probes is illustrated in \cref{fig:figure6}.

Moreover, the reward function is a core component of reinforcement learning. By incorporating $C_D$ and $C_L$ into the reward function, we align it with the primary objectives of reducing drag and suppressing lift.
This reward function clearly defines the behaviors the agent should learn and drives the agent to seek optimal strategies through positive and negative feedback mechanisms that minimize drag and suppress lift.
The specific reward function is defined as follows:
\begin{equation}\label{eq:my_reward}
        r_{T_i}=C_{D,0}-\left(C_D\right)_{T_i}-\omega\left|\left(C_L\right)_{T_i}\right|,
\end{equation}
where \(C_{D,0}\) represents the baseline drag coefficient $C_D$, serving as a reference point. \((C_D)_{T_i}\) denotes the drag coefficient at time step \(T_i\), with the objective being to minimize this value relative to the baseline. \((C_L)_{T_i}\) denotes the lift coefficient at time step \(T_i\). 
The parameter \(\omega\) is a weight factor for the lift coefficient. 
We adopt a penalty factor of \(0.2\) based on \citeauthor{rabault2019artificial}\cite{rabault2019artificial}, balancing drag reduction and lift suppression. 

During DRL training, each episode has a total duration of 1.25 seconds, including 100 training time steps. 
The control action is updated every 0.0125 seconds, equivalent to 25 numerical simulation time steps. 
Tested Reynolds numbers include 100 and 500, with vortex shedding periods approximately between 0.127 and 0.145 seconds. Therefore, an epoch is designed to last 1.25 seconds, corresponding to approximately 8.6 to 9.8 vortex shedding cycles. 
In each episode of DRL training, the CFD simulation spans multiple vortex shedding cycles, allowing the agent to observe flow characteristics over several shedding periods.

\section{RESULTS}\label{sec:results}

\subsection{Baseline flow simulation}\label{sec:Baseline}

\begin{figure*}[htbp]
    \centering
    \includegraphics{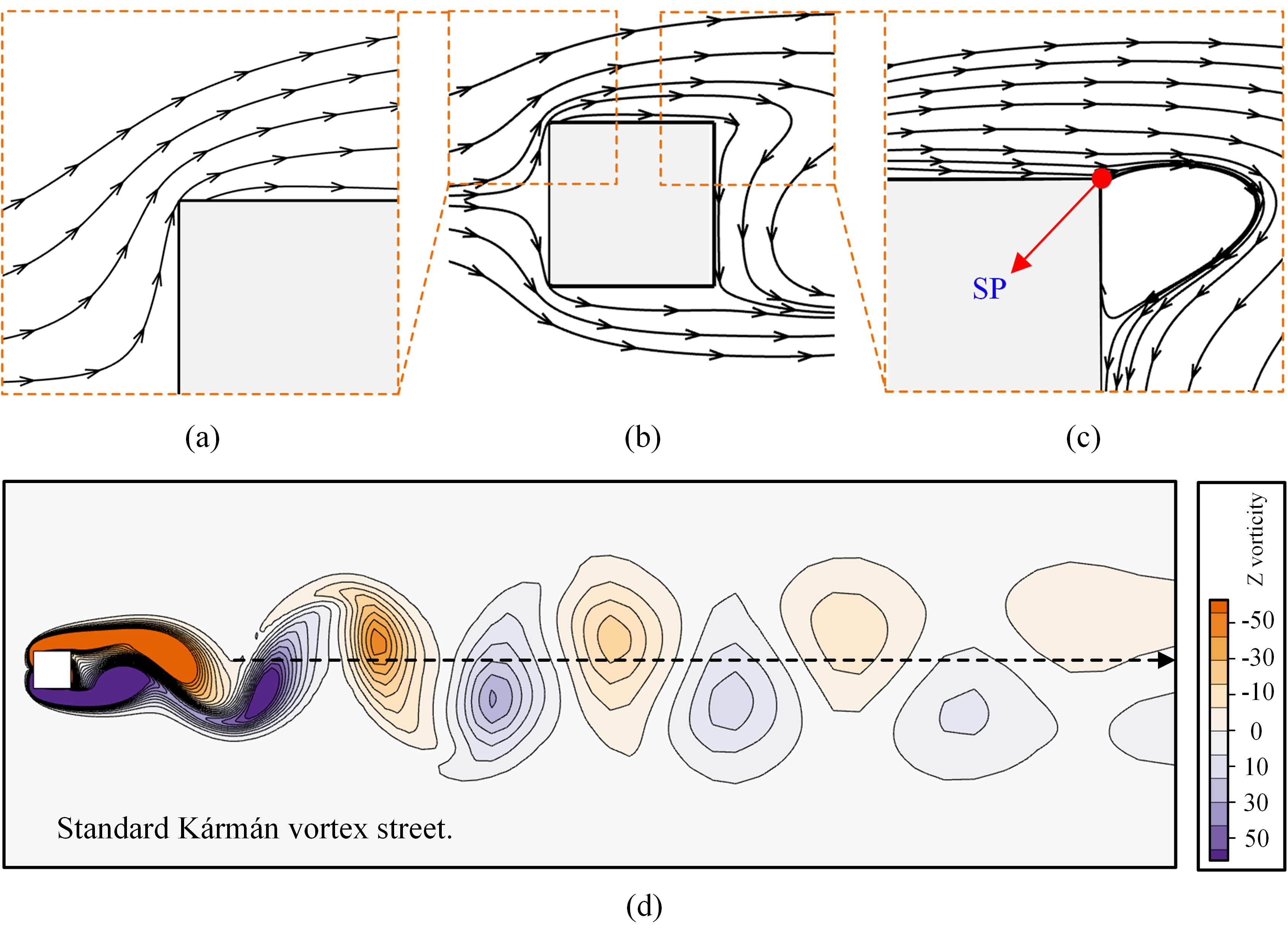}
    \caption{The streamlines and instantaneous vorticity around a square cylinder at $Re = 100$. (a) Streamlines near the leading corner of the square cylinder. (b) Streamlines of the square cylinder. (c) Streamlines near the trailing corner of the square cylinder. (d) Snapshots of the spanwise vorticity field. SP represents the separation point.}
    \label{fig:figure7}
\end{figure*}

We define the flow state without implemented flow control as the baseline.
As the flow a square cylinder exhibits many intriguing phenomena depending on the Reynolds number, we present two studies at $Re$ of 100 and 500 to investigate the mechanism of flow control, with a focus on the phenomena of vortex shedding and flow separation.
Active flow control commences from the baseline state, hence, it also serves as a basis for comparison with the controlled flow.
As illustrated in \cref{fig:figure7}(a), at a $Re$ of 100, the flow separates right at the leading edge corner of the square cylinder. However, the separation bubble is almost negligible as the flow reattaches immediately behind the corners. The boundary layer remains largely attached to the side surface of the square cylinder, as depicted in \cref{fig:figure7}(b). The primary flow separation occurs at the trailing edge of the cylinder, resulting in a wake formation predominantly at the leading (\cref{fig:figure7}(c)). 
The wake region of the square cylinder exhibits regular vortex shedding, as shown in \cref{fig:figure7}(d). 
Vortices shed from the trailing edge of the cylinder form a regular pattern, characterized by alternating vortices along the centerline, creating a street-like structure. 

\begin{figure*}[htbp]
    \centering
    \includegraphics{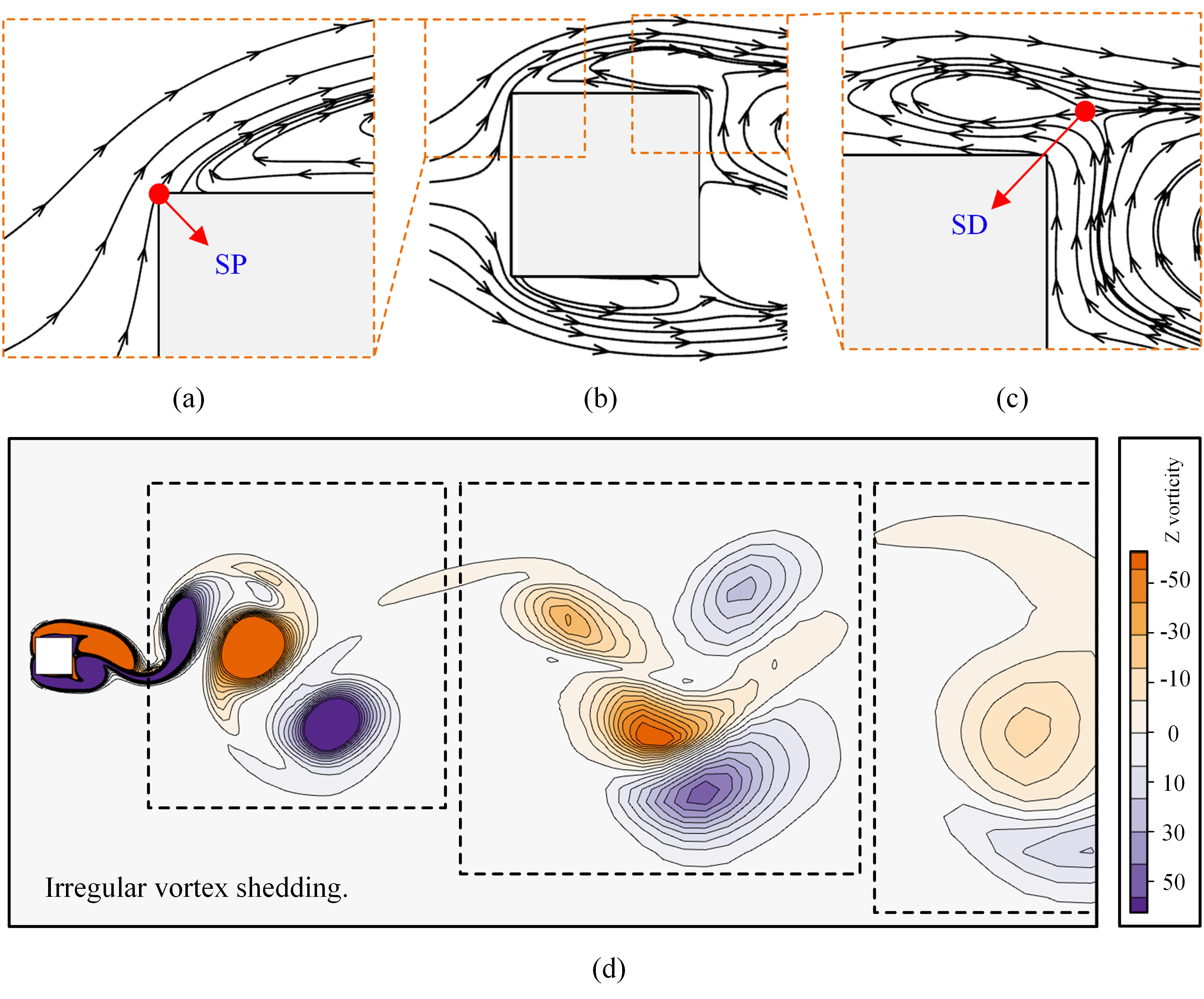}
    \caption{The streamlines and instantaneous vorticity around a square cylinder at $Re = 500$. (a) Streamlines near the leading corner of the square cylinder. (b) Streamlines of the square cylinder. (c) Streamlines near the trailing corner of the square cylinder. (d) Snapshots of the spanwise vorticity field. SP represents the separation point, and SD represents the saddle point.}
    \label{fig:figure8}
\end{figure*}

In contrast, at a $Re$ of 500, the primary flow separation points shift to the leading edge corners of the cylinder. Comparing \cref{fig:figure7}(a) and \cref{fig:figure8}(a), flow separation occurs at the leading edge in both cases. However, there is a significant separation bubble behind the corner. As seen in \cref{fig:figure8}(b) and \cref{fig:figure8}(c), the separation bubble extends beyond the length of the square cylinder and engulfs the trailing edge corner, submerging it in the recirculation region.
This shift suggests that at higher Reynolds numbers, the boundary layer detaches earlier due to increased inertial forces, resulting in a more turbulent wake and potentially different drag characteristics.
Understanding these variations in flow separation mechanisms with changing Reynolds numbers is essential for designing effective flow control strategies for drag reduction and enhanced flow stability.
At $Re = 500 $, the shedding pattern in the wake changes, as depicted in \cref{fig:figure8}(d).
After flow separation occurs at the leading edge of the square cylinder, some separated fluid reattaches to the cylinder surface. The vortex shedding in the wake region of the square cylinder becomes increasingly unstable, with vortices squeezing, stretching, and deforming as they develop downstream in the form of vortex pairs. 
Compared to $Re = 100 $, the flow exhibits greater instability and complexity at $Re = 500$ , indicating an increased difficulty in flow control. In this state, implementing AFC based on DRL requires a more sophisticated control algorithm.

To achieve our objective of suppressing unsteady vortex shedding for drag reduction, we incorporate an additional test case that simulates the conditions of a steady flow. 
This is achieved by simulating a half-body with symmetric boundary conditions to enforce symmetry of the wake flow.
The base flow represents the ideal state without vortex shedding and is used solely to calculate the drag coefficient in this idealized condition. This serves as an asymptote to measure the maximum possible drag reduction.
The base flow corresponds to a stable equilibrium state of the controlled Navier-Stokes equations where the velocity and pressure fields of the fluid remain unchanged with respect to time, constituting a steady solution of the Navier-Stokes equations.
This state can be considered a potential fixed point in the configuration space of fluid flow. 
However, despite the apparent stability of the baseline flow, it is, in fact, unstable to infinitesimal perturbations, leading to the occurrence of vortex shedding phenomena.
Vortex shedding is one of the sources of drag experienced by the cylinder, and our active flow control aims to mitigate this resistance. Therefore, assessing the effectiveness of the control strategy can be done by using a modified baseline drag value, assuming the absence of vortex shedding.
To simulate the scenario without vortex shedding, the strategy involves simulating only the upper half of the computational domain along the centerline. Symmetric boundary conditions are applied at the lower boundary of the computational domain (where \(y = 0\)). These symmetric boundary conditions are explicitly defined as follows: the velocity component \(v = 0\), and the derivatives of the velocity \(u\) and the pressure \(p\) with respect to \(y\) are zero, i.e., \(\frac{\partial u}{\partial y} = 0\) and \(\frac{\partial p}{\partial y} = 0\). The boundary conditions for the rest of the domain remain unchanged. This approach effectively models an idealized flow scenario.

\begin{figure*}[htbp]
    \centering
    \includegraphics{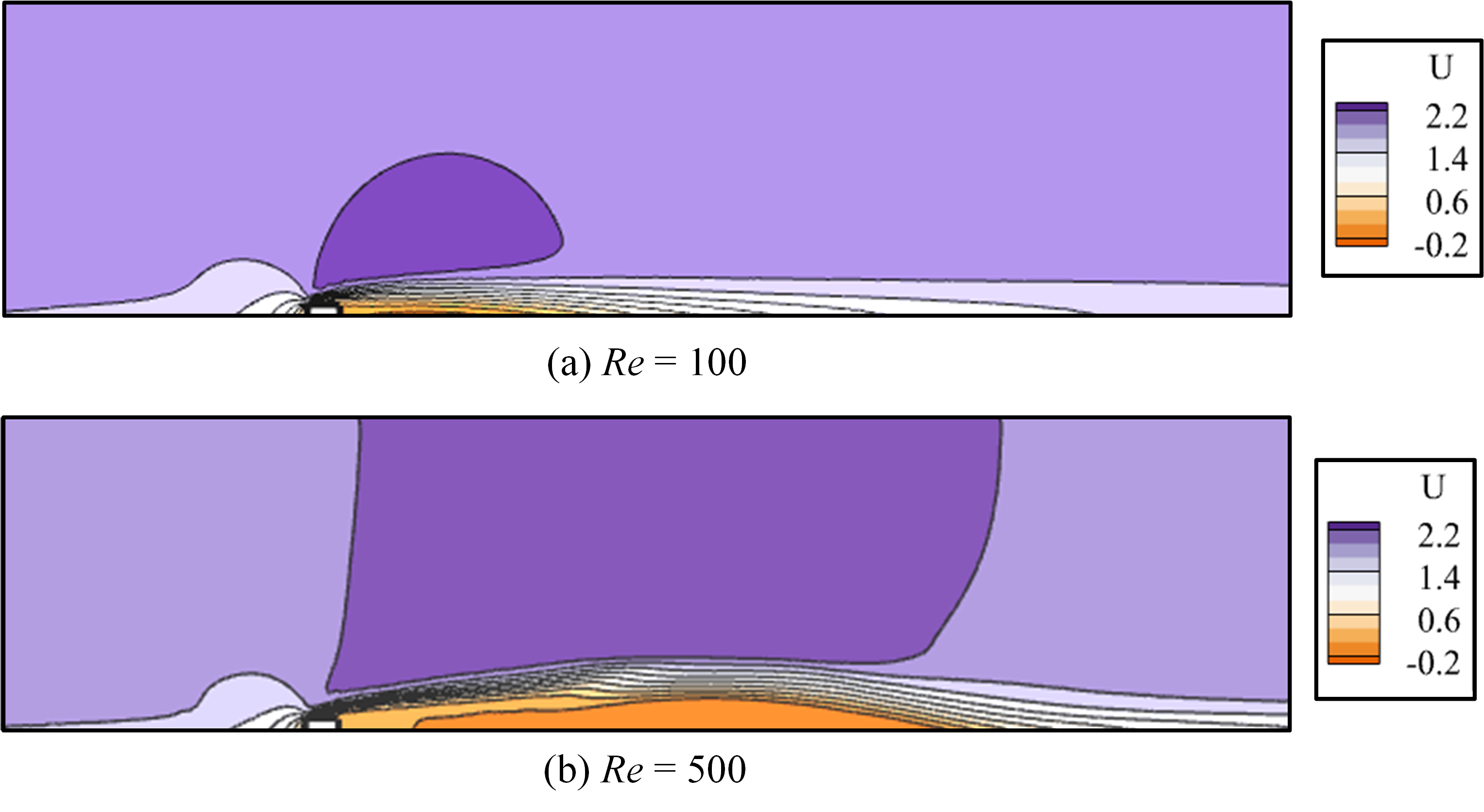}
    \caption{The base flow is obtained through a simulation of a half-domain, where the $x$-coordinate ranges from -10$D$ to 30$D$, and the $y$-coordinate ranges from 0 to 10$D$. We apply symmetric boundary conditions on the $y$ = 0 boundary, and the computational domain's mesh is consistent with \cref{fig:figure3}. This configuration results in a scenario without vortex shedding, serving as a hypothetical baseline for comparison with the results of active flow control. (a) $Re = 100$; (b) $Re = 500$.}
    \label{fig:figure9}
\end{figure*}

In \cref{fig:figure9}, simulations are performed for the cases of $Re=100$ and $Re=500$, where the wake flow behind the semi-square cylinder is in a steady state without vortex shedding, and the only reason for pressure drag is flow separation. The total drag coefficient encapsulates the intrinsic drag associated with the baseline flow and the correctable component, with only the second part of drag being adjustable through strategic manipulation of the von Kármán vortex shedding\cite{protas2002drag}.
We compute the asymptotic drag coefficients, denoted as $C_{Dh}$, for the semi-square cylinder at $Re=100$ and $Re=500$, which are found to be 0.668 and 0.498, respectively. These values are extended to obtain the baseline drag coefficients, denoted as $C_{Db}$, for a fully square cylinder in a steady-state flow, resulting in values of 1.335 and 0.995 for $Re=100$ and $Re=500$, respectively. The result of $Re = 100$ is consistent with the study of \citeauthor{Xia2024}\cite{Xia2024}.
The baseline drag coefficient, $C_{Db}$, is compared with the drag obtained during active flow control to evaluate the effectiveness of the control strategy in reducing drag.

\subsection{DRL-based control performance}

\begin{figure*}[htbp]
    \centering
    \includegraphics{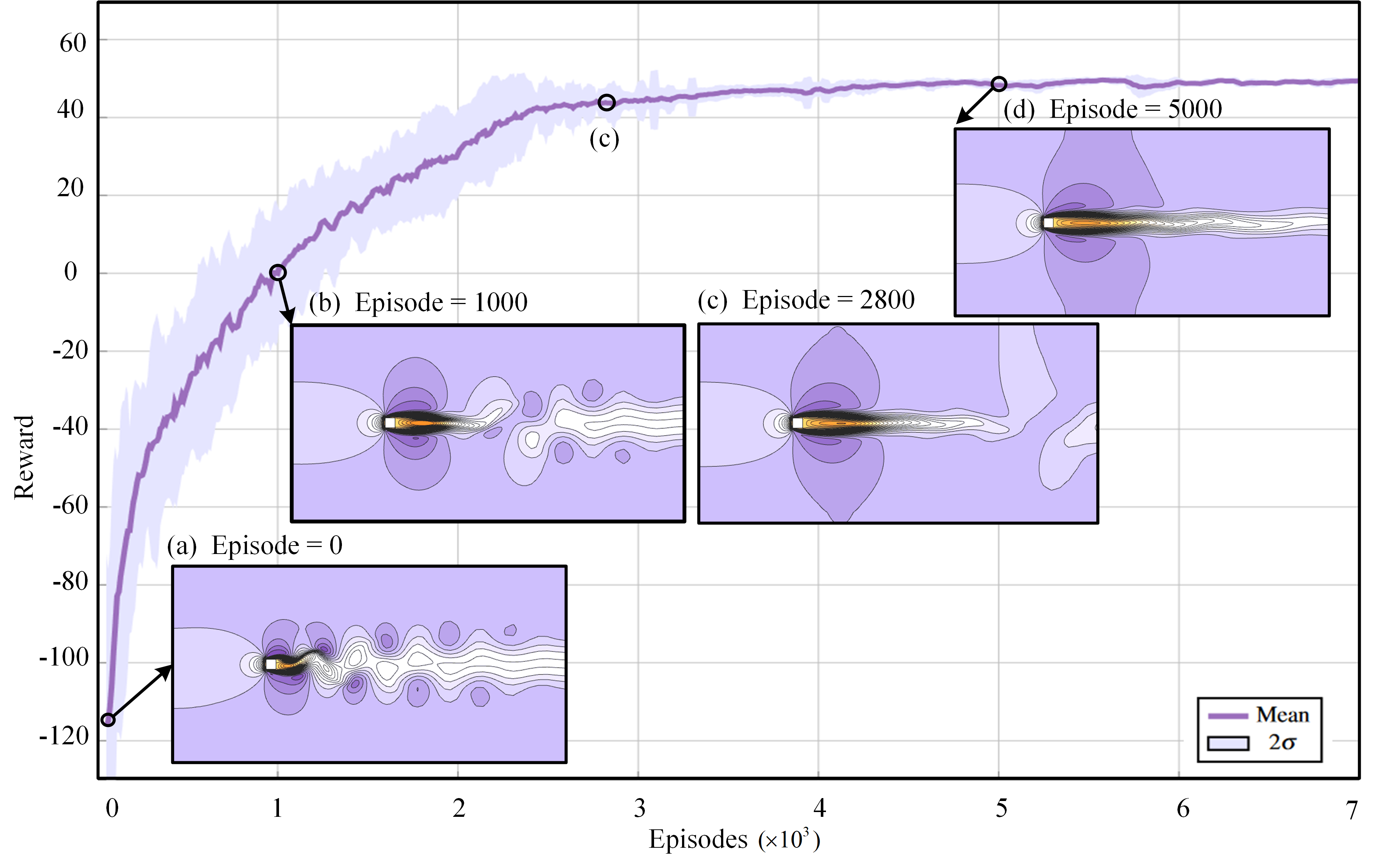}
    \caption{When $Re=100$, observed the evolution of the instantaneous velocity magnitude when DRL training. Specifically, focused on four key moments: (a) Episode = 0, (b) Episode = 1000, (c) Episode = 2800, and (d) Episode = 5000.}
    \label{fig:figure10}
\end{figure*}

\begin{figure*}[htbp]
    \centering
    \includegraphics{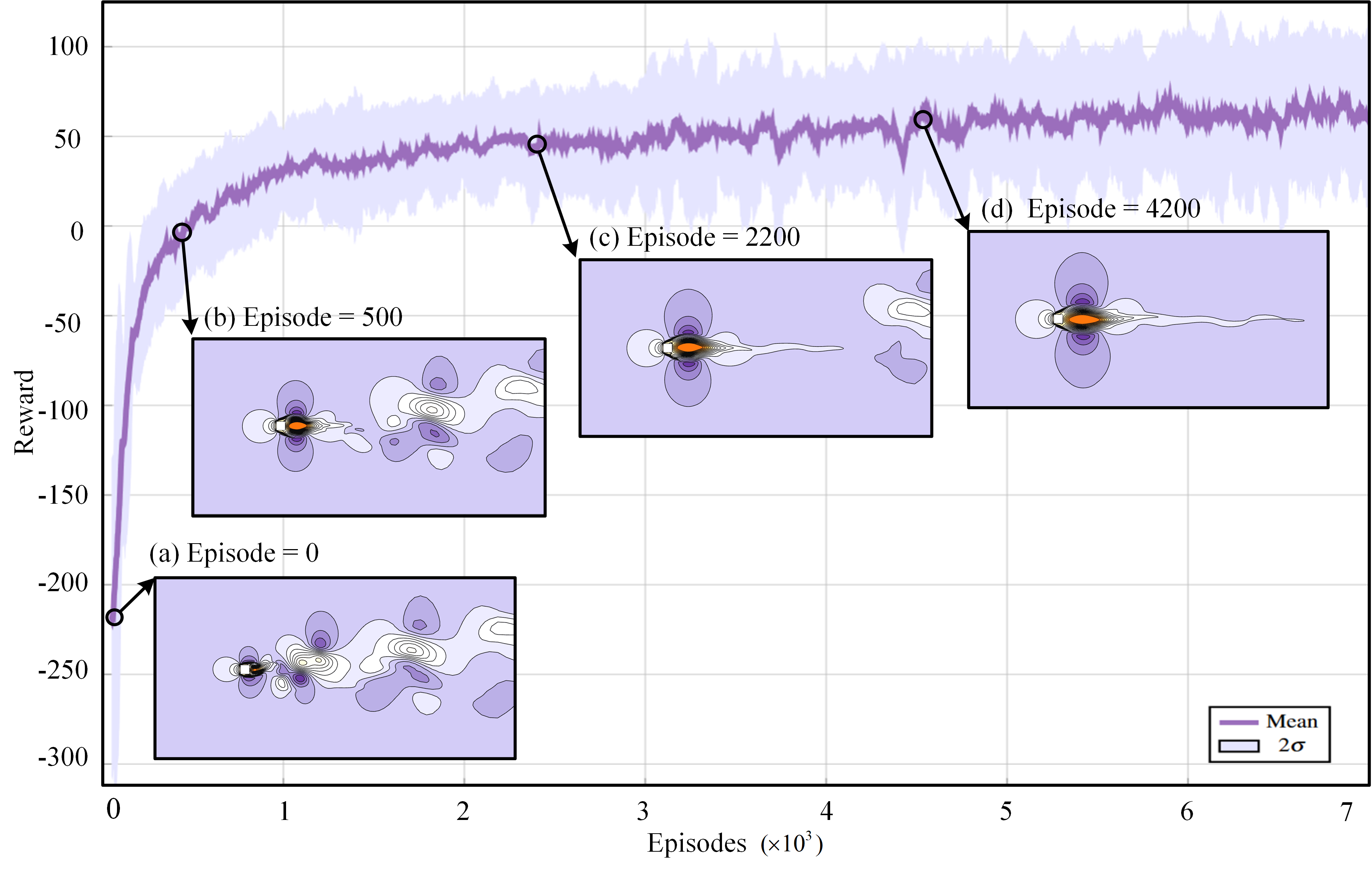}
    \caption{When $Re=500$, observed the evolution of the instantaneous velocity magnitude when DRL training. Specifically, focused on four key moments: (a) Episode = 0, (b) Episode = 500, (c) Episode = 2200, and (d) Episode = 4200.}
    \label{fig:figure11}
\end{figure*}

In \cref{sec:Baseline}, we have observed that the vortex shedding patterns are completely different at $Re=100$ and $Re=500$. 
This section primarily evaluates the performance of DRL-based AFC control, focusing on its effectiveness in suppressing vortex shedding in the wake flow field, as well as its performance in drag reduction and lift suppression.
The synthetic jets are positioned near the trailing edge corners of the square cylinder, with a width of $D/25$.
In this section, we will evaluate the control performance of the DRL-based AFC technique at $Re=100$ and $Re=500$ separately.
To visually observe the effect of vortex shedding suppression during the training process, \cref{fig:figure10} ($Re=100$) and \cref{fig:figure11} ($Re=500$) capture snapshots of the instantaneous velocity magnitude at four representative episodes on the learning curve.

For $Re=100$, AFC is applied to the flow around a square cylinder using DRL, and the learning curve along with snapshots of the instantaneous velocity magnitude of the square cylinder are depicted in \cref{fig:figure10}. The DRL training commences with the baseline flow field, characterized by alternating vortex shedding in the wake region. Flow control is then executed based on this initial condition.
The cumulative reward function curve of the DRL training exhibits a sharp increase during the initial stages. 
After 1000 episodes of training, the cumulative reward becomes positive, indicating that the SAC agent has discovered a control strategy that meets the objective function. 
At this point, the onset of vortex shedding is shifted downstream in the wake region, and the previously oscillating recirculation bubble behind the square cylinder is elongated. 
As training progresses to 2800 episodes, the growth of the cumulative reward begins to level off, indicating that the DRL training is converging. 
By this stage, the vortex shedding phenomenon has been largely suppressed, with virtually no visible vortices being shed downstream. 
The recirculation zone has been greatly elongated and remains relatively stable, with only the initial shed vortex continuing to move towards the outlet.
After 5000 episodes of training, the cumulative reward reaches a plateau, signifying that the DRL training has stabilized and fully converged. 
At this point, vortex shedding in the wake region of the square cylinder is completely suppressed.

Similarly, under a Reynolds number of 500, we employ DRL to execute active flow control around a square cylinder, with \cref{fig:figure11} depicting its learning curve and the instantaneous velocity magnitude during characteristic episodes. In the RL training process, the cumulative reward value of the first episode is considered the starting point of the learning process. The initial episode reward value is -300, which is much lower compared to the case with a Reynolds number of 100. This indicates that the training task at $Re=500$ is significantly more challenging.
At the beginning of the control, the shedding of vortices in the wake of the baseline flow no longer follows a regular alternating pattern but evolves in the form of more complex “vortex pairs.” After 500 episodes of training, the cumulative reward increases to a positive value, indicating that the SAC agent has explored effective control strategies that guide the reward function towards positivity. Currently, the recirculation area in the wake of the cylinder has enlarged and developed into an approximately symmetric and stable state. The two sets of vortices originally shed by the baseline flow gradually develop downstream in the computational domain.
As DRL training progresses to 2200 episodes, the growth rate of the learning curve slows down towards convergence. Upon reaching 4200 episodes of training, the learning curve enters a plateau phase, and the reward function tends to converge. Furthermore, the recirculation area in the wake of the cylinder has stabilized to its maximum extent, and the phenomenon of vortex shedding has been completely suppressed. 
To sum up, at $Re=100$ and $Re=500$, we completely suppress vortex shedding in the wake, demonstrating that the control strategy can significantly eliminate flow field instability. The control performance of DRL-based AFC is validated in the flow around a square cylinder at $Re=100$ and $Re=500$.

\subsection{Parameterization and Sensitivity Analysis of Synthetic Jets}

In the previous section, we observed the suppression process of vortex shedding in the wake of a square cylinder during DRL training at $Re = 100$ and $Re = 500$. The training results indicate that when the synthetic jets are positioned near the trailing edge corners of the square cylinder, DRL-based AFC can completely suppress vortex shedding in the wake at both $Re = 100$ and $Re = 500$. Additionally, \citeauthor{chen2023deep}\cite{chen2023deep} and \citeauthor{yan2023stabilizing}\cite{yan2023stabilizing} analyzed the impact of synthetic jets positioning on flow control efficacy, yielding some intriguing conclusions.
Building on these findings, we integrate the optimization of jet actuator positioning with the optimization of feedback laws using DRL. This study tests $Re$ of 100 and 500, representing two distinctly different vortex shedding modes.
First, we perform flow control using symmetric zero-mass-flux jets positioned at the leading and trailing edges of the square cylinder. The objective is to investigate how the positioning of jets actuators influences flow control performance. 
econdly, we design synthetic jets with four different widths to study how the width of the synthetic jets affects flow control performance.
Then, we analyze the impact of jet parameters on control performance by examining the physical mechanisms of flow control achieved through synthetic jets.

\subsubsection{Location of Synthetic Jets}

In \cref{tab:tab3}, the training results indicate that at $Re = 100$, placing the synthetic jets near the leading edge of the square cylinder reduces the average drag coefficient $C_D$ by 12.7\% compared to the baseline. The standard deviation of $C_D$ is suppressed by 99.9\%, and the standard deviation of $C_L$ is controlled by 81.5\%.
When the synthetic jets is positioned near the trailing edge of the square cylinder, the average $C_D$ decreases by 14.4\% compared to the baseline. The standard deviation of $C_D$ is suppressed by 99.9\%, and $C_L$ is controlled by 93.1\%.
At $Re = 500$, placing the synthetic jets near the leading corner of the cylinder achieves a drag reduction rate of 65.5\%. The standard deviation of $C_D$ decreases by 86.7\%, and $C_L$ decreases by 44.4\%.
Placing the synthetic jets near the trailing corner of the cylinder achieves a drag reduction rate of 51.4\% at $Re = 500$. The standard deviation of $C_D$ decreases by 90.5\%, and $C_L$ decreases by 74.9\%.
Our training results demonstrate that at $Re = 100$, positioning the synthetic jets at the trailing corner of the square cylinder is more beneficial for reducing drag. However, for $Re = 500$, placing the synthetic jets near the leading corner of the cylinder shows a more pronounced drag reduction effect.
Regarding the influence of synthetic jets position on drag reduction, the findings of \citeauthor{chen2023deep} and \citeauthor{yan2023stabilizing} align with our results.

\begin{table}[htbp]
\centering
\caption{Impact of Synthetic Jets Actuator Placement on AFC Control Performance: Optimizing Position Selection.}
\label{tab:tab3}
\begin{tabular}{ccccccccccccc}
\toprule
\toprule
\multicolumn{4}{c}{} & \multicolumn{3}{c}{Mean $C_D$} & \multicolumn{3}{c}{Std of $C_D$} & \multicolumn{3}{c}{Std of $C_L$}  \\ 
\cmidrule(lr){5-7} \cmidrule(lr){8-10} \cmidrule(lr){11-13}
Reference & $Re$ & Jet Location & Suppress vortex& $C_{D,Baseline}$ & $C_{D,DRL}$ & Reduction & $C_{D,Baseline}$ & $C_{D,DRL}$ & Reduction & $C_{L,Baseline}$ & $C_{L,DRL}$ & Reduction \\
\midrule
\citeauthor{chen2023deep}\cite{chen2023deep} & 100 & Leading & NO & 1.480 & 1.370 & 7.20 \% & - & - & - &  0.530 & 0.190 & 64.2 \% \\
\citeauthor{chen2023deep}\cite{chen2023deep} & 100 & Trailing & \textbf{YES} & 1.480 & 1.280 & 13.3 \% & - & - & - & 0.530 & 0.080 & 84.9 \% \\
Our study & 100 & Leading & NO & 1.549 & 1.353 & 12.7 \% & 2.020 & 0.001 & 99.9 \% & 0.179 & 0.033 & 81.5 \% \\
Our study & 100 & Trailing & \textbf{YES} & 1.549 & 1.325 & 14.4 \% & 2.020 & 0.001 & 99.9 \% & 0.179 & 0.025 & 86.1 \% \\
\citeauthor{yan2023stabilizing}\cite{yan2023stabilizing} & 500 & Leading &  NO & 1.980 & 1.100 & 44.4 \% & 0.470 & 0.070 & 84.3 \% & 1.260 & 0.170 & 85.9 \% \\
\citeauthor{yan2023stabilizing}\cite{yan2023stabilizing} & 500 & Trailing & NO & 1.980 & 1.950 & 1.40 \% & 0.470 & 0.520 & 10.7 \% & 1.260 & 1.070 & 13.2 \% \\
Our study & 500 & Leading & NO & 2.060 & 0.711 & 65.5 \% & 0.441 & 0.268 & 86.7 \% & 1.173 & 0.650 & 44.4 \% \\
Our study & 500 & Trailing & \textbf{YES} & 2.060 & 1.001 & 51.4 \% & 0.441 & 0.042 & 90.5 \% & 1.173 & 0.294 & 74.9 \% \\
\bottomrule
\bottomrule
\end{tabular}
\end{table}

\begin{figure*}[htbp]
    \centering
    \includegraphics[width=\textwidth]{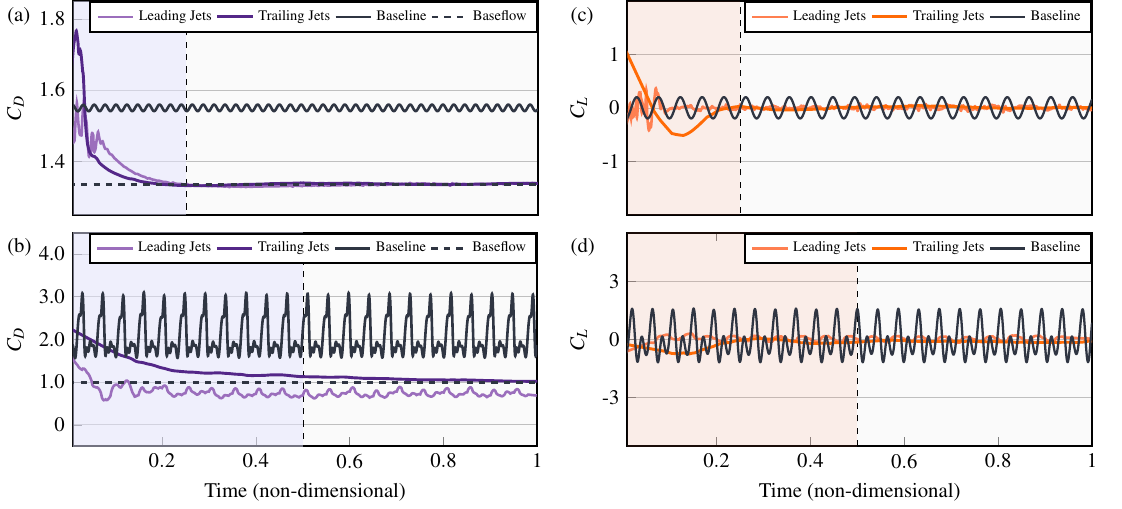}
    \caption{Effect of synthetic jets positioning on flow control at $Re=100$ and $Re=500$. The leading jets are positioned near the leading corners of the square cylinder, while the trailing jets are positioned near the trailing corner points. The precise locations of the synthetic jets are detailed in \cref{fig:figure2}. (a) $Re=100$, $C_D$; (b) $Re=100$, $C_L$; (c) $Re=500$, $C_D$; (d) $Re=500$, $C_L$.}
    \label{fig:figure12}
\end{figure*}

The time history curves of the controlled flow's $C_D$ and $C_L$ are illustrated in \cref{fig:figure12}. At $Re = 100$, we observe periodic fluctuations in the baseline flow's lift coefficient. When the synthetic jets are positioned near the leading and trailing corners of the cylinder, both drag and lift coefficients exhibit transient changes upon activation of the synthetic jets, quickly reducing to minimum levels.
Notably, positioning the synthetic jets near the trailing corner results in more stable $C_D$ and $C_L$ coefficients. This difference becomes more pronounced at $Re = 500$, where positioning the synthetic jet near the leading corner still results in periodic fluctuations in both coefficients. Conversely, positioning the synthetic jets near the trailing corner stabilizes both $C_D$ and $C_L$ coefficients.
In both $Re = 100$ and $Re = 500$ cases, positioning the synthetic jets near the trailing corner stabilizes the flow's $C_D$ and $C_L$ coefficients effectively. This observation is consistent with the findings in \cref{tab:tab3}, which describe how only positioning the synthetic jets near the trailing corner suppresses vortex shedding in the wake completely.
Similar control effects were observed by \citeauthor{chen2023deep} at $Re = 100$. To investigate the reasons behind these results, we compare streamline plots of the baseline and controlled flows in \cref{fig:figure12}, providing a detailed analysis of how the position of the synthetic jets influences control effectiveness.

\begin{figure*}[htbp]
    \centering
    \includegraphics{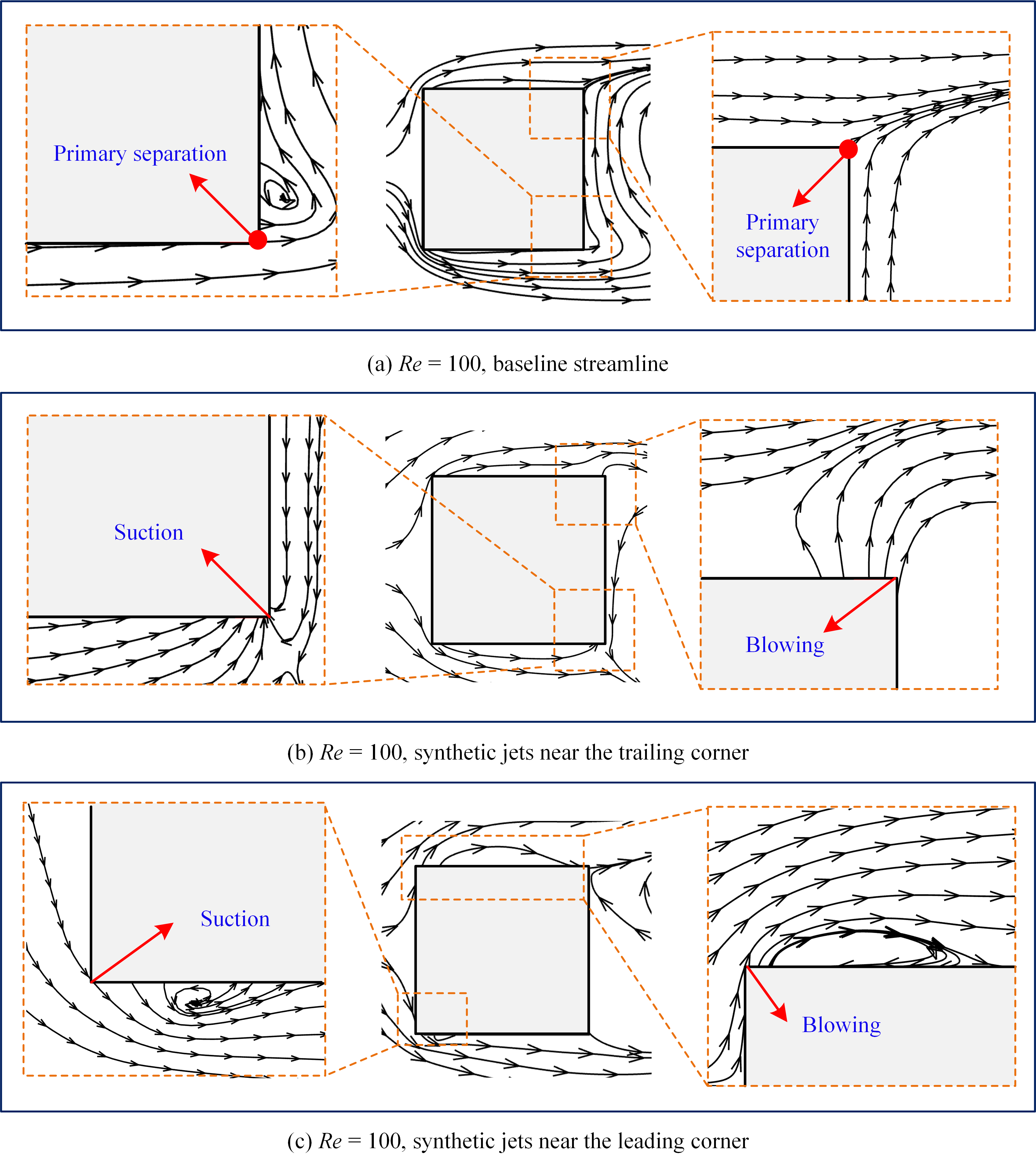}
    \caption{ $Re = 100$, Streamlines of a square cylinder during baseline and flow control. (a) baseline flow. (b) the controlled flow with the synthetic jets positioned near the trailing edge. (c) for the controlled flow with the synthetic jets positioned near the leading edge.}
    \label{fig:figure13}
\end{figure*}

\cref{fig:figure13} illustrates the streamlines of the baseline flow at $Re = 100$, as well as the streamlines with synthetic jets positioned near either the leading or trailing corners. In the baseline flow, separation primarily occurs near the trailing corners. When synthetic jets are placed near the trailing corners, performing suction or blowing control directly disrupts the vortices near the trailing corners, delaying the separation point.
In the baseline flow, there is no reverse flow on the sidewalls of the square cylinder, and the fluid separates near the trailing corners. However, when synthetic jets are placed near the leading corners, performing suction or blowing control creates recirculation regions on both sides of the cylinder, leading to separation again near the trailing corners. In this scenario, regardless of how the jet speed is adjusted, flow separation reoccurs at the trailing corners.

Our research indicates that for the baseline flow at $ Re = 100 $, positioning the synthetic jets near the trailing corner yields a maximum drag reduction of 14.4\%. For the baseline flow at $Re = 500$, the optimal drag reduction efficiency is achieved by placing the synthetic jet near the leading corner, with a maximum drag reduction rate of 65.5\%. This phenomenon is related to the differing locations of flow separation around the square cylinder at $ Re = 100 $ and $Re = 500$. At $ Re = 100 $, flow separation occurs near the trailing corner, whereas at $Re = 500$, flow separation occurs near the leading corner, with the separated bubble reattaching to the sidewall and subsequently reattaching near the trailing corner. Based on the flow control effects given by DRL, it is found that placing the synthetic jets at the location of flow separation is more conducive to achieving control objectives. Similarly, \cite{KIM2009172} and \citeauthor{KIM2009172} observed that when controlling flow separation on an airfoil using synthetic jets, the control effectiveness is optimal when the separation point coincides with the location of the synthetic jets.

\begin{figure*}[htbp]
    \centering
    \includegraphics{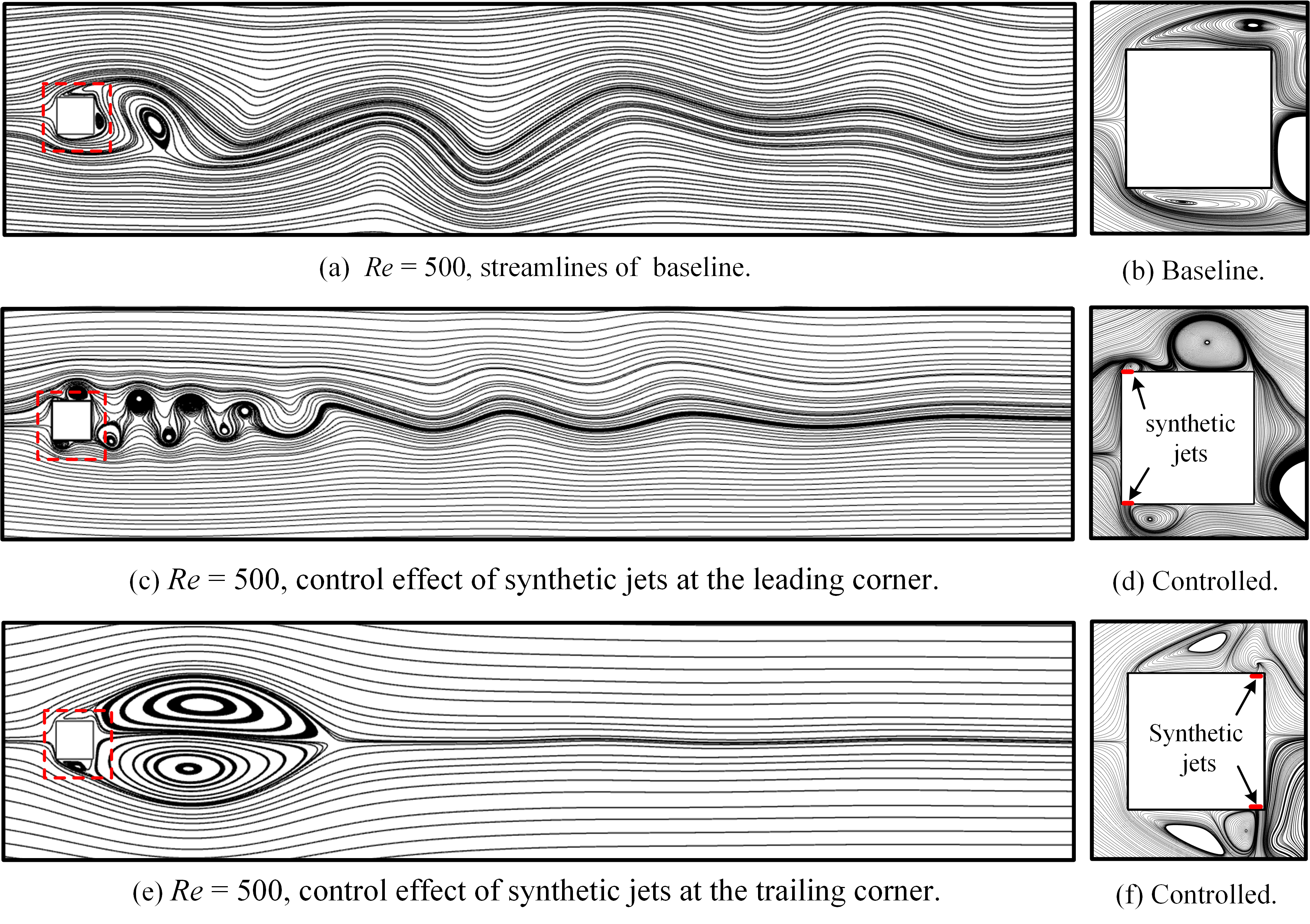}
    \caption{At $Re = 500$, the visualizations depict streamlines under various configurations: baseline flow, the controlled flow with the synthetic jets near the leading edge, and synthetic jets near the trailing edge. Specifically: (a) and (b) for baseline flow. (c) and (d) for the controlled flow with the synthetic jets positioned near the leading edge. (e) and (f) for the controlled flow with the synthetic jets positioned near the trailing edge.}
    \label{fig:figure14}
\end{figure*}

As an illustration at $Re=500$, we present the streamline visualizations for the baseline flow, controlled flow with the synthetic jets positioned near the leading edge, and controlled flow with the synthetic jets positioned near the trailing edge.
Both the baseline and controlled flows represent the stable flow states attained after sufficient flow development. In \cref{fig:figure14}, streamline plots are provided for these three scenarios, along with magnified views focusing on the flow patterns around the square cylinder.
In \cref{fig:figure14}(a), flow separation occurs at the leading edge of the square cylinder in the baseline flow, forming recirculation regions on both sides of the cylinder, with significant oscillations in the wake flow streamlines. The streamlines around the square cylinder, along with separation bubbles near the sidewalls, are depicted in \cref{fig:figure14}(b). The stabilized flow field achieved after implementing control with the synthetic jets positioned near the leading edge of the square cylinder is illustrated in \cref{fig:figure14}(c). In \cref{fig:figure14}(d), the placement of the synthetic jets and the resulting streamlines around the cylinder are displayed.
On one side of the cylinder, the influence of the jet's blowing action enlarges the flow separation region, even forming a smaller separation bubble ahead of a larger one. 
Conversely, on the other side of the cylinder, the suction action of the jet creates a relatively smaller separation bubble near the leading corner. The separated flow near the leading edge quickly reattaches to the sidewall, undergoes flow separation again at the trailing edge, and sheds a vortex.
Under the control of the synthetic jets, the near-field vortices continuously shed alternately, while the far-field wake gradually stabilizes.

Moreover, the synthetic jets are positioned near the trailing corner of the square cylinder, resulting in a stabilized wake flow after control in \cref{fig:figure14}(e).
\cref{fig:figure14}(f) illustrates the placement of the synthetic jets and the streamlines around the square cylinder. Due to the influence of the synthetic jets, a stable and symmetric separation bubble forms on the leeward side of the cylinder, with minimal oscillation observed in the wake region's streamlines.
Under the flow conditions at $Re = 500$, the stability of the controlled flow is effectively enhanced when the synthetic jets are positioned near the leading edge. When the synthetic jets are placed near the trailing edge, the controlled flow achieves a stable state without vortex shedding.
In the work of \citeauthor{yan2023stabilizing}\cite{yan2023stabilizing}, the DRL control framework they designed focuses on reducing $C_D$ and $C_L$, and the control strategy produces obvious control effects at multiple Reynolds numbers.
In our research, we not only focus on reducing the lift and drag coefficients, but also on controlling the stability of the wake region and the detachment of the vortex behind the square cylinder. 
The control strategy we obtained completely suppresses the shedding of vortices in the square cylinder wake area under the flow of $Re=500$.
The first reason for this discrepancy is that, in our research, the agent is capable of observing not only the flow field information around the square cylinder but also, more importantly, the flow field information in the most pulsating regions of the wake field.
\citeauthor{liReinforcementlearning}\cite{liReinforcementlearning} employed a similar probe design approach and developed a control strategy capable of completely suppressing the Kármán vortex street phenomenon in the wake field of a cylinder.
More importantly, near the trailing corners, the synthetic jets, by executing blowing or suction actions, introduce small-scale vortex structures, leading to the formation of stable and symmetric separation vortices behind the square cylinder. At $Re = 500$, positioning the synthetic jets near the trailing corners can directly influence the development of the boundary layer, effectively disrupting flow separation and enhancing flow stability.

\begin{figure*}[htbp]
    \centering
    \includegraphics{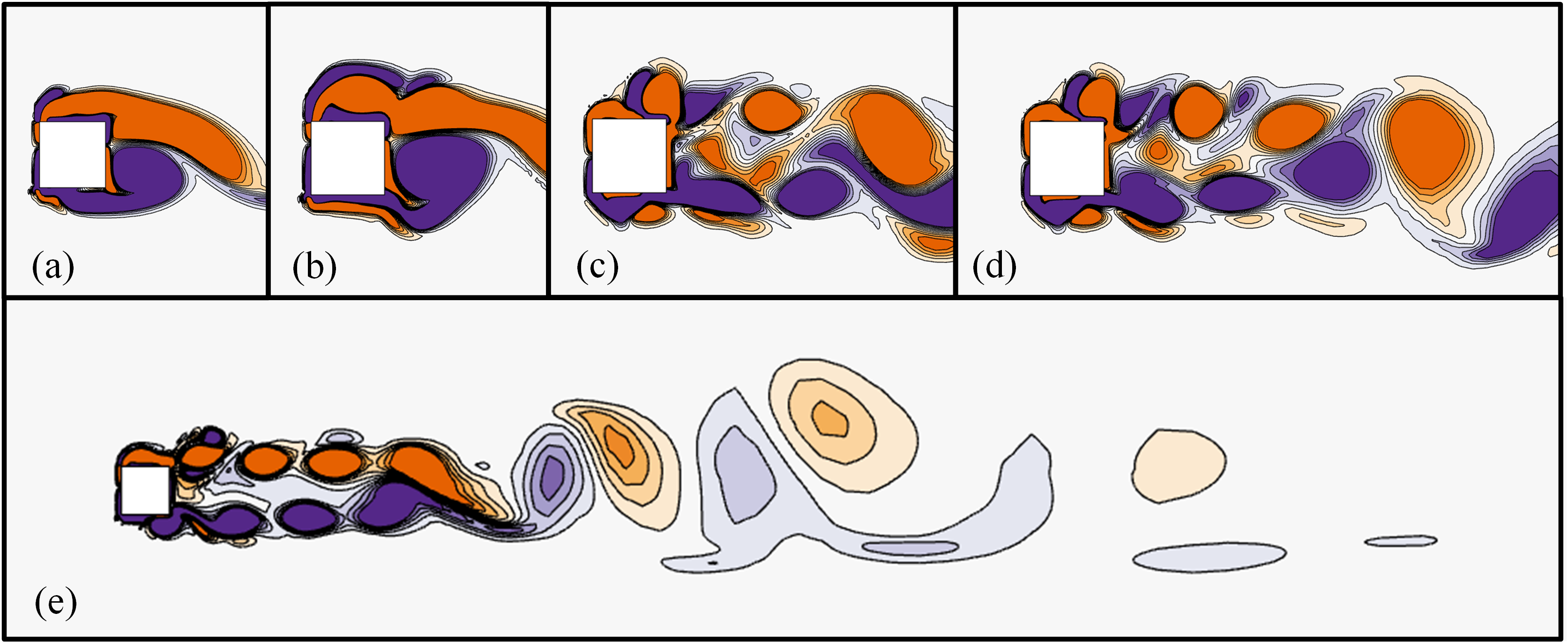}
    \caption{At $Re=500$, the instantaneous vorticity at characteristic moments during the control process are displayed for synthetic jets positioned near the leading corner points of the square cylinder. (a) $t_1$; (b) $t_2$; (c) $t_3$; (d) $t_4$; (e) $t_5$.}
    \label{fig:figure15}
\end{figure*}

To comprehensively illustrate the control process of synthetic jets positioned near the leading and trailing corner points on the wake flow field around a square cylinder, we present the case of $Re=500$. We display the contours of instantaneous vorticity from the initiation to the end of the control process. Initially, the activation of the synthetic jets affects only the flow near the square cylinder. We focus on this region initially, and as the control progresses, the influence extends to the wake field, ultimately showcasing the instantaneous vorticity values across the entire computational domain.
When the synthetic jets are positioned near the leading edge of the square cylinder, the evolution of the instantaneous vorticity field during the control process is described using five representative time snapshots, as depicted in \cref{fig:figure15}. 
In \cref{fig:figure15}(a), the control action initiates near the leading edge of the square cylinder, causing a slight disruption in the originally continuous vorticity field. Moving to \cref{fig:figure15}(b), as the synthetic jets near the leading edge continue their control action, their influence on the vorticity around the square cylinder gradually extends trailingward. By the time \cref{fig:figure15}(c) is reached, vortices generated from flow separation near the leading edge reattach to the cylinder's sidewalls, elongating along the walls towards the trailing corner of the cylinder.
As the flow control progresses to \cref{fig:figure15}(d), the originally continuous large vortex around the cylinder is disrupted and divided into several smaller vortices. These smaller vortices develop towards the trailing corner, and after separating again, form alternating vortex shedding in the wake region. When the synthetic jets are located near the leading edge of the square cylinder, after the flow control reaches a stable state in \cref{fig:figure15}(e), alternating vortex shedding occurs on the leeward side of the cylinder, and the shed vortices gradually dissipate in the far field due to the dissipation effect.

\begin{figure*}[htbp]
    \centering
    \includegraphics{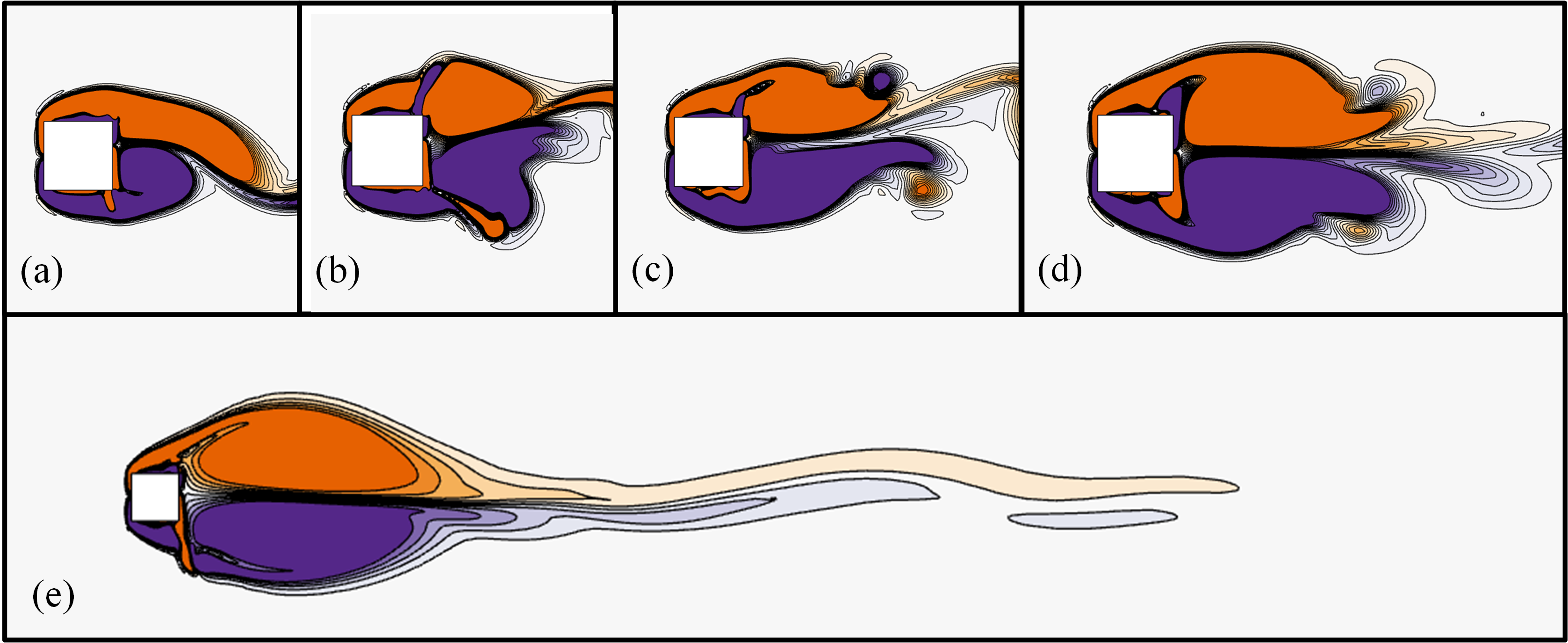}
    \caption{At $Re=500$, the instantaneous vorticity at characteristic moments during the control process are displayed for synthetic jets positioned near the trailing corner points of the square cylinder. (a) $t_1$; (b) $t_2$; (c) $t_3$; (d) $t_4$; (e) $t_5$.}
    \label{fig:figure16}
\end{figure*}

When the synthetic jets are positioned near the trailing edge of the square cylinder, as shown in \cref{fig:figure16}(a), the control process begins with one side of the square cylinder initiating a suction action while the other side performs a blowing action.
In \cref{fig:figure16}(b), at time $t_1$, the control action disrupts the vorticity field downstream, and by \cref{fig:figure16}(c), a new pair of vortices has formed downstream of the square cylinder. 
The originally periodic oscillation of positive and negative vortices on the trailing side of the square cylinder gradually develops into a symmetrical pattern under the influence of the synthetic jets, as depicted in \cref{fig:figure16}(d). 
In \cref{fig:figure16}(e), as the flow control progresses to its conclusion, the recirculation region on the backside of the square cylinder appears symmetrical and stable, with the phenomenon of vortex shedding completely suppressed in the wake flow field.

\subsubsection{Width of Synthetic Jets}

\begin{figure*}[htbp]
    \centering
    \includegraphics[width=\textwidth]{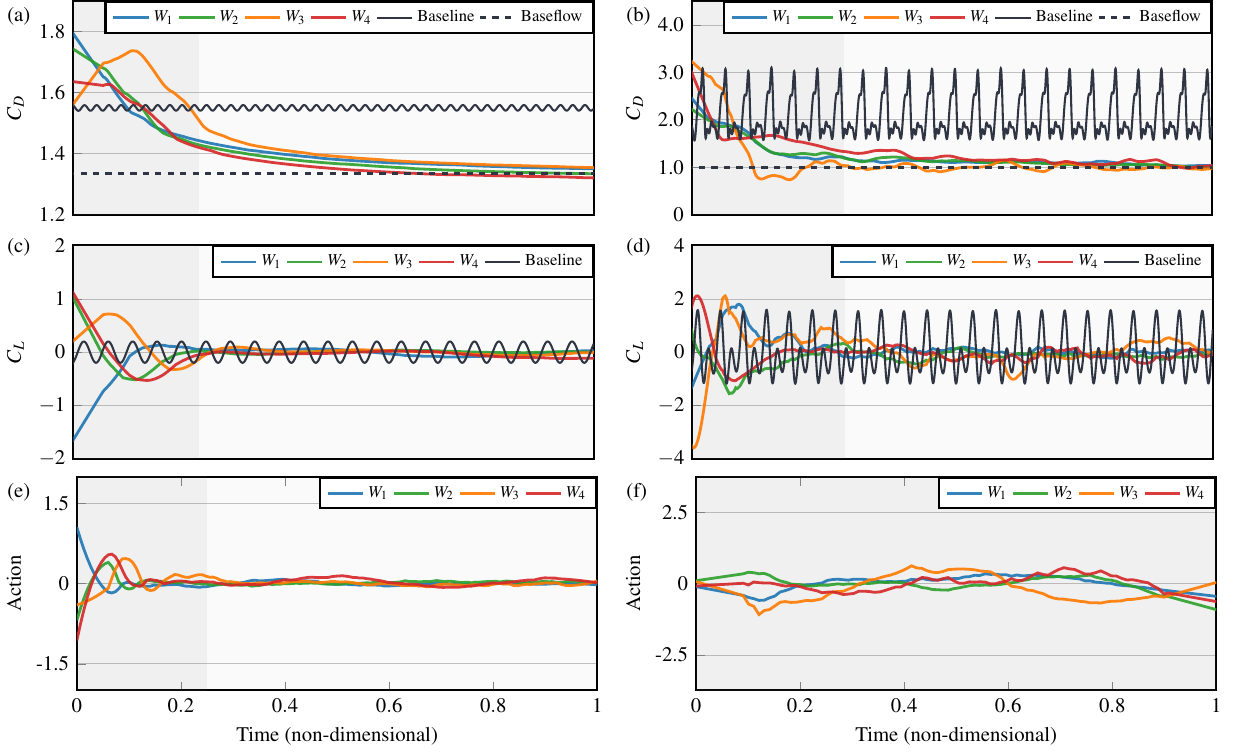}
    \caption{Effect of synthetic jets width on flow control, for $Re=100$ and $Re=500$. $W_1$, $W_2$, $W_3$, and $W_4$ represent jets width of $D/20$, $D/25$, $D/30$, and $D/35$, respectively. (a) $Re=100$, $C_D$; (b) $Re=500$, $C_D$; (c) $Re=100$, $C_L$; (d) $Re=500$, $C_L$; (e) $Re=100$, Action; (f) $Re=500$, Action.}
    \label{fig:figure17}
\end{figure*}

Synthetic jets parameters, particularly the width of the jets, significantly influence flow control performance alongside its positioning. In this section, we focus on studying how the width of synthetic jets impacts flow control effectiveness.
To achieve the complete suppression of vortices in the wake flow, we position the synthetic jets near the trailing corner of the square cylinder.
The width of the synthetic jets under study ranges from $D/20$, $D/25$, $D/30$ to $D/35$, where D represents the characteristic dimension of the square cylinder. 
Performance metrics for flow control include the reduction rate of the average drag coefficient, and the standard deviations of lift and drag coefficients, as well as the average and standard deviation of action. The stability of training results is evaluated using the standard deviations of $C_D$, $C_L$, and action. We conduct experiments under two flow conditions, $Re=100$ and $Re=500$, to test the training results for the four jets width. The summarized performance metrics are presented in \cref{tab:tab4}, and the time-history curves of $C_D$, $C_L$, and action are illustrated in \cref{fig:figure17}.

\begin{table}[h]
\centering
\begin{threeparttable}
\caption{Impact of Synthetic Jets Width on AFC Control Performance: Optimizing Width Selection.}
\label{tab:tab4}
\begin{tabular}{ccccccccccccc}
\toprule
\toprule
\multicolumn{2}{c}{} & \multicolumn{3}{c}{Mean $C_D$} & \multicolumn{3}{c}{Std of $C_D$} & \multicolumn{3}{c}{Std of $C_L$} & \multicolumn{2}{c}{Action} \\ 
\cmidrule(lr){3-5} \cmidrule(lr){6-8} \cmidrule(lr){9-11} \cmidrule(lr){12-13}
$Re$ & Jet Width & $C_{D,Baseline}$ & $C_{D,DRL}$ & Reduction & $C_{D,Baseline}$ & $C_{D,DRL}$ & Reduction & $C_{L,Baseline}$ & $C_{L,DRL}$ & Reduction & Action Mean & Action Std \\
\midrule
100 & $D/20$ & 1.549 & 1.346 & 13.1 \% & 2.020 & 0.001 & 99.9 \% & 0.179 & 0.028 & 84.3 \% & -0.016 & 0.146 \\
100 & $D/25$ & 1.549 & 1.325 & 14.4 \% & 2.020 & 0.001 & 99.9 \% & 0.179 & 0.025 & 86.1 \% & 0.037 & 0.164 \\
100 & $D/30$ & 1.549 & 1.345 & 13.2 \% & 2.020 & 0.003 & 99.9 \% & 0.179 & 0.037 & 79.0 \% & 0.063 & 0.316 \\
100 & $D/35$ & 1.549 & 1.314 & 15.2 \% & 2.020 & 0.001 & 99.9 \% & 0.179 & 0.056 & 68.5 \% & 0.072 & 0.290 \\
500 & $D/20$ & 2.060 & 1.021 & 50.4 \% & 0.441 & 0.023 & 94.8 \% & 1.173 & 0.186 & 84.2 \% & -0.384 & 1.048 \\
500 & $D/25$ & 2.060 & 1.001 & 51.4 \% & 0.441 & 0.042 & 90.5 \% & 1.173 & 0.294 & 74.9 \% & -0.702 & 1.223 \\
500 & $D/30$ & 2.060 & 0.968 & 53.0 \% & 0.441 & 0.080 & 81.8 \% & 1.173 & 0.291 & 75.2 \% & -2.068 & 1.857 \\
500 & $D/35$ & 2.060 & 1.053 & 48.9 \% & 0.441 & 0.219 & 50.3 \% & 1.173 & 0.583 & 50.3 \% & -0.789 & 4.333 \\
\bottomrule
\bottomrule
\end{tabular}
\begin{footnotesize}
\begin{tablenotes}[flushleft]
\small
\item[] a. $C_{D,Baseline}$ represents the time-averaged drag coefficient of the square cylinder when the baseline flow is fully developed.
\item[] b. $C_{D,DRL}$ represents the time-averaged drag coefficient of the square cylinder after AFC based on DRL. 
\item[] c. The reduction is used to quantify the reduction ratio of the DRL control result relative to the baseline, expressed in percentage (\%).
\item[] d. Action represents the ratio of the mass flow rate of the synthetic jets to the flow rate from the inlet.
\end{tablenotes}
\end{footnotesize}
\end{threeparttable}
\end{table}

\cref{tab:tab4} presents the results of training the jets at four different widths.
At $Re = 100$, the average drag coefficient reduction ranges from 13.1\% to 15.2\% as the width of the synthetic jets transitions from $D/20$, $D/25$, $D/30$ to $D/35$, with the standard deviations of the $C_D$ all approaching 99.9\%.
The impact on the drag reduction rate and the stability of the $C_D$ is minor across the four widths of the synthetic jets. 
However, the reduction in the standard deviation of the $C_L$ under $D/30$ and $D/35$ widths is smaller than that under $D/20$ and $D/25$.
Further reduction in the width of the synthetic jets adversely affects the stability of the $C_L$. 
The average and standard deviation of actions are greater under $D/30$ and $D/35$ compared to $D/20$ and $D/25$.
Decreasing the width of the synthetic jets leads to an increase in both the average and standard deviation of actions utilized.
This shows that as the jet width decreases, the control system requires larger amplitude and varying range of actions to achieve the same control effect.
At $Re=500$, the reduction rate in the average drag coefficient ranges from 48.9\% to 53.0\%, with the width of the synthetic jets having minimal impact on the drag reduction rate. However, the reduction rates in the standard deviations of the $C_D$ and $C_L$ decrease with decreasing width of the synthetic jets.
Additionally, the average value and standard deviation of action show consistency with $Re=100$, where narrowing the width of the synthetic jets results in an increase in both the average value and standard deviation of action.

The time curves of $C_D$, $C_L$, and control actions for the square cylinder at $Re = 100$ and $Re = 500$ are depicted in \cref{fig:figure17}. For the $Re = 100$ case, at four different synthetic jets widths, the initial stages show significant fluctuations in $C_D$, $C_L$, and control actions. After this period of intense fluctuation, $C_L$ and control actions stabilize near zero, while $C_D$ stabilizes near its minimum value. Upon observing the stabilized results, it is evident that the width of the synthetic jets has a minimal impact on drag reduction, although narrower jets width adversely affect the stability of $C_L$ and control actions.
In the case of $Re = 500$, $C_D$, $C_L$, and control actions also experience significant fluctuations during the initial stages across all four synthetic jets widths. However, the stability of $C_D$, $C_L$, and control actions is generally worse compared to the $Re = 100$ case. This is attributed to the inherently more unstable flow field at $Re = 500$, making control more challenging. Comparing the control results for different synthetic jets widths, further narrowing the jet width has a more significant negative impact on the stability of $C_D$, $C_L$, and control actions.

\section{CONCLUSIONS}\label{sec:Conclusions} 

This study employs CFD as the training environment for a DRL framework to conduct an AFC investigation on the mass flow rate of synthetic jets on the upper and lower surfaces of a square cylinder, with a focus on examining the influence of the position and width of the synthetic jets on the flow control performance. Through iterative optimization, the DRL agent, specifically a SAC agent, is found to be capable of achieving significant reductions in the lift and drag coefficients of a square cylinder, as well as completely suppress vortex shedding in the wake. The main findings of the study are:

\begin{enumerate}

    \item DRL-based control is shown to effectively control flow around square cylinders at two different Reynolds numbers of 100 and 500. At $Re$ = 100, synthetic jets located near the trailing edge corners achieve the maximum drag reduction effect, with the mean drag coefficient reduced by 14.4\%, the standard deviation of the drag coefficient reduced by 99\%, and the standard deviation of the lift coefficient reduced by 86.1\%. 
    At $Re$ = 500, synthetic jets located near the leading edge corners achieve the maximum drag reduction effect, with the mean drag coefficient reduced by 51.4\%, the standard deviation of the drag coefficient reduced by 90.5\%, and the standard deviation of the lift coefficient reduced by 74.9\%.
    
    \item In all numerical experiments, the DRL-based flow control strategy enables dynamic adjustments to variations in the physical field. In the beginning, a large jet velocity is often used to rapidly intervene and adjust the flow state. Subsequently, a smaller synthetic jet velocity is employed to sustain the flow control effect and reduce energy consumption. The overall control strategy demonstrates the ability of DRL control in achieving drag reduction while minimizing energy consumption under various flow conditions.

    \item The control effectiveness is found to be highly sensitive to the location of synthetic jets Positioning synthetic jets near the trailing corners is advantageous for drag reduction at Re = 100. In contrast, at Re = 500, positioning synthetic jets near the leading corners yields a more pronounced drag reduction effect. Detailed investigations of control mechanisms reveal that optimal control is achieved when the synthetic jets are located at the primary flow separation points. At primary flow separation points, the synthetic jet can most effectively modify the behavior of the boundary layers, directly influencing the location and process of flow separation. By executing blowing or suction actions, the synthetic jets introduce small-scale vortical structures that interfere with the separation bubble to achieve flow control. Without the feeding flow physics into the DRL algorithm, the DRL agent is capable of learning the precise intervention that optimizes the flow control objectives adaptively.
    
    \item The control effectiveness of drag reduction is found to be relatively insensitive to the width of synthetic jets. However, narrower jets can lead to decreased flow stability, particularly at higher Reynolds numbers. A narrower jet concentrates the injected or extracted fluid in a smaller area, resulting in uneven momentum distribution within the boundary layer, which affects the overall flow stability.    
    At $Re = 100$, when the jets width decreases from  $D/20$  to  $D/35$, the standard deviation of the lift coefficient is reduced by 15.8\%. 
    When the jet width decreases from  $D/20$  to  $D/35$ at $R e= 500$, the standard deviation of the drag coefficient is reduced by 44.5\%, and the standard deviation of the lift coefficient is reduced by 33.9\%.
\end{enumerate}

DRL-based AFC offers significant advantages in handling complex nonlinear systems and high-dimensional state spaces, providing powerful tools for flow control. DRL employs a reward mechanism to identify efficient control strategies that significantly reduce lift and drag coefficients and suppress vortex shedding. Investigating the impact of the position and width of synthetic jets on flow control performance can further optimize control strategies, enhance flow stability, and provide valuable insights for practical engineering applications.

\section*{ACKNOWLEDGMENTS}

The authors would like to express their gratitude to Dr. Jean Rabault (University of Oslo, Oslo, Norway), Dr. Jichao Li (National University of Singapore, Singapore), and Mr. Qiulei Wang (The University of Hong Kong, Hong Kong SAR, China) for making their open-source codes for deep reinforcement learning and numerical simulation available online. These resources have significantly contributed to the research presented in this paper. 
The relevant code repositories can be accessed at \url{https://github.com/jerabaul29/Cylinder2DFlowControlDRLParallel} \cite{rabault2024cylinder2dflowcontroldrlparallel,rabault2019artificial},
\url{https://github.com/npuljc/RL_control_Nek5000} \cite{Nek5000,liReinforcementlearning},
and \url{https://github.com/venturi123/DRLinFluids} \cite{DRLinFluids,wangDRLinFluids}.

\section*{AUTHOR DECLARATIONS}

\subsection{Conflict of Interest}

The authors report no conflict of interest.

\subsection{Author Contributions}

\section*{DATA AVAILABILITY}

The data that support the findings of this study are available from the corresponding author upon reasonable request.

\section*{AUTHOR ORCIDs}

\noindent Wang Jia \href{https://orcid.org/0009-0008-2786-397X}{https://orcid.org/0009-0008-2786-397X}\\
Hang Xu \href{https://orcid.org/0000-0003-4176-0738}{https://orcid.org/0000-0003-4176-0738}.

\section*{References}

\bibliography{aipsamp}

\begin{thebibliography}{62}%
\makeatletter
\providecommand \@ifxundefined [1]{%
 \@ifx{#1\undefined}
}%
\providecommand \@ifnum [1]{%
 \ifnum #1\expandafter \@firstoftwo
 \else \expandafter \@secondoftwo
 \fi
}%
\providecommand \@ifx [1]{%
 \ifx #1\expandafter \@firstoftwo
 \else \expandafter \@secondoftwo
 \fi
}%
\providecommand \natexlab [1]{#1}%
\providecommand \enquote  [1]{``#1''}%
\providecommand \bibnamefont  [1]{#1}%
\providecommand \bibfnamefont [1]{#1}%
\providecommand \citenamefont [1]{#1}%
\providecommand \href@noop [0]{\@secondoftwo}%
\providecommand \href [0]{\begingroup \@sanitize@url \@href}%
\providecommand \@href[1]{\@@startlink{#1}\@@href}%
\providecommand \@@href[1]{\endgroup#1\@@endlink}%
\providecommand \@sanitize@url [0]{\catcode `\\12\catcode `\$12\catcode
  `\&12\catcode `\#12\catcode `\^12\catcode `\_12\catcode `\%12\relax}%
\providecommand \@@startlink[1]{}%
\providecommand \@@endlink[0]{}%
\providecommand \url  [0]{\begingroup\@sanitize@url \@url }%
\providecommand \@url [1]{\endgroup\@href {#1}{\urlprefix }}%
\providecommand \urlprefix  [0]{URL }%
\providecommand \Eprint [0]{\href }%
\providecommand \doibase [0]{http://dx.doi.org/}%
\providecommand \selectlanguage [0]{\@gobble}%
\providecommand \bibinfo  [0]{\@secondoftwo}%
\providecommand \bibfield  [0]{\@secondoftwo}%
\providecommand \translation [1]{[#1]}%
\providecommand \BibitemOpen [0]{}%
\providecommand \bibitemStop [0]{}%
\providecommand \bibitemNoStop [0]{.\EOS\space}%
\providecommand \EOS [0]{\spacefactor3000\relax}%
\providecommand \BibitemShut  [1]{\csname bibitem#1\endcsname}%
\let\auto@bib@innerbib\@empty
\bibitem [{\citenamefont {Manohar}\ \emph {et~al.}(2018)\citenamefont
  {Manohar}, \citenamefont {Brunton}, \citenamefont {Kutz},\ and\ \citenamefont
  {Brunton}}]{DataDriven}%
  \BibitemOpen
  \bibfield  {author} {\bibinfo {author} {\bibfnamefont {K.}~\bibnamefont
  {Manohar}}, \bibinfo {author} {\bibfnamefont {B.~W.}\ \bibnamefont
  {Brunton}}, \bibinfo {author} {\bibfnamefont {J.~N.}\ \bibnamefont {Kutz}}, \
  and\ \bibinfo {author} {\bibfnamefont {S.~L.}\ \bibnamefont {Brunton}},\
  }\bibfield  {title} {\enquote {\bibinfo {title} {Data-driven sparse sensor
  placement for reconstruction: Demonstrating the benefits of exploiting known
  patterns},}\ }\href {\doibase 10.1109/MCS.2018.2810460} {\bibfield  {journal}
  {\bibinfo  {journal} {IEEE Control Systems Magazine}\ }\textbf {\bibinfo
  {volume} {38}},\ \bibinfo {pages} {63--86} (\bibinfo {year}
  {2018})}\BibitemShut {NoStop}%
\bibitem [{\citenamefont {Luchtenburg}(2021)}]{Datadrivenscience}%
  \BibitemOpen
  \bibfield  {author} {\bibinfo {author} {\bibfnamefont {D.~M.}\ \bibnamefont
  {Luchtenburg}},\ }\bibfield  {title} {\enquote {\bibinfo {title} {Data-driven
  science and engineering: machine learning, dynamical systems, and control
  (brunton, steven l. and kutz, j. nathan; 2020) [bookshelf]},}\ }\href
  {\doibase 10.1109/MCS.2021.3076544} {\bibfield  {journal} {\bibinfo
  {journal} {IEEE Control Systems Magazine}\ }\textbf {\bibinfo {volume}
  {41}},\ \bibinfo {pages} {95--102} (\bibinfo {year} {2021})}\BibitemShut
  {NoStop}%
\bibitem [{\citenamefont {Brunton}, \citenamefont {Noack},\ and\ \citenamefont
  {Koumoutsakos}(2020)}]{annurevfluid}%
  \BibitemOpen
  \bibfield  {author} {\bibinfo {author} {\bibfnamefont {S.~L.}\ \bibnamefont
  {Brunton}}, \bibinfo {author} {\bibfnamefont {B.~R.}\ \bibnamefont {Noack}},
  \ and\ \bibinfo {author} {\bibfnamefont {P.}~\bibnamefont {Koumoutsakos}},\
  }\bibfield  {title} {\enquote {\bibinfo {title} {Machine learning for fluid
  mechanics},}\ }\href {\doibase 10.1146/annurev-fluid-010719-060214}
  {\bibfield  {journal} {\bibinfo  {journal} {Annual Review of Fluid
  Mechanics}\ }\textbf {\bibinfo {volume} {52}},\ \bibinfo {pages} {477--508}
  (\bibinfo {year} {2020})}\BibitemShut {NoStop}%
\bibitem [{\citenamefont {Brunton}\ and\ \citenamefont
  {Noack}(2015)}]{Brunton2015}%
  \BibitemOpen
  \bibfield  {author} {\bibinfo {author} {\bibfnamefont {S.~L.}\ \bibnamefont
  {Brunton}}\ and\ \bibinfo {author} {\bibfnamefont {B.~R.}\ \bibnamefont
  {Noack}},\ }\bibfield  {title} {\enquote {\bibinfo {title} {Closed-loop
  turbulence control: Progress and challenges},}\ }\href {\doibase
  10.1115/1.4031175} {\bibfield  {journal} {\bibinfo  {journal} {ASME Applied
  Mechanics Reviews}\ }\textbf {\bibinfo {volume} {67}},\ \bibinfo {pages}
  {050801} (\bibinfo {year} {2015})}\BibitemShut {NoStop}%
\bibitem [{\citenamefont {Brunton}, \citenamefont {Proctor},\ and\
  \citenamefont {Kutz}(2016)}]{Stevendynamical}%
  \BibitemOpen
  \bibfield  {author} {\bibinfo {author} {\bibfnamefont {S.~L.}\ \bibnamefont
  {Brunton}}, \bibinfo {author} {\bibfnamefont {J.~L.}\ \bibnamefont
  {Proctor}}, \ and\ \bibinfo {author} {\bibfnamefont {J.~N.}\ \bibnamefont
  {Kutz}},\ }\bibfield  {title} {\enquote {\bibinfo {title} {Discovering
  governing equations from data by sparse identification of nonlinear dynamical
  systems},}\ }\href {\doibase 10.1073/pnas.1517384113} {\bibfield  {journal}
  {\bibinfo  {journal} {Proceedings of the National Academy of Sciences}\
  }\textbf {\bibinfo {volume} {113}},\ \bibinfo {pages} {3932--3937} (\bibinfo
  {year} {2016})}\BibitemShut {NoStop}%
\bibitem [{\citenamefont {Mahesh}(2020)}]{mahesh2020machine}%
  \BibitemOpen
  \bibfield  {author} {\bibinfo {author} {\bibfnamefont {B.}~\bibnamefont
  {Mahesh}},\ }\bibfield  {title} {\enquote {\bibinfo {title} {Machine learning
  algorithms-a review},}\ }\href {\doibase 10.21275/ART20203995} {\bibfield
  {journal} {\bibinfo  {journal} {International Journal of Science and Research
  (IJSR).[Internet]}\ }\textbf {\bibinfo {volume} {9}},\ \bibinfo {pages}
  {381--386} (\bibinfo {year} {2020})}\BibitemShut {NoStop}%
\bibitem [{\citenamefont {{Scott Collis}}\ \emph {et~al.}(2004)\citenamefont
  {{Scott Collis}}, \citenamefont {Joslin}, \citenamefont {Seifert},\ and\
  \citenamefont {Theofilis}}]{SCOTTCOLLIS2004237}%
  \BibitemOpen
  \bibfield  {author} {\bibinfo {author} {\bibfnamefont {S.}~\bibnamefont
  {{Scott Collis}}}, \bibinfo {author} {\bibfnamefont {R.~D.}\ \bibnamefont
  {Joslin}}, \bibinfo {author} {\bibfnamefont {A.}~\bibnamefont {Seifert}}, \
  and\ \bibinfo {author} {\bibfnamefont {V.}~\bibnamefont {Theofilis}},\
  }\bibfield  {title} {\enquote {\bibinfo {title} {Issues in active flow
  control: theory, control, simulation, and experiment},}\ }\href {\doibase
  https://doi.org/10.1016/j.paerosci.2004.06.001} {\bibfield  {journal}
  {\bibinfo  {journal} {Progress in Aerospace Sciences}\ }\textbf {\bibinfo
  {volume} {40}},\ \bibinfo {pages} {237--289} (\bibinfo {year}
  {2004})}\BibitemShut {NoStop}%
\bibitem [{\citenamefont {Perdikaris}, \citenamefont {Venturi},\ and\
  \citenamefont {Karniadakis}(2016)}]{Multifidelity}%
  \BibitemOpen
  \bibfield  {author} {\bibinfo {author} {\bibfnamefont {P.}~\bibnamefont
  {Perdikaris}}, \bibinfo {author} {\bibfnamefont {D.}~\bibnamefont {Venturi}},
  \ and\ \bibinfo {author} {\bibfnamefont {G.~E.}\ \bibnamefont
  {Karniadakis}},\ }\bibfield  {title} {\enquote {\bibinfo {title}
  {Multifidelity information fusion algorithms for high-dimensional systems and
  massive data sets},}\ }\href {\doibase 10.1137/15M1055164} {\bibfield
  {journal} {\bibinfo  {journal} {SIAM Journal on Scientific Computing}\
  }\textbf {\bibinfo {volume} {38}},\ \bibinfo {pages} {B521--B538} (\bibinfo
  {year} {2016})}\BibitemShut {NoStop}%
\bibitem [{\citenamefont {Bewley}, \citenamefont {Moin},\ and\ \citenamefont
  {Temam}(2001)}]{Bewley2001}%
  \BibitemOpen
  \bibfield  {author} {\bibinfo {author} {\bibfnamefont {T.~R.}\ \bibnamefont
  {Bewley}}, \bibinfo {author} {\bibfnamefont {P.}~\bibnamefont {Moin}}, \ and\
  \bibinfo {author} {\bibfnamefont {R.}~\bibnamefont {Temam}},\ }\bibfield
  {title} {\enquote {\bibinfo {title} {Dns-based predictive control of
  turbulence: An optimal benchmark for feedback algorithms},}\ }\href {\doibase
  10.1017/S0022112001005821} {\bibfield  {journal} {\bibinfo  {journal}
  {Journal of Fluid Mechanics}\ }\textbf {\bibinfo {volume} {447}},\ \bibinfo
  {pages} {179--225} (\bibinfo {year} {2001})}\BibitemShut {NoStop}%
\bibitem [{\citenamefont {Cattafesta}\ and\ \citenamefont
  {Sheplak}(2011)}]{annurevCattafesta}%
  \BibitemOpen
  \bibfield  {author} {\bibinfo {author} {\bibfnamefont {L.~N.}\ \bibnamefont
  {Cattafesta}}\ and\ \bibinfo {author} {\bibfnamefont {M.}~\bibnamefont
  {Sheplak}},\ }\bibfield  {title} {\enquote {\bibinfo {title} {Actuators for
  active flow control},}\ }\href {\doibase
  https://doi.org/10.1146/annurev-fluid-122109-160634} {\bibfield  {journal}
  {\bibinfo  {journal} {Annual Review of Fluid Mechanics}\ }\textbf {\bibinfo
  {volume} {43}},\ \bibinfo {pages} {247--272} (\bibinfo {year}
  {2011})}\BibitemShut {NoStop}%
\bibitem [{\citenamefont {Aram}\ \emph {et~al.}(2018)\citenamefont {Aram},
  \citenamefont {Lee}, \citenamefont {Shan},\ and\ \citenamefont
  {Vargas}}]{Aram2018}%
  \BibitemOpen
  \bibfield  {author} {\bibinfo {author} {\bibfnamefont {S.}~\bibnamefont
  {Aram}}, \bibinfo {author} {\bibfnamefont {Y.-T.}\ \bibnamefont {Lee}},
  \bibinfo {author} {\bibfnamefont {H.}~\bibnamefont {Shan}}, \ and\ \bibinfo
  {author} {\bibfnamefont {A.}~\bibnamefont {Vargas}},\ }\bibfield  {title}
  {\enquote {\bibinfo {title} {Computational fluid dynamic analysis of fluidic
  actuator for active flow control applications},}\ }\href {\doibase
  10.2514/1.J056255} {\bibfield  {journal} {\bibinfo  {journal} {AIAA Journal}\
  }\textbf {\bibinfo {volume} {56}},\ \bibinfo {pages} {111--120} (\bibinfo
  {year} {2018})}\BibitemShut {NoStop}%
\bibitem [{\citenamefont {Glezer}\ and\ \citenamefont
  {Amitay}(2002)}]{annurevfluidGlezer}%
  \BibitemOpen
  \bibfield  {author} {\bibinfo {author} {\bibfnamefont {A.}~\bibnamefont
  {Glezer}}\ and\ \bibinfo {author} {\bibfnamefont {M.}~\bibnamefont
  {Amitay}},\ }\bibfield  {title} {\enquote {\bibinfo {title} {Synthetic
  jets},}\ }\href {\doibase
  https://doi.org/10.1146/annurev.fluid.34.090501.094913} {\bibfield  {journal}
  {\bibinfo  {journal} {Annual Review of Fluid Mechanics}\ }\textbf {\bibinfo
  {volume} {34}},\ \bibinfo {pages} {503--529} (\bibinfo {year}
  {2002})}\BibitemShut {NoStop}%
\bibitem [{\citenamefont {Jahanmiri}(2010)}]{jahanmiri2010active}%
  \BibitemOpen
  \bibfield  {author} {\bibinfo {author} {\bibfnamefont {M.}~\bibnamefont
  {Jahanmiri}},\ }\bibfield  {title} {\enquote {\bibinfo {title} {Active flow
  control: a review},}\ }\href {\doibase 10.1016/j.flowmeasinst.2009.11.001}
  {\bibfield  {journal} {\bibinfo  {journal} {Flow Measurement and
  Instrumentation}\ }\textbf {\bibinfo {volume} {21}},\ \bibinfo {pages}
  {7--28} (\bibinfo {year} {2010})}\BibitemShut {NoStop}%
\bibitem [{\citenamefont {Smith}\ and\ \citenamefont
  {Glezer}(1998)}]{Smith1998}%
  \BibitemOpen
  \bibfield  {author} {\bibinfo {author} {\bibfnamefont {B.~L.}\ \bibnamefont
  {Smith}}\ and\ \bibinfo {author} {\bibfnamefont {A.}~\bibnamefont {Glezer}},\
  }\bibfield  {title} {\enquote {\bibinfo {title} {{The formation and evolution
  of synthetic jets}},}\ }\href {\doibase 10.1063/1.869828} {\bibfield
  {journal} {\bibinfo  {journal} {Physics of Fluids}\ }\textbf {\bibinfo
  {volume} {10}},\ \bibinfo {pages} {2281--2297} (\bibinfo {year}
  {1998})}\BibitemShut {NoStop}%
\bibitem [{\citenamefont {François-Lavet}\ \emph {et~al.}(2018)\citenamefont
  {François-Lavet}, \citenamefont {Henderson}, \citenamefont {Islam},
  \citenamefont {Bellemare},\ and\ \citenamefont {Pineau}}]{MAL071}%
  \BibitemOpen
  \bibfield  {author} {\bibinfo {author} {\bibfnamefont {V.}~\bibnamefont
  {François-Lavet}}, \bibinfo {author} {\bibfnamefont {P.}~\bibnamefont
  {Henderson}}, \bibinfo {author} {\bibfnamefont {R.}~\bibnamefont {Islam}},
  \bibinfo {author} {\bibfnamefont {M.~G.}\ \bibnamefont {Bellemare}}, \ and\
  \bibinfo {author} {\bibfnamefont {J.}~\bibnamefont {Pineau}},\ }\bibfield
  {title} {\enquote {\bibinfo {title} {An introduction to deep reinforcement
  learning},}\ }\href {\doibase 10.1561/2200000071} {\bibfield  {journal}
  {\bibinfo  {journal} {Foundations and Trends® in Machine Learning}\ }\textbf
  {\bibinfo {volume} {11}},\ \bibinfo {pages} {219--354} (\bibinfo {year}
  {2018})}\BibitemShut {NoStop}%
\bibitem [{\citenamefont {Xie}\ \emph {et~al.}(2023)\citenamefont {Xie},
  \citenamefont {Zheng}, \citenamefont {Ji}, \citenamefont {Zhang},
  \citenamefont {Bi}, \citenamefont {Zhou},\ and\ \citenamefont
  {Zheng}}]{Aerospace2023}%
  \BibitemOpen
  \bibfield  {author} {\bibinfo {author} {\bibfnamefont {F.}~\bibnamefont
  {Xie}}, \bibinfo {author} {\bibfnamefont {C.}~\bibnamefont {Zheng}}, \bibinfo
  {author} {\bibfnamefont {T.}~\bibnamefont {Ji}}, \bibinfo {author}
  {\bibfnamefont {X.}~\bibnamefont {Zhang}}, \bibinfo {author} {\bibfnamefont
  {R.}~\bibnamefont {Bi}}, \bibinfo {author} {\bibfnamefont {H.}~\bibnamefont
  {Zhou}}, \ and\ \bibinfo {author} {\bibfnamefont {Y.}~\bibnamefont {Zheng}},\
  }\bibfield  {title} {\enquote {\bibinfo {title} {Deep reinforcement learning:
  A new beacon for intelligent active flow control},}\ }\href {\doibase
  10.3389/arc.2023.11130} {\bibfield  {journal} {\bibinfo  {journal} {Aerospace
  Research Communications}\ }\textbf {\bibinfo {volume} {1}} (\bibinfo {year}
  {2023}),\ 10.3389/arc.2023.11130}\BibitemShut {NoStop}%
\bibitem [{\citenamefont {Smith}\ and\ \citenamefont
  {Swift}(2003)}]{smith2003comparison}%
  \BibitemOpen
  \bibfield  {author} {\bibinfo {author} {\bibfnamefont {B.}~\bibnamefont
  {Smith}}\ and\ \bibinfo {author} {\bibfnamefont {G.}~\bibnamefont {Swift}},\
  }\bibfield  {title} {\enquote {\bibinfo {title} {A comparison between
  synthetic jets and continuous jets},}\ }\href {\doibase
  10.1007/s00348-002-0577-6} {\bibfield  {journal} {\bibinfo  {journal}
  {Experimental Fluids}\ }\textbf {\bibinfo {volume} {34}},\ \bibinfo {pages}
  {467--472} (\bibinfo {year} {2003})}\BibitemShut {NoStop}%
\bibitem [{\citenamefont {Mnih}\ \emph {et~al.}(2013)\citenamefont {Mnih},
  \citenamefont {Kavukcuoglu}, \citenamefont {Silver}, \citenamefont {Graves},
  \citenamefont {Antonoglou}, \citenamefont {Wierstra},\ and\ \citenamefont
  {Riedmiller}}]{mnih2013playing}%
  \BibitemOpen
  \bibfield  {author} {\bibinfo {author} {\bibfnamefont {V.}~\bibnamefont
  {Mnih}}, \bibinfo {author} {\bibfnamefont {K.}~\bibnamefont {Kavukcuoglu}},
  \bibinfo {author} {\bibfnamefont {D.}~\bibnamefont {Silver}}, \bibinfo
  {author} {\bibfnamefont {A.}~\bibnamefont {Graves}}, \bibinfo {author}
  {\bibfnamefont {I.}~\bibnamefont {Antonoglou}}, \bibinfo {author}
  {\bibfnamefont {D.}~\bibnamefont {Wierstra}}, \ and\ \bibinfo {author}
  {\bibfnamefont {M.}~\bibnamefont {Riedmiller}},\ }\href@noop {} {\enquote
  {\bibinfo {title} {Playing atari with deep reinforcement learning},}\ }
  (\bibinfo {year} {2013}),\ \Eprint {http://arxiv.org/abs/1312.5602}
  {arXiv:1312.5602 [cs.LG]} \BibitemShut {NoStop}%
\bibitem [{\citenamefont {LeCun}, \citenamefont {Bengio},\ and\ \citenamefont
  {Hinton}(2015)}]{lecun2015deep}%
  \BibitemOpen
  \bibfield  {author} {\bibinfo {author} {\bibfnamefont {Y.}~\bibnamefont
  {LeCun}}, \bibinfo {author} {\bibfnamefont {Y.}~\bibnamefont {Bengio}}, \
  and\ \bibinfo {author} {\bibfnamefont {G.}~\bibnamefont {Hinton}},\
  }\bibfield  {title} {\enquote {\bibinfo {title} {Deep learning},}\ }\href
  {\doibase 10.1038/nature14539} {\bibfield  {journal} {\bibinfo  {journal}
  {Nature}\ }\textbf {\bibinfo {volume} {521}},\ \bibinfo {pages} {436--444}
  (\bibinfo {year} {2015})}\BibitemShut {NoStop}%
\bibitem [{\citenamefont {Mnih}\ \emph {et~al.}(2015)\citenamefont {Mnih},
  \citenamefont {Kavukcuoglu}, \citenamefont {Silver} \emph
  {et~al.}}]{mnih2015human}%
  \BibitemOpen
  \bibfield  {author} {\bibinfo {author} {\bibfnamefont {V.}~\bibnamefont
  {Mnih}}, \bibinfo {author} {\bibfnamefont {K.}~\bibnamefont {Kavukcuoglu}},
  \bibinfo {author} {\bibfnamefont {D.}~\bibnamefont {Silver}},  \emph
  {et~al.},\ }\bibfield  {title} {\enquote {\bibinfo {title} {Human-level
  control through deep reinforcement learning},}\ }\href {\doibase
  10.1038/nature14236} {\bibfield  {journal} {\bibinfo  {journal} {Nature}\
  }\textbf {\bibinfo {volume} {518}},\ \bibinfo {pages} {529--533} (\bibinfo
  {year} {2015})}\BibitemShut {NoStop}%
\bibitem [{\citenamefont {Lillicrap}\ \emph {et~al.}(2019)\citenamefont
  {Lillicrap}, \citenamefont {Hunt}, \citenamefont {Pritzel}, \citenamefont
  {Heess}, \citenamefont {Erez}, \citenamefont {Tassa}, \citenamefont
  {Silver},\ and\ \citenamefont {Wierstra}}]{lillicrap2019continuous}%
  \BibitemOpen
  \bibfield  {author} {\bibinfo {author} {\bibfnamefont {T.~P.}\ \bibnamefont
  {Lillicrap}}, \bibinfo {author} {\bibfnamefont {J.~J.}\ \bibnamefont {Hunt}},
  \bibinfo {author} {\bibfnamefont {A.}~\bibnamefont {Pritzel}}, \bibinfo
  {author} {\bibfnamefont {N.}~\bibnamefont {Heess}}, \bibinfo {author}
  {\bibfnamefont {T.}~\bibnamefont {Erez}}, \bibinfo {author} {\bibfnamefont
  {Y.}~\bibnamefont {Tassa}}, \bibinfo {author} {\bibfnamefont
  {D.}~\bibnamefont {Silver}}, \ and\ \bibinfo {author} {\bibfnamefont
  {D.}~\bibnamefont {Wierstra}},\ }\href@noop {} {\enquote {\bibinfo {title}
  {Continuous control with deep reinforcement learning},}\ } (\bibinfo {year}
  {2019}),\ \Eprint {http://arxiv.org/abs/1509.02971} {arXiv:1509.02971
  [cs.LG]} \BibitemShut {NoStop}%
\bibitem [{\citenamefont {Henderson}\ \emph {et~al.}(2019)\citenamefont
  {Henderson}, \citenamefont {Islam}, \citenamefont {Bachman}, \citenamefont
  {Pineau}, \citenamefont {Precup},\ and\ \citenamefont
  {Meger}}]{henderson2019deep}%
  \BibitemOpen
  \bibfield  {author} {\bibinfo {author} {\bibfnamefont {P.}~\bibnamefont
  {Henderson}}, \bibinfo {author} {\bibfnamefont {R.}~\bibnamefont {Islam}},
  \bibinfo {author} {\bibfnamefont {P.}~\bibnamefont {Bachman}}, \bibinfo
  {author} {\bibfnamefont {J.}~\bibnamefont {Pineau}}, \bibinfo {author}
  {\bibfnamefont {D.}~\bibnamefont {Precup}}, \ and\ \bibinfo {author}
  {\bibfnamefont {D.}~\bibnamefont {Meger}},\ }\href@noop {} {\enquote
  {\bibinfo {title} {Deep reinforcement learning that matters},}\ } (\bibinfo
  {year} {2019}),\ \Eprint {http://arxiv.org/abs/1709.06560} {arXiv:1709.06560
  [cs.LG]} \BibitemShut {NoStop}%
\bibitem [{\citenamefont {Li}(2018)}]{li2018deep}%
  \BibitemOpen
  \bibfield  {author} {\bibinfo {author} {\bibfnamefont {Y.}~\bibnamefont
  {Li}},\ }\href@noop {} {\enquote {\bibinfo {title} {Deep reinforcement
  learning: An overview},}\ } (\bibinfo {year} {2018}),\ \Eprint
  {http://arxiv.org/abs/1701.07274} {arXiv:1701.07274 [cs.LG]} \BibitemShut
  {NoStop}%
\bibitem [{\citenamefont {Kaiser}\ \emph {et~al.}(2024)\citenamefont {Kaiser},
  \citenamefont {Babaeizadeh}, \citenamefont {Milos}, \citenamefont {Osinski},
  \citenamefont {Campbell}, \citenamefont {Czechowski}, \citenamefont {Erhan},
  \citenamefont {Finn}, \citenamefont {Kozakowski}, \citenamefont {Levine},
  \citenamefont {Mohiuddin}, \citenamefont {Sepassi}, \citenamefont {Tucker},\
  and\ \citenamefont {Michalewski}}]{kaiser2024modelbased}%
  \BibitemOpen
  \bibfield  {author} {\bibinfo {author} {\bibfnamefont {L.}~\bibnamefont
  {Kaiser}}, \bibinfo {author} {\bibfnamefont {M.}~\bibnamefont {Babaeizadeh}},
  \bibinfo {author} {\bibfnamefont {P.}~\bibnamefont {Milos}}, \bibinfo
  {author} {\bibfnamefont {B.}~\bibnamefont {Osinski}}, \bibinfo {author}
  {\bibfnamefont {R.~H.}\ \bibnamefont {Campbell}}, \bibinfo {author}
  {\bibfnamefont {K.}~\bibnamefont {Czechowski}}, \bibinfo {author}
  {\bibfnamefont {D.}~\bibnamefont {Erhan}}, \bibinfo {author} {\bibfnamefont
  {C.}~\bibnamefont {Finn}}, \bibinfo {author} {\bibfnamefont {P.}~\bibnamefont
  {Kozakowski}}, \bibinfo {author} {\bibfnamefont {S.}~\bibnamefont {Levine}},
  \bibinfo {author} {\bibfnamefont {A.}~\bibnamefont {Mohiuddin}}, \bibinfo
  {author} {\bibfnamefont {R.}~\bibnamefont {Sepassi}}, \bibinfo {author}
  {\bibfnamefont {G.}~\bibnamefont {Tucker}}, \ and\ \bibinfo {author}
  {\bibfnamefont {H.}~\bibnamefont {Michalewski}},\ }\href@noop {} {\enquote
  {\bibinfo {title} {Model-based reinforcement learning for atari},}\ }
  (\bibinfo {year} {2024}),\ \Eprint {http://arxiv.org/abs/1903.00374}
  {arXiv:1903.00374 [cs.LG]} \BibitemShut {NoStop}%
\bibitem [{\citenamefont {Arulkumaran}\ \emph {et~al.}(2017)\citenamefont
  {Arulkumaran}, \citenamefont {Deisenroth}, \citenamefont {Brundage},\ and\
  \citenamefont {Bharath}}]{8103164}%
  \BibitemOpen
  \bibfield  {author} {\bibinfo {author} {\bibfnamefont {K.}~\bibnamefont
  {Arulkumaran}}, \bibinfo {author} {\bibfnamefont {M.~P.}\ \bibnamefont
  {Deisenroth}}, \bibinfo {author} {\bibfnamefont {M.}~\bibnamefont
  {Brundage}}, \ and\ \bibinfo {author} {\bibfnamefont {A.~A.}\ \bibnamefont
  {Bharath}},\ }\bibfield  {title} {\enquote {\bibinfo {title} {Deep
  reinforcement learning: A brief survey},}\ }\href {\doibase
  10.1109/MSP.2017.2743240} {\bibfield  {journal} {\bibinfo  {journal} {IEEE
  Signal Processing Magazine}\ }\textbf {\bibinfo {volume} {34}},\ \bibinfo
  {pages} {26--38} (\bibinfo {year} {2017})}\BibitemShut {NoStop}%
\bibitem [{\citenamefont {Rabault}\ \emph {et~al.}(2019)\citenamefont
  {Rabault}, \citenamefont {Kuchta}, \citenamefont {Jensen}, \citenamefont
  {R{\'e}glade},\ and\ \citenamefont {Cerardi}}]{rabault2019artificial}%
  \BibitemOpen
  \bibfield  {author} {\bibinfo {author} {\bibfnamefont {J.}~\bibnamefont
  {Rabault}}, \bibinfo {author} {\bibfnamefont {M.}~\bibnamefont {Kuchta}},
  \bibinfo {author} {\bibfnamefont {A.}~\bibnamefont {Jensen}}, \bibinfo
  {author} {\bibfnamefont {U.}~\bibnamefont {R{\'e}glade}}, \ and\ \bibinfo
  {author} {\bibfnamefont {N.}~\bibnamefont {Cerardi}},\ }\bibfield  {title}
  {\enquote {\bibinfo {title} {Artificial neural networks trained through deep
  reinforcement learning discover control strategies for active flow
  control},}\ }\href {\doibase 10.1017/jfm.2019.62} {\bibfield  {journal}
  {\bibinfo  {journal} {Journal of fluid mechanics}\ }\textbf {\bibinfo
  {volume} {865}},\ \bibinfo {pages} {281--302} (\bibinfo {year}
  {2019})}\BibitemShut {NoStop}%
\bibitem [{\citenamefont {Tang}\ \emph
  {et~al.}(2020{\natexlab{a}})\citenamefont {Tang}, \citenamefont {Rabault},
  \citenamefont {Kuhnle}, \citenamefont {Wang},\ and\ \citenamefont
  {Wang}}]{tang2020robust}%
  \BibitemOpen
  \bibfield  {author} {\bibinfo {author} {\bibfnamefont {H.}~\bibnamefont
  {Tang}}, \bibinfo {author} {\bibfnamefont {J.}~\bibnamefont {Rabault}},
  \bibinfo {author} {\bibfnamefont {A.}~\bibnamefont {Kuhnle}}, \bibinfo
  {author} {\bibfnamefont {Y.}~\bibnamefont {Wang}}, \ and\ \bibinfo {author}
  {\bibfnamefont {T.}~\bibnamefont {Wang}},\ }\bibfield  {title} {\enquote
  {\bibinfo {title} {Robust active flow control over a range of {{Reynolds}}
  numbers using an artificial neural network trained through deep reinforcement
  learning},}\ }\href {\doibase 10.1063/5.0006492} {\bibfield  {journal}
  {\bibinfo  {journal} {Physics of Fluids}\ }\textbf {\bibinfo {volume} {32}},\
  \bibinfo {pages} {053605} (\bibinfo {year} {2020}{\natexlab{a}})}\BibitemShut
  {NoStop}%
\bibitem [{\citenamefont {Paris}, \citenamefont {Beneddine},\ and\
  \citenamefont {Dandois}(2021)}]{parisRobustFlowControl2021}%
  \BibitemOpen
  \bibfield  {author} {\bibinfo {author} {\bibfnamefont {R.}~\bibnamefont
  {Paris}}, \bibinfo {author} {\bibfnamefont {S.}~\bibnamefont {Beneddine}}, \
  and\ \bibinfo {author} {\bibfnamefont {J.}~\bibnamefont {Dandois}},\
  }\bibfield  {title} {\enquote {\bibinfo {title} {Robust flow control and
  optimal sensor placement using deep reinforcement learning},}\ }\href
  {\doibase 10.1017/jfm.2020.1170} {\bibfield  {journal} {\bibinfo  {journal}
  {Journal of Fluid Mechanics}\ }\textbf {\bibinfo {volume} {913}},\ \bibinfo
  {pages} {A25} (\bibinfo {year} {2021})}\BibitemShut {NoStop}%
\bibitem [{\citenamefont {Li}\ and\ \citenamefont
  {Zhang}(2022)}]{liReinforcementlearning}%
  \BibitemOpen
  \bibfield  {author} {\bibinfo {author} {\bibfnamefont {J.}~\bibnamefont
  {Li}}\ and\ \bibinfo {author} {\bibfnamefont {M.}~\bibnamefont {Zhang}},\
  }\bibfield  {title} {\enquote {\bibinfo {title} {Reinforcement-learning-based
  control of confined cylinder wakes with stability analyses},}\ }\href
  {\doibase 10.1017/jfm.2021.1045} {\bibfield  {journal} {\bibinfo  {journal}
  {Journal of Fluid Mechanics}\ }\textbf {\bibinfo {volume} {932}},\ \bibinfo
  {pages} {A44} (\bibinfo {year} {2022})},\ \Eprint
  {http://arxiv.org/abs/2111.07498} {2111.07498 [physics]} \BibitemShut
  {NoStop}%
\bibitem [{\citenamefont {Fan}\ \emph {et~al.}(2020)\citenamefont {Fan},
  \citenamefont {Yang}, \citenamefont {Wang}, \citenamefont {Triantafyllou},\
  and\ \citenamefont {Karniadakis}}]{Dixiapnas}%
  \BibitemOpen
  \bibfield  {author} {\bibinfo {author} {\bibfnamefont {D.}~\bibnamefont
  {Fan}}, \bibinfo {author} {\bibfnamefont {L.}~\bibnamefont {Yang}}, \bibinfo
  {author} {\bibfnamefont {Z.}~\bibnamefont {Wang}}, \bibinfo {author}
  {\bibfnamefont {M.~S.}\ \bibnamefont {Triantafyllou}}, \ and\ \bibinfo
  {author} {\bibfnamefont {G.~E.}\ \bibnamefont {Karniadakis}},\ }\bibfield
  {title} {\enquote {\bibinfo {title} {Reinforcement learning for bluff body
  active flow control in experiments and simulations},}\ }\href {\doibase
  10.1073/pnas.2004939117} {\bibfield  {journal} {\bibinfo  {journal}
  {Proceedings of the National Academy of Sciences}\ }\textbf {\bibinfo
  {volume} {117}},\ \bibinfo {pages} {26091--26098} (\bibinfo {year}
  {2020})}\BibitemShut {NoStop}%
\bibitem [{\citenamefont {Rabault}\ and\ \citenamefont
  {Kuhnle}(2019{\natexlab{a}})}]{rabault2019accelerating}%
  \BibitemOpen
  \bibfield  {author} {\bibinfo {author} {\bibfnamefont {J.}~\bibnamefont
  {Rabault}}\ and\ \bibinfo {author} {\bibfnamefont {A.}~\bibnamefont
  {Kuhnle}},\ }\bibfield  {title} {\enquote {\bibinfo {title} {{Accelerating
  deep reinforcement learning strategies of flow control through a
  multi-environment approach}},}\ }\href {\doibase 10.1063/1.5116415}
  {\bibfield  {journal} {\bibinfo  {journal} {Physics of Fluids}\ }\textbf
  {\bibinfo {volume} {31}},\ \bibinfo {pages} {094105} (\bibinfo {year}
  {2019}{\natexlab{a}})}\BibitemShut {NoStop}%
\bibitem [{\citenamefont {Wang}\ and\ \citenamefont
  {Xu}(2024{\natexlab{a}})}]{jia2024optimal}%
  \BibitemOpen
  \bibfield  {author} {\bibinfo {author} {\bibfnamefont {J.}~\bibnamefont
  {Wang}}\ and\ \bibinfo {author} {\bibfnamefont {H.}~\bibnamefont {Xu}},\
  }\bibfield  {title} {\enquote {\bibinfo {title} {Optimal parallelization
  strategies for active flow control in deep reinforcement learning-based
  computational fluid dynamics},}\ }\href {\doibase 10.1063/5.0204237}
  {\bibfield  {journal} {\bibinfo  {journal} {Physics of Fluids}\ }\textbf
  {\bibinfo {volume} {36}},\ \bibinfo {pages} {043623} (\bibinfo {year}
  {2024}{\natexlab{a}})}\BibitemShut {NoStop}%
\bibitem [{\citenamefont {Wang}\ \emph {et~al.}(2022)\citenamefont {Wang},
  \citenamefont {Yan}, \citenamefont {Hu}, \citenamefont {Li}, \citenamefont
  {Xiao}, \citenamefont {Xiong}, \citenamefont {Rabault},\ and\ \citenamefont
  {Noack}}]{wangDRLinFluids}%
  \BibitemOpen
  \bibfield  {author} {\bibinfo {author} {\bibfnamefont {Q.}~\bibnamefont
  {Wang}}, \bibinfo {author} {\bibfnamefont {L.}~\bibnamefont {Yan}}, \bibinfo
  {author} {\bibfnamefont {G.}~\bibnamefont {Hu}}, \bibinfo {author}
  {\bibfnamefont {C.}~\bibnamefont {Li}}, \bibinfo {author} {\bibfnamefont
  {Y.}~\bibnamefont {Xiao}}, \bibinfo {author} {\bibfnamefont {H.}~\bibnamefont
  {Xiong}}, \bibinfo {author} {\bibfnamefont {J.}~\bibnamefont {Rabault}}, \
  and\ \bibinfo {author} {\bibfnamefont {B.~R.}\ \bibnamefont {Noack}},\
  }\bibfield  {title} {\enquote {\bibinfo {title} {{{DRLinFluids}}: {{An}}
  open-source {{Python}} platform of coupling deep reinforcement learning and
  {{OpenFOAM}}},}\ }\href {\doibase 10.1063/5.0103113} {\bibfield  {journal}
  {\bibinfo  {journal} {Physics of Fluids}\ }\textbf {\bibinfo {volume} {34}},\
  \bibinfo {pages} {081801} (\bibinfo {year} {2022})}\BibitemShut {NoStop}%
\bibitem [{\citenamefont {Castellanos}\ \emph {et~al.}(2022)\citenamefont
  {Castellanos}, \citenamefont {Cornejo~Maceda}, \citenamefont {de~la Fuente},
  \citenamefont {Noack}, \citenamefont {Ianiro},\ and\ \citenamefont
  {Discetti}}]{castellanos}%
  \BibitemOpen
  \bibfield  {author} {\bibinfo {author} {\bibfnamefont {R.}~\bibnamefont
  {Castellanos}}, \bibinfo {author} {\bibfnamefont {G.~Y.}\ \bibnamefont
  {Cornejo~Maceda}}, \bibinfo {author} {\bibfnamefont {I.}~\bibnamefont {de~la
  Fuente}}, \bibinfo {author} {\bibfnamefont {B.~R.}\ \bibnamefont {Noack}},
  \bibinfo {author} {\bibfnamefont {A.}~\bibnamefont {Ianiro}}, \ and\ \bibinfo
  {author} {\bibfnamefont {S.}~\bibnamefont {Discetti}},\ }\bibfield  {title}
  {\enquote {\bibinfo {title} {Machine-learning flow control with few sensor
  feedback and measurement noise},}\ }\href {\doibase 10.1063/5.0087208}
  {\bibfield  {journal} {\bibinfo  {journal} {Physics of Fluids}\ }\textbf
  {\bibinfo {volume} {34}},\ \bibinfo {pages} {047118} (\bibinfo {year}
  {2022})}\BibitemShut {NoStop}%
\bibitem [{\citenamefont {Jia}\ and\ \citenamefont {Xu}(2024)}]{jia2024deep}%
  \BibitemOpen
  \bibfield  {author} {\bibinfo {author} {\bibfnamefont {W.}~\bibnamefont
  {Jia}}\ and\ \bibinfo {author} {\bibfnamefont {H.}~\bibnamefont {Xu}},\
  }\href@noop {} {\enquote {\bibinfo {title} {Deep reinforcement learning-based
  active flow control of an elliptical cylinder: Transitioning from an
  elliptical cylinder to a circular cylinder and a flat plate},}\ } (\bibinfo
  {year} {2024}),\ \Eprint {http://arxiv.org/abs/2404.13003} {arXiv:2404.13003
  [physics.flu-dyn]} \BibitemShut {NoStop}%
\bibitem [{\citenamefont {Ren}, \citenamefont {Wang},\ and\ \citenamefont
  {Tang}(2021)}]{ren2021bluff}%
  \BibitemOpen
  \bibfield  {author} {\bibinfo {author} {\bibfnamefont {F.}~\bibnamefont
  {Ren}}, \bibinfo {author} {\bibfnamefont {C.}~\bibnamefont {Wang}}, \ and\
  \bibinfo {author} {\bibfnamefont {H.}~\bibnamefont {Tang}},\ }\bibfield
  {title} {\enquote {\bibinfo {title} {Bluff body uses
  deep-reinforcement-learning trained active flow control to achieve
  hydrodynamic stealth},}\ }\href {\doibase 10.1063/5.0060690} {\bibfield
  {journal} {\bibinfo  {journal} {Physics of Fluids}\ }\textbf {\bibinfo
  {volume} {33}},\ \bibinfo {pages} {093602} (\bibinfo {year}
  {2021})}\BibitemShut {NoStop}%
\bibitem [{\citenamefont {He}\ \emph {et~al.}(2023)\citenamefont {He},
  \citenamefont {Wang}, \citenamefont {Hua}, \citenamefont {Chen},
  \citenamefont {Li},\ and\ \citenamefont {Wu}}]{he2023policy}%
  \BibitemOpen
  \bibfield  {author} {\bibinfo {author} {\bibfnamefont {X.-J.}\ \bibnamefont
  {He}}, \bibinfo {author} {\bibfnamefont {Y.-Z.}\ \bibnamefont {Wang}},
  \bibinfo {author} {\bibfnamefont {Y.}~\bibnamefont {Hua}}, \bibinfo {author}
  {\bibfnamefont {Z.-H.}\ \bibnamefont {Chen}}, \bibinfo {author}
  {\bibfnamefont {Y.-B.}\ \bibnamefont {Li}}, \ and\ \bibinfo {author}
  {\bibfnamefont {W.-T.}\ \bibnamefont {Wu}},\ }\bibfield  {title} {\enquote
  {\bibinfo {title} {Policy transfer of reinforcement learning-based flow
  control: From two- to three-dimensional environment},}\ }\href {\doibase
  10.1063/5.0147190} {\bibfield  {journal} {\bibinfo  {journal} {Physics of
  Fluids}\ }\textbf {\bibinfo {volume} {35}},\ \bibinfo {pages} {055116}
  (\bibinfo {year} {2023})}\BibitemShut {NoStop}%
\bibitem [{\citenamefont {Tang}\ \emph
  {et~al.}(2020{\natexlab{b}})\citenamefont {Tang}, \citenamefont {Rabault},
  \citenamefont {Kuhnle}, \citenamefont {Wang},\ and\ \citenamefont
  {Wang}}]{tangRobust}%
  \BibitemOpen
  \bibfield  {author} {\bibinfo {author} {\bibfnamefont {H.}~\bibnamefont
  {Tang}}, \bibinfo {author} {\bibfnamefont {J.}~\bibnamefont {Rabault}},
  \bibinfo {author} {\bibfnamefont {A.}~\bibnamefont {Kuhnle}}, \bibinfo
  {author} {\bibfnamefont {Y.}~\bibnamefont {Wang}}, \ and\ \bibinfo {author}
  {\bibfnamefont {T.}~\bibnamefont {Wang}},\ }\bibfield  {title} {\enquote
  {\bibinfo {title} {Robust active flow control over a range of {{Reynolds}}
  numbers using an artificial neural network trained through deep reinforcement
  learning},}\ }\href {\doibase 10.1063/5.0006492} {\bibfield  {journal}
  {\bibinfo  {journal} {Physics of Fluids}\ }\textbf {\bibinfo {volume} {32}},\
  \bibinfo {pages} {053605} (\bibinfo {year} {2020}{\natexlab{b}})}\BibitemShut
  {NoStop}%
\bibitem [{\citenamefont {Ren}, \citenamefont {Rabault},\ and\ \citenamefont
  {Tang}(2021)}]{renApplying}%
  \BibitemOpen
  \bibfield  {author} {\bibinfo {author} {\bibfnamefont {F.}~\bibnamefont
  {Ren}}, \bibinfo {author} {\bibfnamefont {J.}~\bibnamefont {Rabault}}, \ and\
  \bibinfo {author} {\bibfnamefont {H.}~\bibnamefont {Tang}},\ }\bibfield
  {title} {\enquote {\bibinfo {title} {Applying deep reinforcement learning to
  active flow control in turbulent conditions},}\ }\href {\doibase
  10.1063/5.0037371} {\bibfield  {journal} {\bibinfo  {journal} {Physics of
  Fluids}\ }\textbf {\bibinfo {volume} {33}},\ \bibinfo {pages} {037121}
  (\bibinfo {year} {2021})},\ \Eprint {http://arxiv.org/abs/2006.10683}
  {2006.10683 [physics]} \BibitemShut {NoStop}%
\bibitem [{\citenamefont {Chen}\ \emph {et~al.}(2023)\citenamefont {Chen},
  \citenamefont {Wang}, \citenamefont {Yan}, \citenamefont {Hu},\ and\
  \citenamefont {Noack}}]{chen2023deep}%
  \BibitemOpen
  \bibfield  {author} {\bibinfo {author} {\bibfnamefont {W.}~\bibnamefont
  {Chen}}, \bibinfo {author} {\bibfnamefont {Q.}~\bibnamefont {Wang}}, \bibinfo
  {author} {\bibfnamefont {L.}~\bibnamefont {Yan}}, \bibinfo {author}
  {\bibfnamefont {G.}~\bibnamefont {Hu}}, \ and\ \bibinfo {author}
  {\bibfnamefont {B.~R.}\ \bibnamefont {Noack}},\ }\bibfield  {title} {\enquote
  {\bibinfo {title} {Deep reinforcement learning-based active flow control of
  vortex-induced vibration of a square cylinder},}\ }\href {\doibase
  10.1063/5.0152777} {\bibfield  {journal} {\bibinfo  {journal} {Physics of
  Fluids}\ }\textbf {\bibinfo {volume} {35}},\ \bibinfo {pages} {053610}
  (\bibinfo {year} {2023})}\BibitemShut {NoStop}%
\bibitem [{\citenamefont {Xia}\ \emph {et~al.}(2024)\citenamefont {Xia},
  \citenamefont {Zhang}, \citenamefont {Kerrigan},\ and\ \citenamefont
  {Rigas}}]{Xia2024}%
  \BibitemOpen
  \bibfield  {author} {\bibinfo {author} {\bibfnamefont {C.}~\bibnamefont
  {Xia}}, \bibinfo {author} {\bibfnamefont {J.}~\bibnamefont {Zhang}}, \bibinfo
  {author} {\bibfnamefont {E.}~\bibnamefont {Kerrigan}}, \ and\ \bibinfo
  {author} {\bibfnamefont {G.}~\bibnamefont {Rigas}},\ }\bibfield  {title}
  {\enquote {\bibinfo {title} {Active flow control for bluff body drag
  reduction using reinforcement learning with partial measurements},}\ }\href
  {\doibase 10.1017/jfm.2024.69} {\bibfield  {journal} {\bibinfo  {journal}
  {Journal of Fluid Mechanics}\ }\textbf {\bibinfo {volume} {981}},\ \bibinfo
  {pages} {A17} (\bibinfo {year} {2024})}\BibitemShut {NoStop}%
\bibitem [{\citenamefont {Wang}\ and\ \citenamefont
  {Xu}(2024{\natexlab{b}})}]{jia2024robust}%
  \BibitemOpen
  \bibfield  {author} {\bibinfo {author} {\bibfnamefont {J.}~\bibnamefont
  {Wang}}\ and\ \bibinfo {author} {\bibfnamefont {H.}~\bibnamefont {Xu}},\
  }\bibfield  {title} {\enquote {\bibinfo {title} {Robust and adaptive deep
  reinforcement learning for enhancing flow control around a square cylinder
  with varying reynolds numbers},}\ }\href {\doibase 10.1063/5.0207879}
  {\bibfield  {journal} {\bibinfo  {journal} {Physics of Fluids}\ }\textbf
  {\bibinfo {volume} {36}},\ \bibinfo {pages} {054103} (\bibinfo {year}
  {2024}{\natexlab{b}})}\BibitemShut {NoStop}%
\bibitem [{\citenamefont {Yan}\ \emph {et~al.}(2023)\citenamefont {Yan},
  \citenamefont {Li}, \citenamefont {Hu}, \citenamefont {Chen}, \citenamefont
  {Zhong},\ and\ \citenamefont {Noack}}]{yan2023stabilizing}%
  \BibitemOpen
  \bibfield  {author} {\bibinfo {author} {\bibfnamefont {L.}~\bibnamefont
  {Yan}}, \bibinfo {author} {\bibfnamefont {Y.}~\bibnamefont {Li}}, \bibinfo
  {author} {\bibfnamefont {G.}~\bibnamefont {Hu}}, \bibinfo {author}
  {\bibfnamefont {W.-l.}\ \bibnamefont {Chen}}, \bibinfo {author}
  {\bibfnamefont {W.}~\bibnamefont {Zhong}}, \ and\ \bibinfo {author}
  {\bibfnamefont {B.~R.}\ \bibnamefont {Noack}},\ }\bibfield  {title} {\enquote
  {\bibinfo {title} {Stabilizing the square cylinder wake using deep
  reinforcement learning for different jet locations},}\ }\href {\doibase
  10.1063/5.0171188} {\bibfield  {journal} {\bibinfo  {journal} {Physics of
  Fluids}\ }\textbf {\bibinfo {volume} {35}},\ \bibinfo {pages} {115104}
  (\bibinfo {year} {2023})}\BibitemShut {NoStop}%
\bibitem [{\citenamefont {Sohankar}, \citenamefont {Norberg},\ and\
  \citenamefont {Davidson}(1999)}]{Sohankar}%
  \BibitemOpen
  \bibfield  {author} {\bibinfo {author} {\bibfnamefont {A.}~\bibnamefont
  {Sohankar}}, \bibinfo {author} {\bibfnamefont {C.}~\bibnamefont {Norberg}}, \
  and\ \bibinfo {author} {\bibfnamefont {L.}~\bibnamefont {Davidson}},\
  }\bibfield  {title} {\enquote {\bibinfo {title} {{Simulation of
  three-dimensional flow around a square cylinder at moderate Reynolds
  numbers}},}\ }\href {\doibase 10.1063/1.869879} {\bibfield  {journal}
  {\bibinfo  {journal} {Physics of Fluids}\ }\textbf {\bibinfo {volume} {11}},\
  \bibinfo {pages} {288--306} (\bibinfo {year} {1999})}\BibitemShut {NoStop}%
\bibitem [{\citenamefont {You}\ and\ \citenamefont {Moin}(2008)}]{YOU20081349}%
  \BibitemOpen
  \bibfield  {author} {\bibinfo {author} {\bibfnamefont {D.}~\bibnamefont
  {You}}\ and\ \bibinfo {author} {\bibfnamefont {P.}~\bibnamefont {Moin}},\
  }\bibfield  {title} {\enquote {\bibinfo {title} {Active control of flow
  separation over an airfoil using synthetic jets},}\ }\href {\doibase
  https://doi.org/10.1016/j.jfluidstructs.2008.06.017} {\bibfield  {journal}
  {\bibinfo  {journal} {Journal of Fluids and Structures}\ }\textbf {\bibinfo
  {volume} {24}},\ \bibinfo {pages} {1349--1357} (\bibinfo {year} {2008})},\
  \bibinfo {note} {unsteady Separated Flows and their Control}\BibitemShut
  {NoStop}%
\bibitem [{\citenamefont {Yan}\ \emph {et~al.}(2024)\citenamefont {Yan},
  \citenamefont {Li}, \citenamefont {Liu},\ and\ \citenamefont
  {Hu}}]{yan2024aero}%
  \BibitemOpen
  \bibfield  {author} {\bibinfo {author} {\bibfnamefont {L.}~\bibnamefont
  {Yan}}, \bibinfo {author} {\bibfnamefont {Y.}~\bibnamefont {Li}}, \bibinfo
  {author} {\bibfnamefont {B.}~\bibnamefont {Liu}}, \ and\ \bibinfo {author}
  {\bibfnamefont {G.}~\bibnamefont {Hu}},\ }\bibfield  {title} {\enquote
  {\bibinfo {title} {{Aerodynamic force reduction of rectangular cylinder using
  deep reinforcement learning-controlled multiple jets}},}\ }\href {\doibase
  10.1063/5.0189009} {\bibfield  {journal} {\bibinfo  {journal} {Physics of
  Fluids}\ }\textbf {\bibinfo {volume} {36}},\ \bibinfo {pages} {025169}
  (\bibinfo {year} {2024})}\BibitemShut {NoStop}%
\bibitem [{\citenamefont {Jasak}\ \emph {et~al.}(2007)\citenamefont {Jasak},
  \citenamefont {Jemcov}, \citenamefont {Tukovic} \emph
  {et~al.}}]{jasakLibrary}%
  \BibitemOpen
  \bibfield  {author} {\bibinfo {author} {\bibfnamefont {H.}~\bibnamefont
  {Jasak}}, \bibinfo {author} {\bibfnamefont {A.}~\bibnamefont {Jemcov}},
  \bibinfo {author} {\bibfnamefont {Z.}~\bibnamefont {Tukovic}},  \emph
  {et~al.},\ }\bibfield  {title} {\enquote {\bibinfo {title} {Openfoam: A c++
  library for complex physics simulations},}\ }in\ \href
  {http://csabai.web.elte.hu/http/simulationLab/jasakEtAlOpenFoam.pdf} {\emph
  {\bibinfo {booktitle} {International Workshop on Coupled Methods in Numerical
  Dynamics}}},\ Vol.\ \bibinfo {volume} {1000}\ (\bibinfo {year} {2007})\ pp.\
  \bibinfo {pages} {1--20}\BibitemShut {NoStop}%
\bibitem [{\citenamefont {Jasak}(2009)}]{jasak2009}%
  \BibitemOpen
  \bibfield  {author} {\bibinfo {author} {\bibfnamefont {H.}~\bibnamefont
  {Jasak}},\ }\bibfield  {title} {\enquote {\bibinfo {title} {{{OpenFOAM}}:
  {{Open}} source {{CFD}} in research and industry},}\ }\href {\doibase
  10.2478/IJNAOE-2013-0011} {\bibfield  {journal} {\bibinfo  {journal}
  {International Journal of Naval Architecture and Ocean Engineering}\ }\textbf
  {\bibinfo {volume} {1}},\ \bibinfo {pages} {89--94} (\bibinfo {year}
  {2009})}\BibitemShut {NoStop}%
\bibitem [{\citenamefont {Sen}, \citenamefont {Mittal},\ and\ \citenamefont
  {Biswas}(2011)}]{sen2011flow}%
  \BibitemOpen
  \bibfield  {author} {\bibinfo {author} {\bibfnamefont {S.}~\bibnamefont
  {Sen}}, \bibinfo {author} {\bibfnamefont {S.}~\bibnamefont {Mittal}}, \ and\
  \bibinfo {author} {\bibfnamefont {G.}~\bibnamefont {Biswas}},\ }\bibfield
  {title} {\enquote {\bibinfo {title} {Flow past a square cylinder at low
  reynolds numbers},}\ }\href {\doibase 10.1002/fld.2416} {\bibfield  {journal}
  {\bibinfo  {journal} {Int. J. Numer. Meth. Fluids}\ }\textbf {\bibinfo
  {volume} {67}},\ \bibinfo {pages} {1160--1174} (\bibinfo {year}
  {2011})}\BibitemShut {NoStop}%
\bibitem [{\citenamefont {Sharma}\ and\ \citenamefont
  {Eswaran}(2004)}]{Sharma}%
  \BibitemOpen
  \bibfield  {author} {\bibinfo {author} {\bibfnamefont {A.}~\bibnamefont
  {Sharma}}\ and\ \bibinfo {author} {\bibfnamefont {V.}~\bibnamefont
  {Eswaran}},\ }\bibfield  {title} {\enquote {\bibinfo {title} {Heat and fluid
  flow across a square cylinder in the two-dimensional laminar flow regime},}\
  }\href {\doibase 10.1080/10407780490278562} {\bibfield  {journal} {\bibinfo
  {journal} {Numerical Heat Transfer, Part A: Applications}\ }\textbf {\bibinfo
  {volume} {45}},\ \bibinfo {pages} {247--269} (\bibinfo {year}
  {2004})}\BibitemShut {NoStop}%
\bibitem [{\citenamefont {Singh}\ \emph {et~al.}(2009)\citenamefont {Singh},
  \citenamefont {De}, \citenamefont {Carpenter}, \citenamefont {Eswaran},\ and\
  \citenamefont {Muralidhar}}]{Singh}%
  \BibitemOpen
  \bibfield  {author} {\bibinfo {author} {\bibfnamefont {A.~P.}\ \bibnamefont
  {Singh}}, \bibinfo {author} {\bibfnamefont {A.~K.}\ \bibnamefont {De}},
  \bibinfo {author} {\bibfnamefont {V.~K.}\ \bibnamefont {Carpenter}}, \bibinfo
  {author} {\bibfnamefont {V.}~\bibnamefont {Eswaran}}, \ and\ \bibinfo
  {author} {\bibfnamefont {K.}~\bibnamefont {Muralidhar}},\ }\bibfield  {title}
  {\enquote {\bibinfo {title} {Flow past a transversely oscillating square
  cylinder in free stream at low reynolds numbers},}\ }\href {\doibase
  https://doi.org/10.1002/fld.1979} {\bibfield  {journal} {\bibinfo  {journal}
  {International Journal for Numerical Methods in Fluids}\ }\textbf {\bibinfo
  {volume} {61}},\ \bibinfo {pages} {658--682} (\bibinfo {year}
  {2009})}\BibitemShut {NoStop}%
\bibitem [{\citenamefont {Cao}, \citenamefont {Ge},\ and\ \citenamefont
  {Tamura}(2012)}]{Shear}%
  \BibitemOpen
  \bibfield  {author} {\bibinfo {author} {\bibfnamefont {S.}~\bibnamefont
  {Cao}}, \bibinfo {author} {\bibfnamefont {Y.}~\bibnamefont {Ge}}, \ and\
  \bibinfo {author} {\bibfnamefont {Y.}~\bibnamefont {Tamura}},\ }\bibfield
  {title} {\enquote {\bibinfo {title} {Shear effects on flow past a square
  cylinder at moderate reynolds numbers},}\ }\href {\doibase
  10.1061/(ASCE)EM.1943-7889.0000309} {\bibfield  {journal} {\bibinfo
  {journal} {Journal of Engineering Mechanics}\ }\textbf {\bibinfo {volume}
  {138}},\ \bibinfo {pages} {116--123} (\bibinfo {year} {2012})}\BibitemShut
  {NoStop}%
\bibitem [{\citenamefont {Bai}\ and\ \citenamefont {Alam}(2018)}]{Dependence}%
  \BibitemOpen
  \bibfield  {author} {\bibinfo {author} {\bibfnamefont {H.}~\bibnamefont
  {Bai}}\ and\ \bibinfo {author} {\bibfnamefont {M.~M.}\ \bibnamefont {Alam}},\
  }\bibfield  {title} {\enquote {\bibinfo {title} {{Dependence of square
  cylinder wake on Reynolds number}},}\ }\href {\doibase 10.1063/1.4996945}
  {\bibfield  {journal} {\bibinfo  {journal} {Physics of Fluids}\ }\textbf
  {\bibinfo {volume} {30}},\ \bibinfo {pages} {015102} (\bibinfo {year}
  {2018})}\BibitemShut {NoStop}%
\bibitem [{\citenamefont {Haarnoja}\ \emph
  {et~al.}(2018{\natexlab{a}})\citenamefont {Haarnoja}, \citenamefont {Zhou},
  \citenamefont {Abbeel},\ and\ \citenamefont {Levine}}]{haarnoja2018soft}%
  \BibitemOpen
  \bibfield  {author} {\bibinfo {author} {\bibfnamefont {T.}~\bibnamefont
  {Haarnoja}}, \bibinfo {author} {\bibfnamefont {A.}~\bibnamefont {Zhou}},
  \bibinfo {author} {\bibfnamefont {P.}~\bibnamefont {Abbeel}}, \ and\ \bibinfo
  {author} {\bibfnamefont {S.}~\bibnamefont {Levine}},\ }\href@noop {}
  {\enquote {\bibinfo {title} {Soft actor-critic: Off-policy maximum entropy
  deep reinforcement learning with a stochastic actor},}\ } (\bibinfo {year}
  {2018}{\natexlab{a}}),\ \Eprint {http://arxiv.org/abs/1801.01290}
  {arXiv:1801.01290 [cs.LG]} \BibitemShut {NoStop}%
\bibitem [{\citenamefont {Haarnoja}\ \emph
  {et~al.}(2018{\natexlab{b}})\citenamefont {Haarnoja}, \citenamefont {Zhou},
  \citenamefont {Abbeel},\ and\ \citenamefont {Levine}}]{pmlrv80haarnoja18b}%
  \BibitemOpen
  \bibfield  {author} {\bibinfo {author} {\bibfnamefont {T.}~\bibnamefont
  {Haarnoja}}, \bibinfo {author} {\bibfnamefont {A.}~\bibnamefont {Zhou}},
  \bibinfo {author} {\bibfnamefont {P.}~\bibnamefont {Abbeel}}, \ and\ \bibinfo
  {author} {\bibfnamefont {S.}~\bibnamefont {Levine}},\ }\bibfield  {title}
  {\enquote {\bibinfo {title} {Soft actor-critic: Off-policy maximum entropy
  deep reinforcement learning with a stochastic actor},}\ }in\ \href@noop {}
  {\emph {\bibinfo {booktitle} {Proceedings of the 35th International
  Conference on Machine Learning}}},\ \bibinfo {series} {Proceedings of Machine
  Learning Research}, Vol.~\bibinfo {volume} {80},\ \bibinfo {editor} {edited
  by\ \bibinfo {editor} {\bibfnamefont {J.}~\bibnamefont {Dy}}\ and\ \bibinfo
  {editor} {\bibfnamefont {A.}~\bibnamefont {Krause}}}\ (\bibinfo  {publisher}
  {PMLR},\ \bibinfo {year} {2018})\ pp.\ \bibinfo {pages}
  {1861--1870}\BibitemShut {NoStop}%
\bibitem [{\citenamefont {Haarnoja}\ \emph
  {et~al.}(2019{\natexlab{a}})\citenamefont {Haarnoja}, \citenamefont {Zhou},
  \citenamefont {Hartikainen}, \citenamefont {Tucker}, \citenamefont {Ha},
  \citenamefont {Tan}, \citenamefont {Kumar}, \citenamefont {Zhu},
  \citenamefont {Gupta}, \citenamefont {Abbeel},\ and\ \citenamefont
  {Levine}}]{DBLP}%
  \BibitemOpen
  \bibfield  {author} {\bibinfo {author} {\bibfnamefont {T.}~\bibnamefont
  {Haarnoja}}, \bibinfo {author} {\bibfnamefont {A.}~\bibnamefont {Zhou}},
  \bibinfo {author} {\bibfnamefont {K.}~\bibnamefont {Hartikainen}}, \bibinfo
  {author} {\bibfnamefont {G.}~\bibnamefont {Tucker}}, \bibinfo {author}
  {\bibfnamefont {S.}~\bibnamefont {Ha}}, \bibinfo {author} {\bibfnamefont
  {J.}~\bibnamefont {Tan}}, \bibinfo {author} {\bibfnamefont {V.}~\bibnamefont
  {Kumar}}, \bibinfo {author} {\bibfnamefont {H.}~\bibnamefont {Zhu}}, \bibinfo
  {author} {\bibfnamefont {A.}~\bibnamefont {Gupta}}, \bibinfo {author}
  {\bibfnamefont {P.}~\bibnamefont {Abbeel}}, \ and\ \bibinfo {author}
  {\bibfnamefont {S.}~\bibnamefont {Levine}},\ }\href@noop {} {\enquote
  {\bibinfo {title} {Soft actor-critic algorithms and applications},}\ }
  (\bibinfo {year} {2019}{\natexlab{a}}),\ \Eprint
  {http://arxiv.org/abs/1812.05905} {arXiv:1812.05905 [cs.LG]} \BibitemShut
  {NoStop}%
\bibitem [{\citenamefont {Haarnoja}\ \emph
  {et~al.}(2019{\natexlab{b}})\citenamefont {Haarnoja}, \citenamefont {Zhou},
  \citenamefont {Hartikainen}, \citenamefont {Tucker}, \citenamefont {Ha},
  \citenamefont {Tan}, \citenamefont {Kumar}, \citenamefont {Zhu},
  \citenamefont {Gupta}, \citenamefont {Abbeel},\ and\ \citenamefont
  {Levine}}]{haarnoja2019soft}%
  \BibitemOpen
  \bibfield  {author} {\bibinfo {author} {\bibfnamefont {T.}~\bibnamefont
  {Haarnoja}}, \bibinfo {author} {\bibfnamefont {A.}~\bibnamefont {Zhou}},
  \bibinfo {author} {\bibfnamefont {K.}~\bibnamefont {Hartikainen}}, \bibinfo
  {author} {\bibfnamefont {G.}~\bibnamefont {Tucker}}, \bibinfo {author}
  {\bibfnamefont {S.}~\bibnamefont {Ha}}, \bibinfo {author} {\bibfnamefont
  {J.}~\bibnamefont {Tan}}, \bibinfo {author} {\bibfnamefont {V.}~\bibnamefont
  {Kumar}}, \bibinfo {author} {\bibfnamefont {H.}~\bibnamefont {Zhu}}, \bibinfo
  {author} {\bibfnamefont {A.}~\bibnamefont {Gupta}}, \bibinfo {author}
  {\bibfnamefont {P.}~\bibnamefont {Abbeel}}, \ and\ \bibinfo {author}
  {\bibfnamefont {S.}~\bibnamefont {Levine}},\ }\href@noop {} {\enquote
  {\bibinfo {title} {Soft actor-critic algorithms and applications},}\ }
  (\bibinfo {year} {2019}{\natexlab{b}}),\ \Eprint
  {http://arxiv.org/abs/1812.05905} {arXiv:1812.05905 [cs.LG]} \BibitemShut
  {NoStop}%
\bibitem [{\citenamefont {Protas}\ and\ \citenamefont
  {Wesfreid}(2002)}]{protas2002drag}%
  \BibitemOpen
  \bibfield  {author} {\bibinfo {author} {\bibfnamefont {B.}~\bibnamefont
  {Protas}}\ and\ \bibinfo {author} {\bibfnamefont {J.~E.}\ \bibnamefont
  {Wesfreid}},\ }\bibfield  {title} {\enquote {\bibinfo {title} {Drag force in
  the open-loop control of the cylinder wake in the laminar regime},}\ }\href
  {\doibase 10.1063/1.1432695} {\bibfield  {journal} {\bibinfo  {journal}
  {Physics of Fluids}\ }\textbf {\bibinfo {volume} {14}},\ \bibinfo {pages}
  {810--826} (\bibinfo {year} {2002})}\BibitemShut {NoStop}%
\bibitem [{\citenamefont {Kim}\ and\ \citenamefont {Kim}(2009)}]{KIM2009172}%
  \BibitemOpen
  \bibfield  {author} {\bibinfo {author} {\bibfnamefont {S.~H.}\ \bibnamefont
  {Kim}}\ and\ \bibinfo {author} {\bibfnamefont {C.}~\bibnamefont {Kim}},\
  }\bibfield  {title} {\enquote {\bibinfo {title} {Separation control on
  naca23012 using synthetic jet},}\ }\href {\doibase
  https://doi.org/10.1016/j.ast.2008.11.001} {\bibfield  {journal} {\bibinfo
  {journal} {Aerospace Science and Technology}\ }\textbf {\bibinfo {volume}
  {13}},\ \bibinfo {pages} {172--182} (\bibinfo {year} {2009})}\BibitemShut
  {NoStop}%
\bibitem [{\citenamefont {Rabault}\ and\ \citenamefont
  {Kuhnle}(2019{\natexlab{b}})}]{rabault2024cylinder2dflowcontroldrlparallel}%
  \BibitemOpen
  \bibfield  {author} {\bibinfo {author} {\bibfnamefont {J.}~\bibnamefont
  {Rabault}}\ and\ \bibinfo {author} {\bibfnamefont {A.}~\bibnamefont
  {Kuhnle}},\ }\href@noop {} {\enquote {\bibinfo {title}
  {Cylinder2dflowcontroldrlparallel},}\ }\bibinfo {howpublished} {GitHub.
  \url{https://github.com/jerabaul29/Cylinder2DFlowControlDRLParallel}}
  (\bibinfo {year} {2019}{\natexlab{b}})\BibitemShut {NoStop}%
\bibitem [{\citenamefont {Li}(2021)}]{Nek5000}%
  \BibitemOpen
  \bibfield  {author} {\bibinfo {author} {\bibfnamefont {J.}~\bibnamefont
  {Li}},\ }\href@noop {} {\enquote {\bibinfo {title} {Rlcontrolnek5000},}\
  }\bibinfo {howpublished} {GitHub.
  \url{https://github.com/npuljc/RL_control_Nek5000}} (\bibinfo {year}
  {2021})\BibitemShut {NoStop}%
\bibitem [{\citenamefont {Qiulei~Wang}(2022)}]{DRLinFluids}%
  \BibitemOpen
  \bibfield  {author} {\bibinfo {author} {\bibfnamefont {W.~C.}\ \bibnamefont
  {Qiulei~Wang}, \bibfnamefont {Lei~Yan}},\ }\href@noop {} {\enquote {\bibinfo
  {title} {Drlinfluids},}\ }\bibinfo {howpublished} {GitHub.
  \url{https://github.com/venturi123/DRLinFluids}} (\bibinfo {year}
  {2022})\BibitemShut {NoStop}%
\end{thebibliography}%

\end{document}